\documentclass[fleqn,usenatbib]{rasti}

\usepackage{newtxtext,newtxmath}

\usepackage[T1]{fontenc}

\DeclareRobustCommand{\VAN}[3]{#2}
\let\VANthebibliography\thebibliography
\def\thebibliography{\DeclareRobustCommand{\VAN}[3]{##3}\VANthebibliography}

\usepackage{graphicx}	
\usepackage{amsmath}	
\usepackage{graphicx}
\usepackage{subcaption}
\usepackage{comment}
\usepackage{pdflscape}	

\title[Star-forming clump detection in nearby galaxies using Faster R-CNN]{Star-forming clump detection in nearby galaxies using Faster R-CNN and \textit{ugrizy} imaging data from CLAUDS and HSC-SSP}

\author[J. J. Popp et al.]{
Jürgen J. Popp,$^{1}$\thanks{E-mail: jurgen.popp@open.ac.uk}
Hugh Dickinson,$^{1}$
Stephen Serjeant,$^{1}$
Lucy F. Fortson,$^{2}$
Tobias Géron,$^{3}$
\newauthor
Brooke D. Simmons$^{4}$
and Vihang Mehta$^{5}$
\\
$^{1}$School of Physical Sciences, The Open University, Milton Keynes, MK7 6AA, UK\\
$^{2}$School of Physics and Astronomy, University of Minnesota, 116 Church Street SE, Minneapolis, MN 55455, USA\\
$^{3}$Dunlap Institute for Astronomy and Astrophysics, University of Toronto, 50 St. George Street, Toronto, ON M5S 3H4, Canada\\
$^{4}$Physics Department, Lancaster University, Lancaster, LA1 4YB, UK\\
$^{5}$IPAC, Mail Code 314-6, California Institute of Technology, 1200 E. California Blvd., Pasadena, CA, 91125, USA\\
}

\date{Accepted XXX. Received YYY; in original form ZZZ}

\pubyear{\the\year{}}

\begin{document}
\label{firstpage}
\pagerange{\pageref{firstpage}--\pageref{lastpage}}
\maketitle

\begin{abstract}
Giant Star-forming Clumps (GSFCs) are kpc-scale regions of enhanced star-formation with stellar masses of $10^7$ to $10^9\,M_\odot$ that are commonly observed in high-redshift galaxies but are rarely detected in low-redshift ($z\lesssim0.5$) galaxy analogues. However, the availability of wide-field galaxy survey data makes it possible to identify potential star-forming clumps in large samples of low-redshift galaxies using object detection models that are based on Deep Learning (DL) techniques. We apply a novel DL-based object detection model to galaxies observed by the Hyper Suprime-Cam Subaru Strategic Survey (HSC-SSP) and CFHT Large Area U-band Deep Survey (CLAUDS). Our model is based on the the Faster Region-Based Convolutional Neural Network (Faster R-CNN or FRCNN) object detection framework but expanded to process the six \textit{ugrizy} filter band images simultaneously and identify not only clumps and their locations in the host galaxy but also additional contaminants. By adopting the \textsc{Zoobot} foundation DL-model as a feature extraction backbone, we also demonstrate one of the first applications of \textsc{Zoobot} in a downstream task for object detection. Our model achieves a detection completeness of $\gtrsim 0.9$ and purity of $\gtrsim 0.8$ which were validated on a large set of real galaxies into which simulated clumps were injected.
\end{abstract}




\section{Introduction}\label{sec:intro}
Giant Star-Forming Clumps (GSFCs) are kpc-scale morphological substructures with high surface brightness that are frequently observed in star-forming galaxies (SFGs) at the peak of cosmic star-formation ($z\sim2$). These star-forming clumps appear to be much larger and brighter than typical star-forming regions in local SFGs with estimated stellar masses of $10^7$ to $10^9\,M_\odot$ \citep[e.g.][]{FoersterSchreiber2011a,Guo2012,DessaugesZavadsky2017,Soto2017,Guo2018,Zanella2019,HuertasCompany2020,Kalita2025}. The star-formation rates (SFRs) of the clumps range from $10^{-4}$ to $10^2\,M_\odot \,{\mathrm yr}^{-1}$ \citep[e.g.][]{Genzel2011,Guo2012,Elmegreen2013,Guo2018,Soto2017} resulting in specific star-formation rates (sSFRs) that are several times higher than their neighbouring areas in the galaxy.

The fraction of clumpy SFGs appears to peak around $z\sim 2$ \citep{Murata2014,Guo2015,Shibuya2016,Guo2018} with $\gtrsim 60\%$ of galaxies showing clumpy substructures, but this fraction seems to decrease with decreasing redshift \citep[e.g.][]{Adams2022}. However, only a few studies have observed GSFCs in low-redshift or nearby galaxies ($z<0.5$) and only small samples of clumps have been analysed \citep[e.g.][]{Overzier2009,Fisher2014,Messa2019,Mehta2021,Lenkic2021,Adams2022}. Due to the scarcity of observed clumpy galaxies at low redshifts, the evolution and properties of clumps have not been fully studied and robust measurements of the physical properties for large samples of clumps are still missing.

To detect and select potential star-forming clumps in the small samples of a few hundred low- or high-redshift galaxies, most studies have applied different transformations of the observed rest-UV/optical imaging data to obtain high contrast images that emphasise regions in the target galaxy with locally increased surface brightness. These regions have then been identified using source extraction or peak finding algorithms \citep[e.g.][]{Guo2015,Fisher2017,Mehta2021,Lenkic2021,Mestric2022,Kalita2025,Sok2025a}. Also, H$\alpha$-line emission maps have been used to extract clumpy star-forming regions using sophisticated algorithms \citep[e.g.][]{EspinosaPonce2020,EspinosaPonce2022,LugoAranda2022,Werle2024}. Other studies have relied on visual identification by experts \citep[e.g.][]{Elmegreen2007a,Overzier2009,Claeyssens2023}, which is limited to even smaller galaxy samples. Only a few recent studies have analysed clump detections in samples that contain $>1500$ galaxies, either with the help of volunteers participating in citizen science projects like the \textit{Galaxy Zoo: Clump Scout} project \citep[GZCS,][]{Adams2022,Dickinson2022} or by applying Deep Learning (DL) methods to extract the clump locations from the target galaxies \citep{HuertasCompany2020,Popp2024,Adams2025}.

Different DL-model architectures have been developed that can detect and identify instances of semantic objects in images \citep[e.g.][]{Girshick2015,Liu2015,Ren2015,Redmon2016,Shehzadi2025}. The architecture we use in this study is the Faster Region-Based Convolutional Neural Network \citep[Faster R-CNN or FRCNN,][]{Ren2015}. The FRCNN relies on a convolutional neural network (CNN) as a feature extraction backbone that extracts the features from the input images, and provides crucial inputs for the Region Proposal Network (RPN) and the region classifier or detector network. Both sub-networks use the same pretrained CNN, sharing its convolutional layers during an alternating training loop for the region proposal and detection task. The final output of a FRCNN model is a collection of rectangular bounding boxes identifying groups of pixels in the image that contain objects and a classification identifying the type of object that each box contains.

Based on the success in using a pretrained CNN for feature extraction \citep{Popp2024}, we adopted the \textsc{Zoobot} foundation deep learning model \citep{Walmsley2023} in its latest version v2.0 as the feature extraction backbone. \textsc{Zoobot} has advanced to become one of the first foundation models in astronomy \citep{Walmsley2024} and it is used here in downstream tasks for object detection for the first time.

This paper is organised as follows. Section \ref{sec:data} describes the galaxy sample used in this work. In Section \ref{sec:frcnn_development}, we present the development of the DL-based clump detection model and quantify its performance during the training process. We then explain how we postprocessed the model detections (Section \ref{sec:frcnn_model_postprocess}) and how we extracted the locations of potential clumps or clump candidates from the model detections (Section \ref{sec:frcnn_model_peaks}). In Section \ref{sec:results} we describe the evaluation of the detection results using artificially generated star-forming clumps that were injected into the real galaxy images. Based on the artificial clumps and their simulated physical properties we quantify the completeness and purity of the model detections. We discuss the implications of our findings in Section \ref{sec:discussion} and conclude with a short summary in Section \ref{sec:conclusion}.

Throughout this paper, all magnitudes are expressed in the AB system \citep{Oke1983}. For simplicity, we use the terms `low-redshift' for a redshift range of $z\leq 0.5$ and `high-redshift' for a range of $z>0.5$. Logarithmic quantities are either referenced to a base of $10$ using the notation $\log$ or to a base of $e$ using the notation $\ln$. In this work we adopt the Planck 2015 \citep{Ade2016} cosmological parameters with $(\Omega_m, \Omega_{\Lambda}, h) = (0.31, 0.69, 0.68)$.

\section{Data}\label{sec:data}
To train our models we used a combination of the Canada-France-Hawaii Telescope (CHFT) Large Area U-band Deep Survey \citep[CLAUDS,][]{Sawicki2019} and the Hyper-Suprime-Cam (HSC) Subaru Strategic Program \citep[HSC-SSP,][]{Aihara2017} imaging data over their four deep fields. HSC-SSP is a multiband, three-layered imaging survey that covers $1400\,\mathrm{deg}^2$ (Wide Survey), $27\,\mathrm{deg}^2$ (Deep Survey) and $3.5\,\mathrm{deg}^2$ (Ultra-Deep Survey) of the sky. Observations were made with the HSC on the $8.2\,\mathrm{m}$ Subaru Telescope using five broadband (\textit{grizy}) and additional narrowband filters with $5\sigma$ point source depths in the \textit{r}-band of $\sim 26\,m_{\mathrm{AB}}$, $\sim 27\,m_{\mathrm{AB}}$ and $\sim 28\,m_{\mathrm{AB}}$ for the Wide, Deep and Ultra-Deep Survey, respectively. CLAUDS provides \textit{u}-band imaging data for those areas where it overlaps with the HSC-SSP survey for the XMM-LSS, E-COSMOS, ELAIS-N1 and DEEP2-3 fields. The \textit{u}-band imaging data was acquired using two different filters on the CFHT MegaCam. The older $u^\star$ filter was mainly used for the XMM-LSS field and later replaced by the $u$ filter which was used for the E-COSMOS, ELAIS-N1, DEEP2-3 fields. The two different \textit{u}-band filters cover slightly different wavelength ranges and are treated differently throughout the image processing and photometry related tasks described in this paper.


The HSC-SSP imaging data has been processed by the HSC pipeline \citep[HSCpipe,][]{Bosch2017}, which has also been used to process the CLAUDS images \citep{Sawicki2019}. Apart from the imaging data, both surveys provide catalogues with multiple photometry measurements in the different filter bands and inferred physical properties of the detected source objects, which are described in \citet{Desprez2023} for the CLAUDS catalogue and in \citet{Aihara2022} for the HSC-SSP PDR3 catalogue. For consistency, we use and report the published source properties from the HSC-SSP PDR3 catalogue as they are provided for all objects in our selected sample. All object identifiers used in this work refer to the object IDs in the HSC-SSP PDR3 catalogue.

\subsection{Galaxy sample}\label{sec:data_gal_sample}
We started with a preselection of target galaxies by querying the Sloan Digital Sky Survey (SDSS) Data Release 18 \citep[SDSS DR18,][]{Almeida2023} catalogue for galaxies with spectroscopic redshift (spec-z) and photometric redshift (photo-z) using the following criteria:
\begin{itemize}
    \item \textbf{spec-z}:
    \begin{itemize}
        \item redshift: $0.0005 \leq z \leq  0.5$
        \item only clean photometry: \texttt{clean} = 1
        \item only galaxies: \texttt{type} = 3
        \item ensure minimum extent of the galaxy: $\texttt{petroR90\_r}\geq3\,\mathrm{arcsec}$
    \end{itemize}
    \item \textbf{photo-z}:
    \begin{itemize}
        \item redshift: $0.0005 \leq z \leq  0.5$
        \item only best redshift measurement: \texttt{photoErrorClass} = 1
        \item only clean photometry: $\texttt{clean} = 1$
        \item only galaxies: $\texttt{type} = 3$
        \item ensure minimum extent of the galaxy: $\texttt{petroR90\_r}\geq3\,\mathrm{arcsec}$
        \item limit overlap with spec-z sources: \texttt{petroMag\_r} $\geq 17.8\,m_{\mathrm{AB}}$
    \end{itemize}
\end{itemize}

The SDSS photo-z sample was magnitude-limited (apparent SDSS \textit{r}-band magnitudes of $r_{\mathrm{SDSS}} \geq 17.8\,m_{\mathrm{AB}}$) to reduce the overlap with the SDSS spec-z sample which is 90-95\% complete for $r_{\mathrm{SDSS}} < 17.8\,m_{\mathrm{AB}}$ \citep{Strauss2002}. Besides the quality flags, the galaxies were also required to have a minimum extent so that morphological features are resolvable (SDSS \textit{r}-band 90\% Petrosian radius $\texttt{petroR90\_r}\geq3\,\mathrm{arcsec}$).

We crossmatched (within 1.0 arcsec) our SDSS preselection with sources detected in the \textit{i}-band from the latest Public Data Release 3 \citep[PDR3,][]{Aihara2022} of the HSC-SSP Wide survey. We chose to use the data from the HSC-SSP Wide survey instead of the Deep and Ultra-Deep surveys because by training our models on the shallower imaging data with only \textit{grizy} photometry, they can be used to detect potential star-forming clumps over the much wider sky area of the HSC-SSP Wide survey. We applied multiple quality flags \citep[see][for details]{Bosch2017,Aihara2022} to ensure a selection of galaxies with clean photometry:
\begin{itemize}
    \item \texttt{isprimary} = True
    \item \texttt{(g|r|i|z|y)\_inputcount\_value} $\geq 0.5$
    \item \texttt{(g|r|i|z|y)\_extendedness\_value} $\geq 0.5$
    \item \texttt{(g|r|i|z|y)\_cmodel\_flag} = False
    \item \texttt{(g|r|i|z|y)\_pixelflags\_saturatedcenter} = False
    \item \texttt{(g|r|i|z|y)\_inputcount\_flag\_noinputs} = False
    \item \texttt{(g|r|i|z|y)\_sdssshape\_flag\_badcentroid} = False
    \item \texttt{(g|r|i|z|y)\_psfflux\_flag} = False
    \item \texttt{(g|r|i|z|y)\_pixelflags\_crcenter} = False
    \item \texttt{(g|r|i|z|y)\_pixelflags\_edge} = False
    \item \texttt{(g|r|i|z|y)\_pixelflags\_bad} = False
    \item \texttt{(g|r|i|z|y)\_mask\_brightstar\_halo} = False
    \item \texttt{(g|r|i|z|y)\_mask\_brightstar\_ghost} = False
    \item \texttt{(g|r|i|z|y)\_mask\_brightstar\_blooming} = False
\end{itemize}

This step resulted in 710,271 HSC-SSP objects being selected. To add \textit{u}-band photometry data to the selected \textit{grizy} galaxies we then crossmatched the HSC-SSP galaxies with the CLAUDS catalogue using the same crossmatching distance of 1.0 arcsec. For the four deep fields, we were able to match 14,231 galaxies with clean photometry so that the final set of 710,271 galaxies includes 14,231 galaxies with six filter band photometry data (\textit{ugrizy}) and 696,040 galaxies with only five filter band photometry data (\textit{grizy}). The final sample is magnitude-limited by the median $5\sigma$ depth for SDSS photometric observations in the $r_{\mathrm{SDSS}}$-band magnitude of $r_{\mathrm{SDSS}} \leq 22.7\,m_{\mathrm{AB}}$ and limited to a maximum redshift of $z \leq 0.5$.

\subsection{Preprocessing of the galaxy images}\label{sec:data_preprocess}
For each target galaxy we created square image cutouts with an edge length of twice the SDSS \textit{r}-band 90\% Petrosian radius in arcsec ($2 \times \texttt{petroR90\_r}$) of each object to ensure a comparable visual size of the target galaxies in each cutout. The images were rescaled by applying an inverse hyperbolic sine function (asinh):
\begin{equation}\label{eq:hsc_data_preprocess_asinh}
    I^{\prime }_{x} = \frac{ \textrm{asinh}(e^{10}\,I_{x}) / \textrm{asinh}(e^{10}) + 0.05 }{0.72},
\end{equation}
where $I_x$ is the input pixel intensity in band $x$ and $I^{\prime }_{x}$ is the scaled pixel intensity. We adopted this rescaling of the images as it was used to generate the RGB-composite images from the \textit{i}-, \textit{r}- and \textit{g}-filter band images (HSC-colours\footnote{\url{https://hsc-gitlab.mtk.nao.ac.jp/ssp-software/data-access-tools/tree/master/pdr3/colorPostage}}) that were used to pretrain our feature extraction backbone (\textsc{Zoobot}) on HSC-SSP galaxies from the \textit{Galaxy Zoo: Cosmic Dawn} project \citep{Pearson2025}. We expected to better benefit from the already `learned' features using the same rescaling which we applied to all of the (six) five (\textit{u})\textit{grizy}-filter band images of the galaxy sample. The rescaled images were stored separately as greyscale PNG image files for each filter band and were also resized to $400\times 400$ pixels in a final processing step. The resizing of the images changed the original pixel scale of $0.168\,\mathrm{px}/\mathrm{arcsec}$ \citep{Aihara2017,Sawicki2019} of the cutouts but ensured a constant pixel size for faster processing by our object detection model.

We also generated RGB-composite images using the HSC-colours to be used in manually creating the labelled training data during the development process of the clump detector (Section \ref{sec:data_training}). Furthermore, additional RGB-composite images were created that stress the \textit{g}-filter band so that possible star-forming regions are visually emphasised to help in the identification of clump locations in the training data. The \textit{i}-, \textit{r}- and \textit{g}-filter band images were mapped to the red, green and blue channels and scaled using `Lupton'-scaling \citep{Lupton2004}:
\begin{eqnarray}\label{eq:lupton}
  I^{\prime }_{x} = \frac{1}{Q}\mathrm{asinh}\left[Q\cdot \frac{\left(\frac{I_{x}}{\beta _{x}}-m\right)}{\alpha }\right],
\end{eqnarray}
where $Q=7.0$, stretch $\alpha=0.2$, minimum $m=0.0$ and channel scales $\beta_{i}=1.818$, $\beta_{r}=1.17$ and $\beta_{g}=0.7$ that are specific to each input filter band. The RGB-composite images were also resized to $400 \times 400$ pixels, which means that the visual size of each target galaxy is similar but with varying resolutions.

\section{Developing the object detection model}\label{sec:frcnn_development}
The object detection models trained to locate and classify potential star-forming clumps in galaxies observed by CLAUDS and HSC-SSP are based on the FRCNN framework described by \citet{Ren2015}. To enable clump detections on the multi-band data from CLAUDS and HSC-SSP we expanded the FRCNN model architecture to accept five and six channel data as input (Section \ref{sec:frcnn_model_arch}). Furthermore, the detection model was modified to identify not only clumps and their locations in the host galaxy but also additional contaminants. This required training data with labels that match the extended classification scheme (Section \ref{sec:data_training}) which can then be used during the model training process (Section \ref{sec:frcnn_model_training}). By evaluating the model performance at each step of the training process the final model setup and its parameters were determined (Section \ref{sec:frcnn_model_performance}).

\subsection{Model architecture}\label{sec:frcnn_model_arch}
\subsubsection{Multichannel input}
As is the case for most image classification and object detection models applied to astronomical objects, \textsc{Zoobot} is built to process either single-channel images that represent data obtained from a single bandpass filter or RGB-composite images combined from three bandpass filters that are closest to the optical red, green and blue wavelength ranges. However, observational data is available from more than three filter bands. To make as much use of the available data, some studies \citep[e.g.][]{HuertasCompany2020} construct classification or detection models for each available filter band and then combine the results of all models to a single prediction. An alternative approach is to use all available filter bands together as input into a single model. CNNs like \textsc{Zoobot} can be easily adapted to this second approach because each of the convolutional layers can be extended to transform any input image or feature map with an input depth of $d_{\mathrm{depth}}>3$ into a corresponding output feature map. 

\textsc{Zoobot} is available from HuggingFace\footnote{\url{https://huggingface.co/mwalmsley}} as pretrained models for various CNN architectures. We use the \textsc{Zoobot} version based on the ResNet50 architecture \citep{He2016}. Briefly, the ResNet50 architecture is built from 16 convolutional blocks each consisting of three convolutional layers with different specifications, a fully connected layer at the end and an input block at the beginning. This first block takes the imaging data as input and processes it through a first convolutional layer followed by a batch normalisation and max-pooling layer. To be able to use the imaging data from the five \textit{grizy} and six \textit{ugrizy} filter bands as input we adjusted the convolutional, batch normalisation and max-pooling layer of the first block to input depths of $d_{\mathrm{depth}}=5$ and $d_{\mathrm{depth}}=6$, respectively. The resulting output feature map has the same dimensions as the output feature map of the first block from an unmodified ResNet50 model so that no changes to the remaining convolutional blocks and the final fully connected layer were needed.

The convolutional layer from the first block of the original ResNet50 model was also initialised with pretrained weights from \textsc{Zoobot}. However, as pretrained weights only exist for a convolutional layer that takes imaging data with three channels ($d_{\mathrm{depth}}=3$) as input, the additional channels of the modified ResNet50 model could not make use of the existing pretrained weights. We then copied the three pretrained channel weights that are mapped to the \textit{g}-, \textit{r} and \textit{i}-filter band to the additional channels so that the new \textit{u}-filter band channel was initialised with the \textit{g}-filter band weights and the new \textit{z}- and \textit{y}-filter band with the \textit{i}-filter band weights. In doing so we tried to use the pretrained \textsc{Zoobot} weights (at least to some degree) by initialising the \textit{u}-filter band channel with the weights from the `bluest' channel of the original ResNet50 model and the \textit{z}- and \textit{y}-filter band channels with the weights from the `reddest' original channels. 

\subsubsection{Multiclassification scheme}
Increasing the number of object classes for the detection model from a simple clump/non-clump classification to a classification scheme with additional non-clump object classes can improve the clump detections as the model can be trained not only on true positive examples but also on these additional true negative examples. Our previous FRCNN model that was trained to detect clumps in SDSS galaxies \citep{Popp2024} has already shown that having a second object class (`odd' clump) in addition to the main classes (clump and generic background) does help to separate clumps from contaminating objects like foreground stars. We extended this approach to a classification scheme with six classes that include object classes for clumps, `odd' clumps, foreground stars, fore-/background galaxies, bulges and the generic background (see also Section \ref{sec:data_training}). These objects are frequently observed around or blended into the target galaxies and can lead to model detections that are misclassified as clumps and contaminate the detection results. Therefore, we set up the final fully connected layer of the detector network so that the model output contains classification scores for six object classes.

\subsubsection{Six and five channel models}
We constructed two different FRCNN models that use the \textsc{Zoobot} foundation model as the feature extraction backbone and were trained to identify the location of clumps and other contaminating objects: (1) a 6-channel model that can be applied to \textit{ugrizy} data from the combined CLAUDS/HSC-SSP galaxy set and (2) a 5-channel model for the HSC-SSP galaxy set that uses only imaging data from the five \textit{grizy} filter bands and for which the \textit{u}-filter band channel with the copied \textit{g}-filter band channel weights is omitted. Each of the RPNs was initialised with default anchor box sizes of $32 \times 32$, $64 \times 64$, $128 \times 128$, $256 \times 256$ and $512 \times 512$ pixels and aspect ratios of $0.5$, $1.0$ and $2.0$. Also, all images are normalised by the model before being passed to the feature extraction backbone. The normalisation is applied to each image per channel $c$:
\begin{equation}
    x'_{i,c} = \frac{x_{i,c}-\mu_c}{\sigma_c},
\end{equation}
where $x'_{i,c}$ is the normalised pixel value at position $i$, $x_{i,c}$ the original pixel value at the same position, $\mu_c$ the mean and $\sigma_c$ the standard deviation of the pixel values in all images per channel $c$. The mean $\mu_c$ and standard deviation $\sigma_c$ are hyper-parameters to the model and are set during its initialisation.

\subsection{Training data}\label{sec:data_training}
To train our models, we manually labelled and marked star-forming clumps and potential contaminants in a subset of our galaxy images. However, instead of marking clump locations on `blank' galaxy images with no initial annotations, we overlaid the galaxy images with a first set of possible clump detections. This set of possible clump detections were the predictions from a model with the same architecture as the final 6-channel model (Section \ref{sec:frcnn_model_arch}) but only trained on 870 CLAUDS/HSC-SSP galaxies that could be crossmatched with galaxies from the \textit{Galaxy Zoo: Clump Scout} project \citep[GZCS,][]{Adams2022,Dickinson2022} and have existing clump markings from the volunteers. By adopting a limited `correct-a-machine' approach, we expected that the final detection model will benefit from being trained not only on new clump labels but also on its previous errors. Correcting features in the galaxy image that were wrongly classified as a clump will increase the likelihood that the retrained clump detector will better identify the differences between a clump and other features/objects in the galaxy image. This statement also applies to any corrections made to the bounding boxes comprising the first set of clump detections during the manual annotation process as the final models will be better trained to detect the true extent of a clump. However, we note that even this limited `correct-a-machine' approach can lead to biases in the training data as volunteers might accept the machine suggestion more easily if the effort to correct the suggestion is perceived as high \citep[e.g.][]{Beck2026}.

We first selected 3,198 galaxies from our CLAUDS/HSC-SSP galaxy sample as training images. From those, we randomly selected 2,000 star-forming ($\log(\mathrm{sSFR}/\mathrm{yr}^{-1}) \geq -11.0$) galaxies with a redshift of $z\leq 0.5$, a minimum SDSS \textit{r}-band 90\% Petrosian radius of $\texttt{petroR90\_r} \geq 13\,\mathrm{arcsec}$ and a maximum seeing FWHM in the \textit{i}-band of $\leq 0.7\,\mathrm{arcsec}$. These requirements ensure that the selected galaxies were observed with a spatial resolution that is suitable for annotating clumps in the images. In addition to the randomly selected galaxies, we added the 870 galaxies that were crossmatched with the GZCS galaxies and were annotated by volunteers. A final set of 328 galaxies was added following visual comparison of CLAUDS \textit{u}-band and HSC-SSP \textit{g}-band galaxy images. Those galaxies mainly contain clumps that are bright in the \textit{u}-band image but much fainter in the \textit{g}-band image. Such cases, where a clump is faint in the images with only \textit{grizy} photometry, will help train the 5-channel model to learn features that might be faint in the redder filter bands.

We created a private Zooniverse\footnote{\url{https://www.zooniverse.org/}} project similar to the GZCS project so that the annotations of clumps is supported by a well established workflow. Each galaxy is shown as a single subject (i.e. on the same screen) with four different images (see Section \ref{sec:data_preprocess}) to help with the visual identification of potential star-forming clumps: (1) a RGB-composite image using HSC-colours, (2) a RGB-composite image with a different scaling of the \textit{g}-band to emphasise star-forming regions, (3) an \textit{u}-band image with simple asinh-stretch and (3) the same \textit{u}-band image with asinh-stretch but where the maximum level of the pixel values is cut at the 99th percentile to emphasise low-surface-brightness features of the target galaxy. Tools available in the user interface allow the user to add or remove a clump annotation, change the size of the bounding box around a clump or to change the classification of a marked morphological feature.


Furthermore, the annotation process was not limited to the identification of clumps but also included the identification of other contaminating features visible in or around the target galaxy. These contaminants include:
\begin{enumerate}
    \item \textbf{`odd' clumps} - objects that appear like clumps but are likely image artifacts or other anomalies,
    \item \textbf{foreground stars} - stars in the Milky Way that blend into the target galaxy,
    \item \textbf{fore-/background galaxies} - smaller galaxies in front of the target galaxy, distant galaxies behind the target galaxies or satellite galaxies of the target galaxy,
    \item \textbf{(secondary) bulges} - bright bulges of close-by galaxies or merging galaxies (also used to correct the initial clump predictions if the central bulge of the target galaxy is marked as a clump).
\end{enumerate}
We note, however, that the training objective for the object detection model was the accurate detection of star-forming clumps and not the complete detection of all features visible in the galaxy images. Therefore, the focus during the labelling process for the training data was the correct identification of clumps and the corrections made to incorrect classifications among the overlaid model predictions. 

The final annotations consist of 16,165 identified features in 3,198 galaxies. Of those, 3,701 features were identified as clumps that are located in 1,001 (clumpy) galaxies. The `correct-a-machine' approach led to a large sample of true negative objects and galaxies that contain no clumps at all because many objects predicted to be clumps by the initial model were removed or had to be corrected to one of the non-clump objects (contaminants). Together with the rescaled PNG images for each of the \textit{ugrizy}-filter bands (Section \ref{sec:data_preprocess}), the annotations form the final training data set that we used to train the 6-channel and 5-channel FRCNN models.

\subsection{Model training}\label{sec:frcnn_model_training}
The 5-channel and 6-channel FRCNN models were separately trained in fine-tuning and transfer learning modes using the training data sets described in Section \ref{sec:data_training}. In the fine-tuning mode, the weights of some or all convolutional blocks of the feature extraction backbone are allowed to vary during model training whereas in transfer learning mode the weights are fixed to their initial value. The additional layers of the RPN and the detector network are allowed to adjust during model training in both training modes.

For the fine-tuning mode we require the model to be able to adjust the weights of all blocks of the feature extraction backbone during the training process because we specifically extended the first convolutional block of the feature extractor. Therefore, we allow the weights for the input convolutional layer and the four main convolutional blocks of the ResNet50 architecture to vary. 

We trained both models over 150 epochs in each training mode. For each run, we divided the training data into random train and validation splits that contain 80\% and 20\% of the total training data, respectively. The training images were augmented by random horizontal and vertical flips before being passed on to the model itself. Only the five \textit{grizy}-filter band images were passed to the 5-channel model while the 6-channel model was trained on all six \textit{ugrizy}-filter band images. We used the `adaptive moment estimation’ optimiser \citep[\textit{Adam},][]{Kingma2014} with an initial learning rate of $10^{-4}$ and set a batch size of 32. The training runs were executed on a multi-GPU environment with eight NVIDIA A100 GPUs each.

\begin{figure}
    \centering
    \subfloat[\centering Fine-tuning mode (6-channel model). \label{fig:hsc_det_over_model_dev_mod_train_loss_a}]{{\includegraphics[width=0.49\columnwidth]{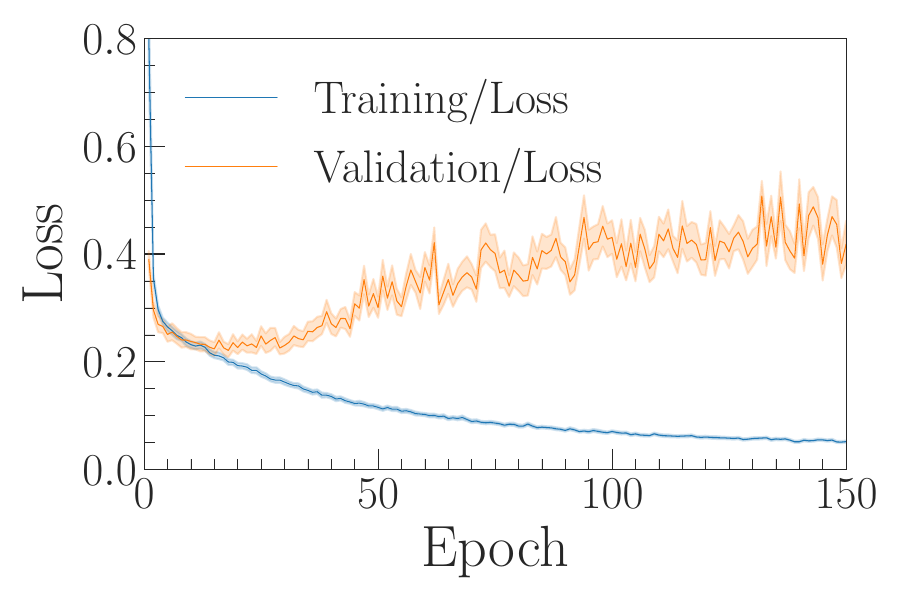} }}
    \subfloat[\centering Transfer mode (6-channel model). \label{fig:hsc_det_over_model_dev_mod_train_loss_b}]{{\includegraphics[width=0.49\columnwidth]{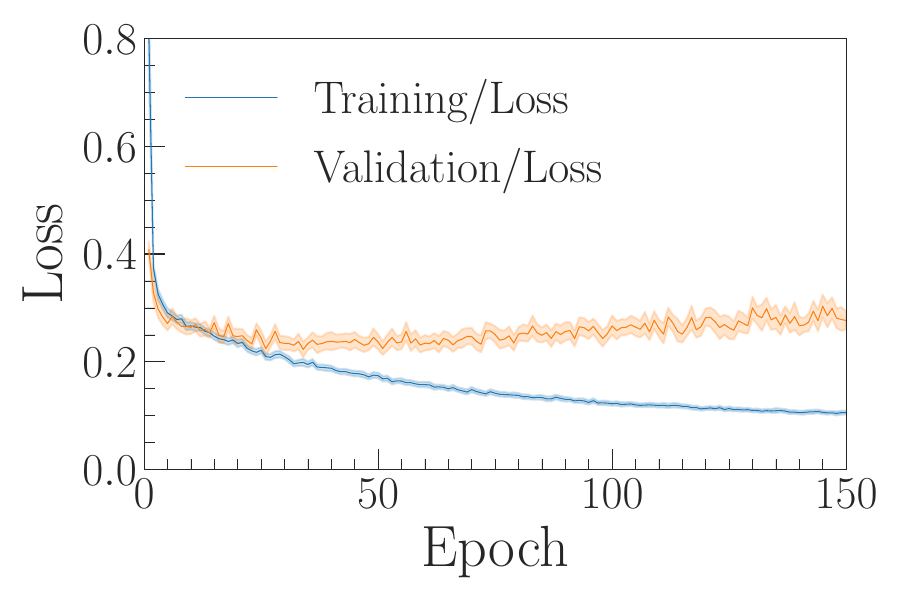} }}
    \\
    \subfloat[\centering Fine-tuning mode (5-channel model). \label{fig:hsc_det_over_model_dev_mod_train_loss_c}]{{\includegraphics[width=0.49\columnwidth]{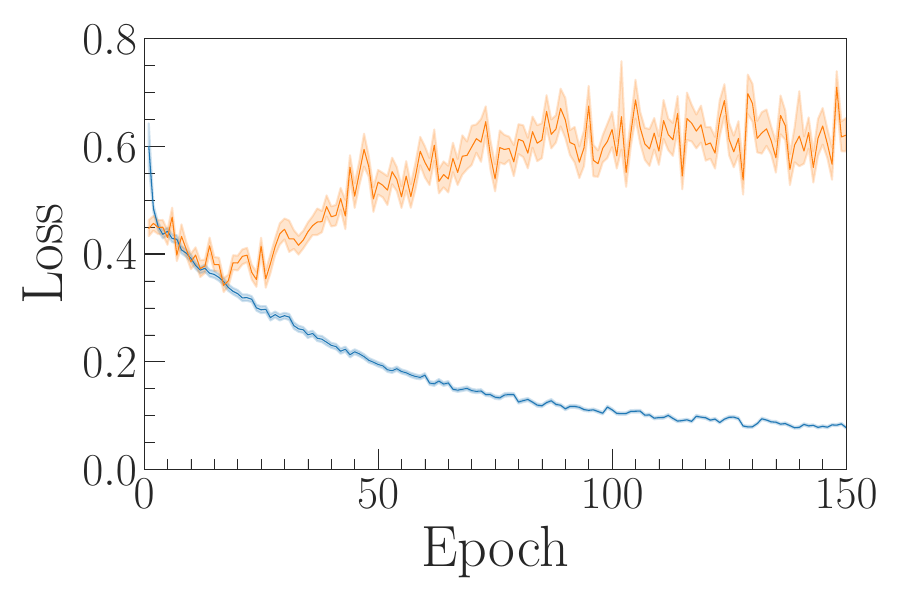} }}
    \subfloat[\centering Transfer mode (5-channel model). \label{fig:hsc_det_over_model_dev_mod_train_loss_d}]{{\includegraphics[width=0.49\columnwidth]{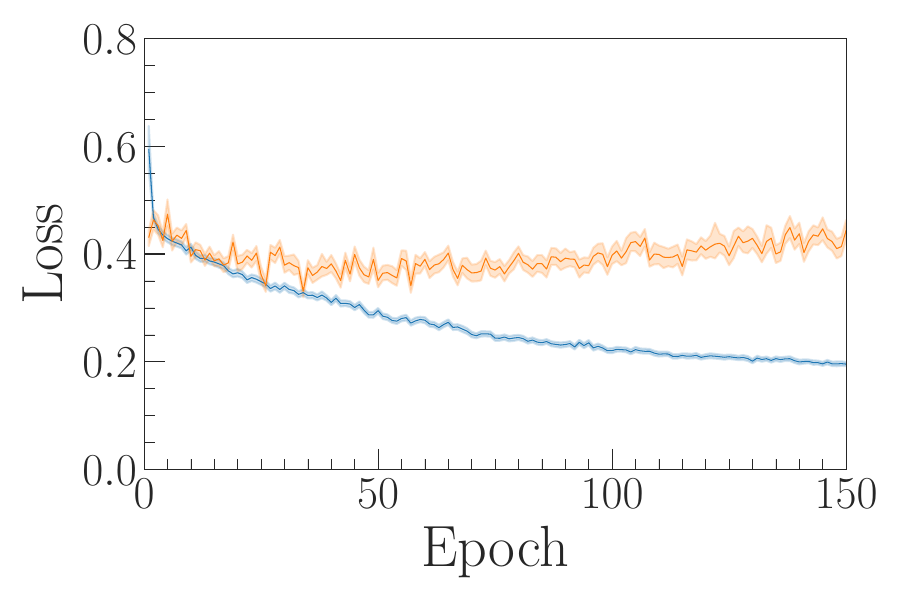} }}
    \caption[Training and validation losses of the FRCNN models for different training modes.]{Training and validation losses of the FRCNN models for different training modes. The training and validation losses for the 6-channel model are shown in the plots at the top with the model trained in fine-tuning mode on the left and in transfer mode on the right. The bottom plots show the losses of the 5-channel model for the same training modes. The loss determined on the validation data is plotted in orange, the loss determined from the training data in blue. Shaded areas show the standard error of the losses.}
    \label{fig:hsc_det_over_model_dev_mod_train_loss}
\end{figure}

The training and validation loss at each epoch was monitored during the model training process and is shown in Figure \ref{fig:hsc_det_over_model_dev_mod_train_loss}. Both losses converge quickly and the validation loss reaches a minimum after $\sim20$ epochs for all models and training modes. However, the loss determined on the validation data set tends to be generally lower for the 6-channel model compared to 5-channel model. This is likely the effect of the additional \textit{u}-band data that is available to the 6-channel model. Both FRCNN models that are trained in fine-tuning mode start to show over-fitting roughly after epoch 30 (Figs. \ref{fig:hsc_det_over_model_dev_mod_train_loss_a} and \ref{fig:hsc_det_over_model_dev_mod_train_loss_c}). The validation loss also starts to rise after $\sim 60$ epochs for the models that are trained in transfer learning mode, but over-fitting is less obvious and appears to start later in the training process (Figs. \ref{fig:hsc_det_over_model_dev_mod_train_loss_b} and \ref{fig:hsc_det_over_model_dev_mod_train_loss_d}).

To test whether the observed training and validation loss are stable or are at risk of over-fitting due to a particular data split, We also evaluated the training process using $k$-fold cross-validation as a resampling procedure. We split the total training data into $k=5$ folds of equal size and used each of the folds in turn as the validation set while the remaining four folds were used to train the models. The observed training and validation losses from each of the five training runs with the resampled train and validation sets are indistinguishable from each other for each of the model/training mode combinations. They are also indistinguishable from the losses shown in Figure \ref{fig:hsc_det_over_model_dev_mod_train_loss} which is a strong indication that the models are not biased by a particular data split and generalise equally well over the given data.

\subsection{Model performance}\label{sec:frcnn_model_performance}
\begin{figure}
    \centering
    \includegraphics[width=0.8\columnwidth]{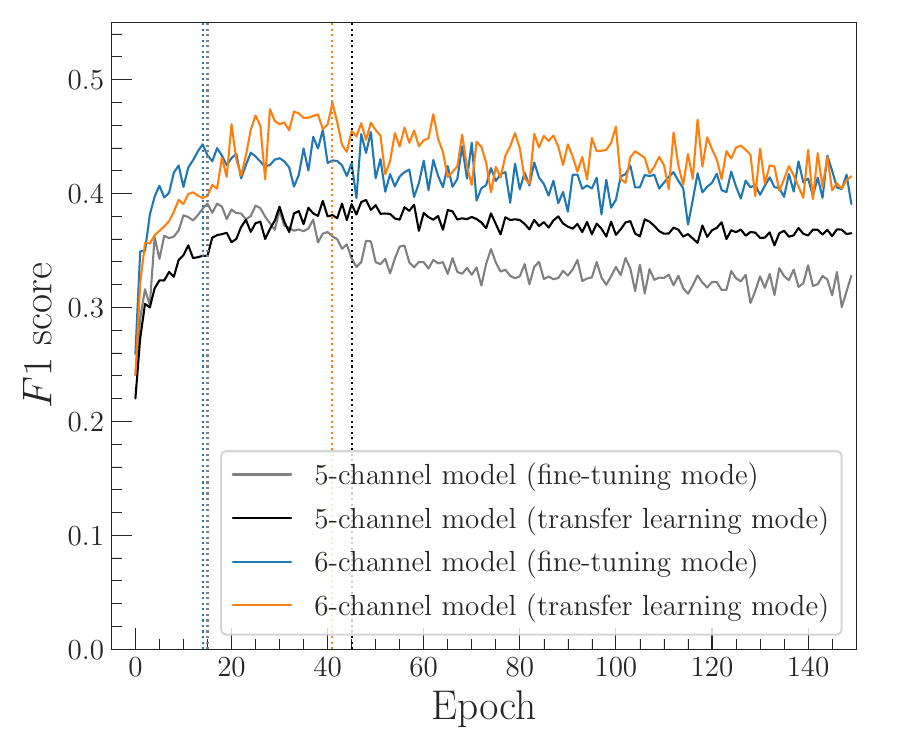}
    \caption[$F1$ scores of the FRCNN models for each epoch.]{Plot showing the $F1$ scores at each epoch for the 6-channel model that are trained in fine-tuning mode (blue) and in transfer learning mode (orange) as well as for the 5-channel models that are trained in fine-tuning mode (grey) and in transfer learning mode (black). The epochs with the highest $F1$ score before the models start to over-fit are indicated by dotted vertical lines in the same colours.}
    \label{fig:hsc_det_over_model_dev_eval_f1}
\end{figure}

During the model training process we monitored the detection performance of the FRCNN models for each epoch. The performance metrics from the COCO Object Detection Challenge \citep[COCO metrics\footnote{\url{https://cocodataset.org/\#detection-eval}},][]{Lin2014} were calculated on the validation set that contains 20\% or 640 galaxies of the total training set during the training process for each model and at each epoch. For each model, we determined the epoch where the maximum $F1$ score is reached (Figure \ref{fig:hsc_det_over_model_dev_eval_f1}) before the model starts to over-fit (see Figure \ref{fig:hsc_det_over_model_dev_mod_train_loss}). The $F1$ score is a single performance metric that combines the mean average precision $\overline{\text{AP}}$ and the mean average recall $\overline{\text{AR}}$:
\begin{equation}\label{eq:dl_metrics_f1}
    F1 = 2 \times \frac{\overline{\text{AP}}\,\overline{\text{AR}}}{\overline{\text{AP}} + \overline{\text{AR}}}
\end{equation}
and is typically used to compare different models, especially if they vary in precision and recall, and if both metrics are equally important for evaluating detection performance.

The model checkpoints with the trained weights at these epochs are stored as the final model weights. For the 6-channel models, the epochs with the highest $F1$ scores are epoch 14 (fine-tuning mode) and epoch 41 (transfer learning mode). For the 5-channel models, epoch 15 (fine-tuning mode) and epoch 45 (transfer learning mode) resulted in the highest $F1$ scores.

Figure \ref{fig:hsc_det_over_model_dev_eval_f1} also shows that the performance of the 6-channel models is better compared to the 5-channel models with regards to the $F1$ score. The additional \textit{u}-band imaging data available to the 6-channel model is clearly beneficial for clump detection. The performance differences are less prominent between the different training modes. As the models trained in transfer learning mode start to show first signs of over-fitting later in the training process than the models trained in fine-tuning mode, the maximum observed $F1$ scores are also more likely to represent a global performance maximum for those models. This is different for the 6-channel model trained in fine-tuning mode. Here, higher $F1$ scores are reached later in the training process but the validation loss has already diverged from the training loss (Figure \ref{fig:hsc_det_over_model_dev_mod_train_loss_a}) indicating that the model is showing signs of over-fitting at that stage.

To select the best FRCNN model we evaluated the trained models primarily based on their recall and subsequently on their precision as we plan to exclude non-clump contaminants in the detection sample based on photometric and inferred physical properties for subsequent analyses. We first calculated the Intersection over Union (IoU) of the bounding boxes of the model predictions with the bounding box labels of the validation data set. The IoU is calculated using the Jaccard index $J(A,B)$ \citep{Jaccard1912} of the bounding box areas $A$ and $B$:
\begin{equation}\label{eq:dl_rpn_jaccard}
    J(A, B) = \frac{A \cap B}{A \cup B} \,.
\end{equation}
We set an IoU threshold of $0.5$ to define bounding boxes that are common to both data sets. Only bounding boxes for which the predicted object class equals the object class of the validation set were compared.

\begin{figure}
    \centering
    \subfloat[\centering Precision and recall for ranges between 0.0 and 1.0. \label{fig:hsc_det_over_model_dev_eval_prec_recall_a}]{{\includegraphics[width=1.0\columnwidth]{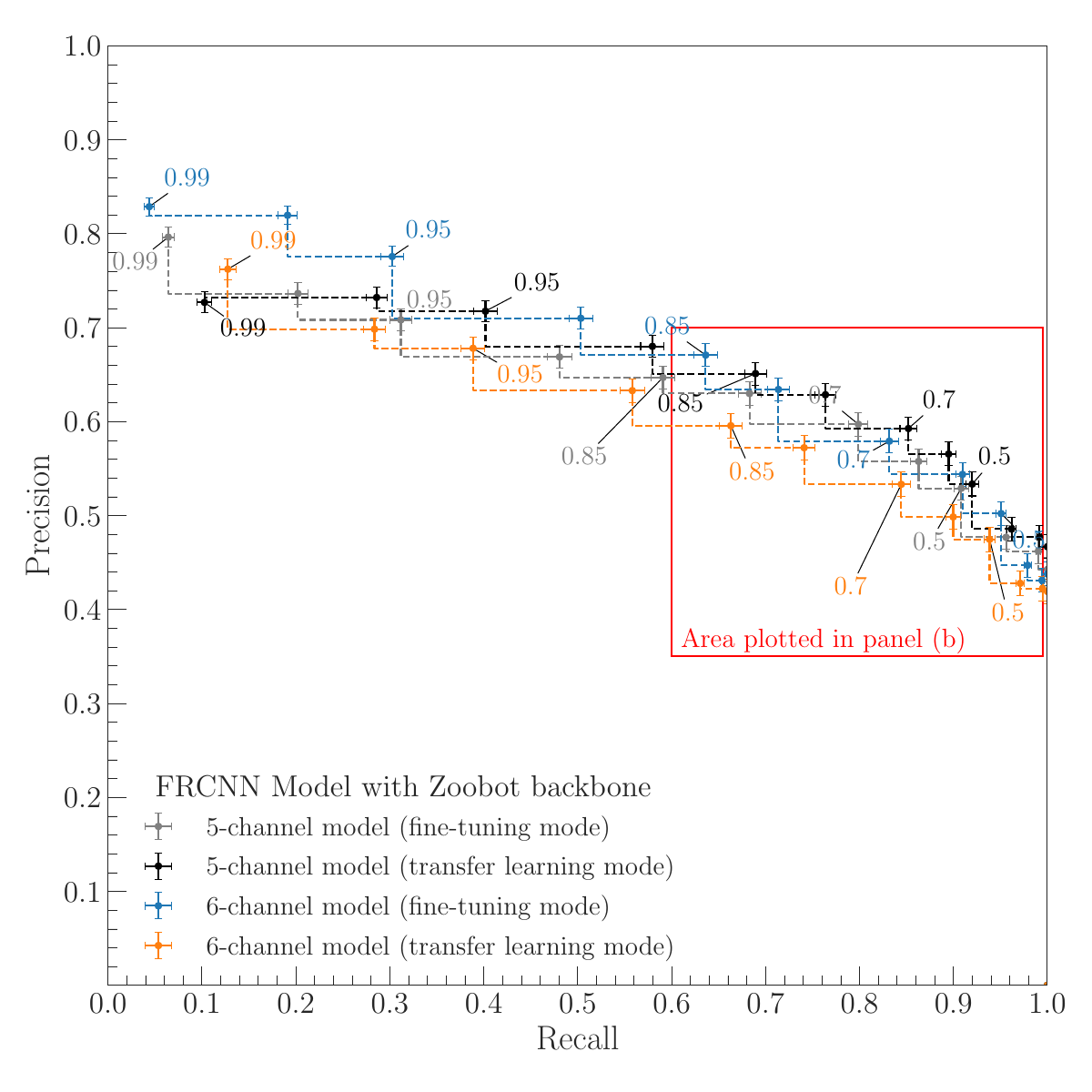} }}
    \\
    \subfloat[\centering Separate plot for precision between 0.35 and 0.7, recall between 0.6 and 1.0. \label{fig:hsc_det_over_model_dev_eval_prec_recall_b}]{{\includegraphics[width=1.0\columnwidth]{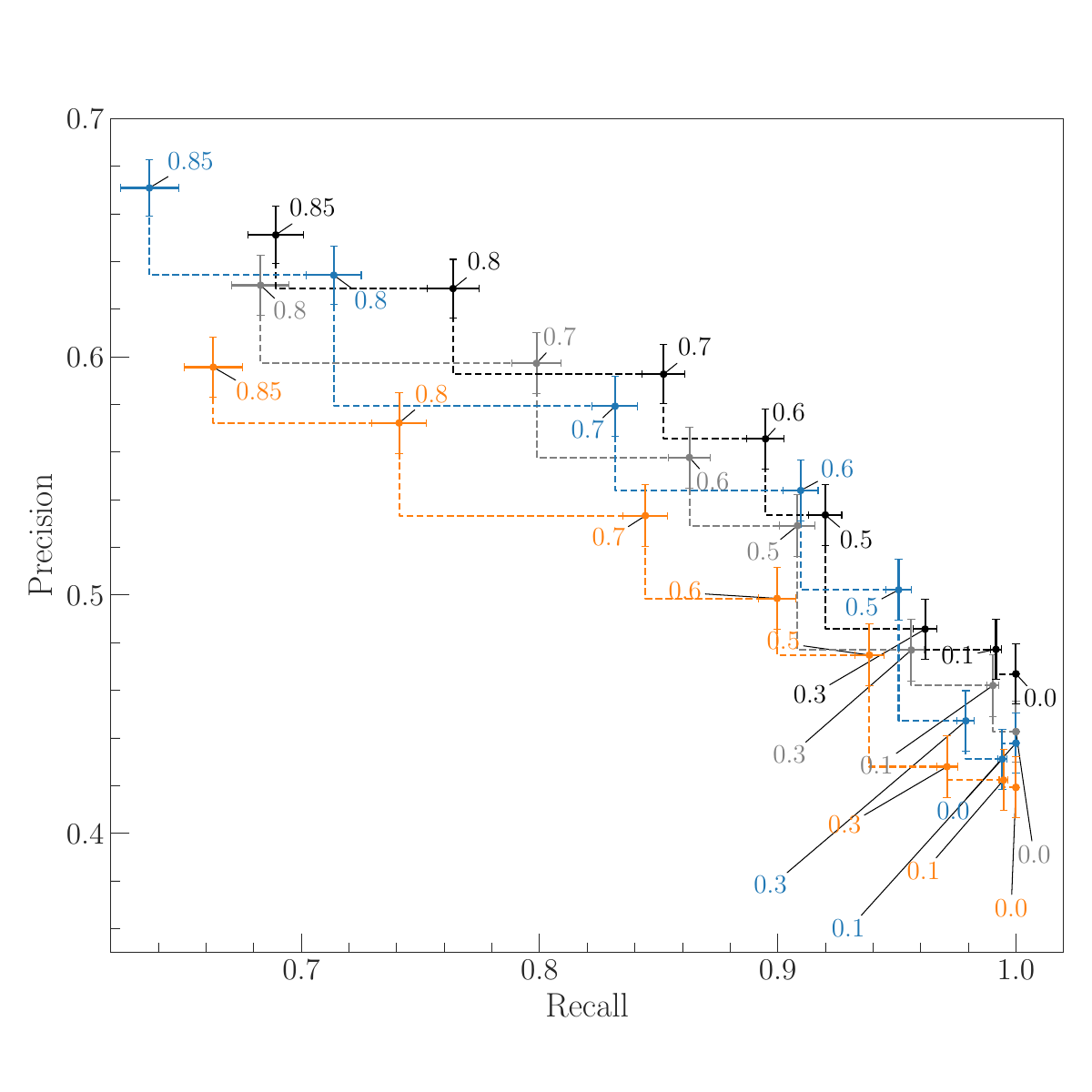} }}
    \caption[Precision and recall of the FRCNN models for different objectness thresholds.]{Precision and recall of the FRCNN models for different objectness thresholds. The objectness threshold $c_n$ is increasing from $0.0$ (right) to $0.99$ (left) as indicated by the annotations. Error bars show the $95\%$ confidence interval. The red square in panel (a) marks the zoomed-in area shown in panel (b).}
    \label{fig:hsc_det_over_model_dev_eval_prec_recall}
\end{figure}

The output of the FRCNN models not only consists of bounding boxes around detected objects with a prediction for the object class but also a score that is called `objectness' and indicates how certain the model is whether the bounding box contains an object or not. If, for a given objectness threshold, the IoU $\geq 0.5$, the detection is a true positive (TP), otherwise it is a false positive (FP) detection. On the other hand, if a bounding box has a low or no overlap with a ground-truth object (IoU $< 0.5$), then this is called a false negative (FN). True negative (TN) detections are not defined for object detection tasks as any patch of the image that is neither marked as a ground-truth object nor marked by a predicted bounding box would be a TN detection. For each of the individual objectness thresholds $c_n$, a different pair of recall and precision values can be calculated and these are shown in Figure \ref{fig:hsc_det_over_model_dev_eval_prec_recall} for the four models. 

Generally, the precision increases for higher thresholds of the objectness score while the recall increases for lower objectness thresholds as more detections are included in the detection sample for which the model predicts object detections with a higher uncertainty. All four models show a similar trend in that the model detections are almost complete (or a recall of $\sim 1.0$) up to an objectness score threshold of $\lesssim 0.3$. In Table \ref{tab:hsc_det_over_model_dev_eval}, we also list the precision, recall and $F1$ score of the 5- and 6-channel models that were trained either in fine-tuning or transfer learning mode for three objectness thresholds of $0.0$, $0.3$ and $0.6$. All models show a recall of 1.0 if no objectness threshold is applied and are close to 1.0 for a threshold of $0.3$ but the recall drops noticeably for the higher threshold of $0.6$. Precision is $\sim 0.01$ to $0.04$ higher for detections with an objectness $\geq 0.3$ compared to the detections with no objectness threshold and increases again by $\sim 0.07$ to $0.10$ for an objectness threshold of $0.6$. The differences in precision and recall between the models are also small, although the 5-channel models tend to show a slightly increased precision while having also a slightly lower recall than the 6-channel models.

\begin{table}
	\centering
	\caption[Performance metrics for the FRCNN models.]{Performance metrics for the FRCNN models. Precision, recall and the $F1$ score were determined for an objectness threshold $c_n \geq 0.0$, $\geq 0.3$ and $\geq 0.6$ on the validation data set with an IoU threshold of $\geq 0.5$.}
    \label{tab:hsc_det_over_model_dev_eval}
	\footnotesize
        \begin{tabular}{lrrr}
		\hline
		Model & \multicolumn{1}{c}{Precision} & \multicolumn{1}{c}{Recall} & \multicolumn{1}{c}{$F1$ score} \\
        \hline
        \multicolumn{4}{l}{\textbf{$c_n \geq 0.0$}} \\
        \hline
        \textit{grizy}, fine-tune  & $0.443 \pm 0.013$ & $1.000 \pm 0.000$ & $0.614 \pm 0.037$ \\
        \textit{grizy}, transfer   & $0.448 \pm 0.013$ & $1.000 \pm 0.000$ & $0.619 \pm 0.036$ \\
        \textit{ugrizy}, fine-tune & $0.424 \pm 0.013$ & $1.000 \pm 0.000$ & $0.595 \pm 0.036$ \\
        \textit{ugrizy}, transfer  & $0.419 \pm 0.013$ & $1.000 \pm 0.000$ & $0.591 \pm 0.037$ \\
        \hline
        \multicolumn{4}{l}{\textbf{$c_n \geq 0.3$}} \\
        \hline
        \textit{grizy}, fine-tune  & $0.477 \pm 0.013$ & $0.956 \pm 0.005$ & $0.636 \pm 0.037$ \\
        \textit{grizy}, transfer   & $0.486 \pm 0.013$ & $0.962 \pm 0.005$ & $0.645 \pm 0.036$ \\
        \textit{ugrizy}, fine-tune & $0.447 \pm 0.013$ & $0.979 \pm 0.004$ & $0.614 \pm 0.036$ \\
        \textit{ugrizy}, transfer  & $0.428 \pm 0.013$ & $0.971 \pm 0.004$ & $0.594 \pm 0.037$ \\
		\hline
        \multicolumn{4}{l}{\textbf{$c_n \geq 0.6$}} \\
        \hline
        \textit{grizy}, fine-tune  & $0.558 \pm 0.013$ & $0.863 \pm 0.009$ & $0.678 \pm 0.037$ \\
        \textit{grizy}, transfer   & $0.566 \pm 0.013$ & $0.895 \pm 0.008$ & $0.693 \pm 0.035$ \\
        \textit{ugrizy}, fine-tune & $0.544 \pm 0.013$ & $0.910 \pm 0.007$ & $0.681 \pm 0.036$ \\
        \textit{ugrizy}, transfer  & $0.499 \pm 0.013$ & $0.900 \pm 0.008$ & $0.642 \pm 0.037$ \\
        \hline
	\end{tabular}
\end{table}

Based on the detections that are not limited by an objectness threshold we chose the FRCNN models that were trained in fine-tuning mode and used the detections from those models for the further analysis. The 6-channel model that was trained in fine-tuning mode has a $0.5\%$ higher precision than the 6-channel model that was trained in transfer learning mode. We also chose the 5-channel model that was trained in fine-tuning mode for consistency, even though the 5-channel model that was trained in transfer learning mode shows a $0.5\%$ higher precision.

\section{Model predictions and postprocessing of the detection results}\label{sec:frcnn_model_postprocess}
\begin{figure*}
    \centering
    \includegraphics[width=1.0\textwidth]{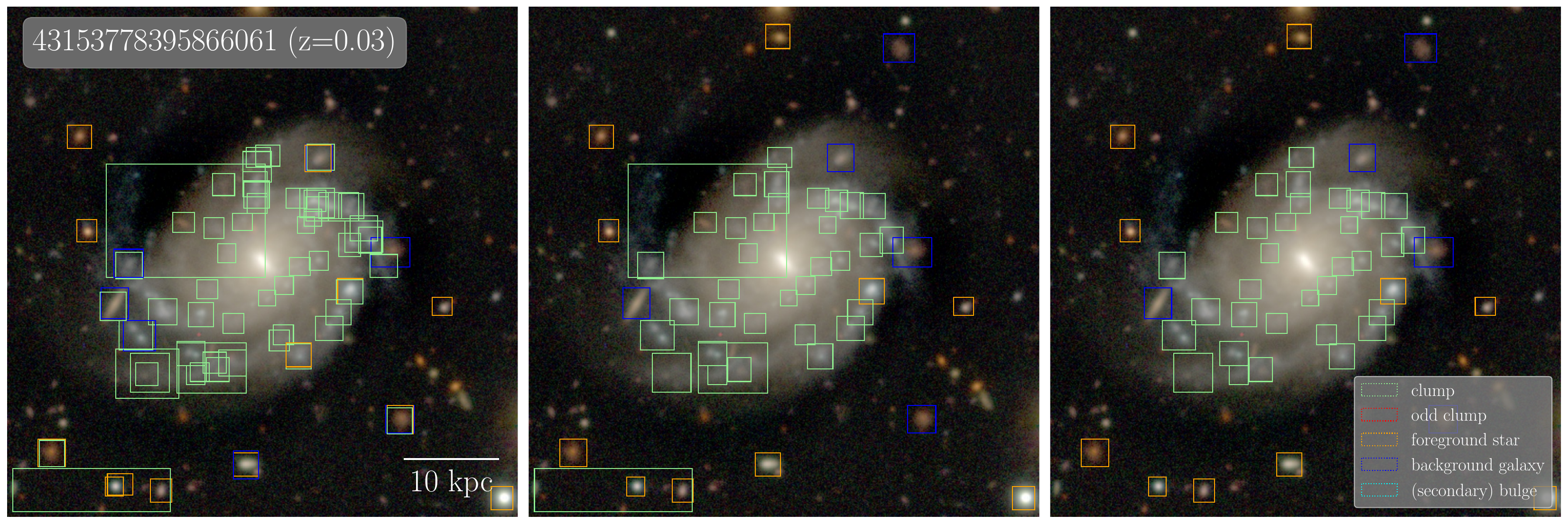}
    \caption[Postprocessing of the FRCNN model detections.]{Example galaxy (object 43153778395866061, z=0.03) showing the different postprocessing steps applied to the FRCNN model detections. The left image shows the raw detection results from the model, the central image the detection results after the non-maximum suppression process was applied and the right image after bounding boxes with size $>7.30\,\mathrm{kpc}$ were removed.}
    \label{fig:hsc_det_over_model_dev_postprocess}
\end{figure*}

After the models have been trained and evaluated, the final 5-channel and 6-channel models were applied to the full sample from the CLAUDS/HSC-SSP and HSC-SSP sets of galaxies. The raw output of the FRCNN models contains bounding boxes that overlap and often mark detections of the same object but with different class predictions and/or objectness scores (e.g. left image from Figure \ref{fig:hsc_det_over_model_dev_postprocess}). We applied a non-maximum suppression (NMS) to all detections from each galaxy using a threshold for the Intersection over Union of $\mathrm{IoU} \geq 0.2$ and kept only those detections and the corresponding object class predictions that have the highest objectness from each subset of overlapping bounding boxes. Bounding boxes that do not overlap with any other bounding boxes in a galaxy image were left unchanged. The central image in Figure \ref{fig:hsc_det_over_model_dev_postprocess} illustrates the result of the NMS process after it was applied to the raw detections that are shown in the left image of Figure \ref{fig:hsc_det_over_model_dev_postprocess}. Note that for some detections in the left image of Figure \ref{fig:hsc_det_over_model_dev_postprocess} the predicted object class also changes due to a higher objectness score of the final bounding box.

After the NMS was applied, the detections still contain a few bounding boxes that are too large to mark plausible detections of clumps or any of the contaminating features (e.g. centre image of Figure \ref{fig:hsc_det_over_model_dev_postprocess}). These large bounding boxes were not removed by the NMS as the IoU with much smaller bounding boxes can be $<0.2$ and lower than our applied threshold. We removed these bounding boxes in the next postprocessing step. 

First, we measured the apparent angular size of the longest edge of each rectangular bounding box that is predicted to mark a clump. The angular size (in arcsec) was converted into a linear size (in kpc) assuming a standard $\Lambda\mathrm{CDM}$ cosmology and depending on the redshift of the host galaxy. After visually inspecting the model detections of a few hundred galaxies we set a maximum bounding box size of $\leq 7.30\,\mathrm{kpc}$, which corresponds to the 95th percentile of the size distribution. All bounding boxes that exceed this size threshold were removed from each sample of model detections.

Then, we removed all remaining bounding boxes that fully contain other, smaller bounding boxes. Such cases still exist because the areas of some bounding boxes that are fully contained in a larger bounding box are so small that the IoU with the larger bounding box is less than $0.2$. In the galaxies we inspected visually, the smaller bounding boxes mark the location of the detected clump candidates much better than the bigger bounding box in which they are contained in. An example of a galaxy with the postprocessed model detections at this stage is shown in the right image in Figure \ref{fig:hsc_det_over_model_dev_postprocess}.

As a final step we further removed all bounding boxes with clump detections that lie outside the target galaxy's segmentation mask (see Appendix \ref{sec:hsc_data_gal_extent}) and also discarded those bounding boxes that are close to or coincide with the central bulge of a galaxy. 


\section{Extracting the clump positions from the model detections}\label{sec:frcnn_model_peaks}
\begin{figure*}
    \centering
    \includegraphics[width=1.0\textwidth]{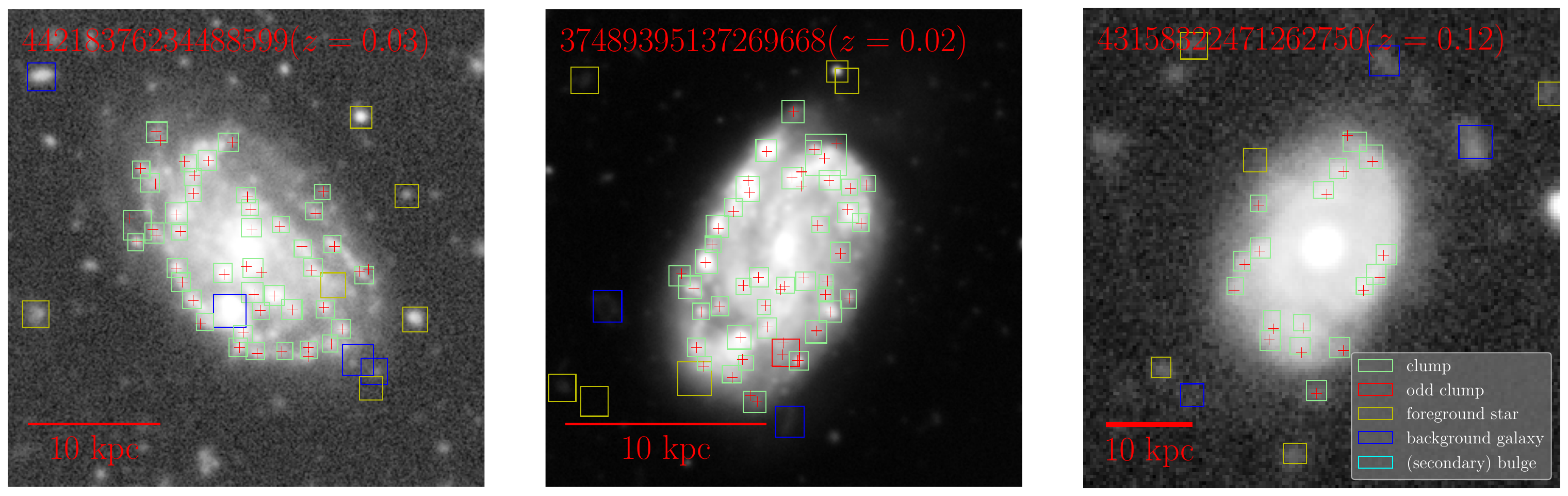}
    \caption[Galaxy examples showing the postprocessed FRCNN model detections with extracted flux peaks.]{Three galaxy examples showing the postprocessed FRCNN model detections with extracted flux peaks. The model detections are shown as boxes where the colour indicates the object class. Flux peaks are marked with red crosses. The galaxies are shown with their \textit{u}-band images.}
    \label{fig:hsc_det_over_peaks_example}
\end{figure*}

The object detection models have been trained to detect clumpy regions that represent overdensities of star-forming regions within the host galaxy. Although, the training data was created with a focus on GSFCs that are unresolved with no visible substructure or sub-clumps, some training galaxy images show resolved clumps. However, the output of the FRCNN models are bounding boxes around those star-forming regions and the bounding boxes occasionally include multiple flux peaks (see the first two galaxy examples in Figure \ref{fig:hsc_det_over_peaks_example}, for example). This suggests that clumps might be built up from smaller (sub-)clumps that together form a `clump complex'. The potential clump complexes with two or more clump candidates are more likely to be detected at very low redshifts and in high-mass galaxies (Figure \ref{fig:hsc_det_over_model_dev_eval_sims_peaks_per_clump}) as the detection of potential clump complexes with resolved substructure is strongly dependent on the spatial resolution and the extent of the host galaxy.

In our analysis, we treated every bounding box from the FRCNN model detections as a potential clump complex and every significant flux peak within a bounding box as a potential clump or clump candidate. As these objects are unresolved even at the lowest redshifts of the galaxy sample, they are treated as point-like objects. To determine the position of each flux peak we applied multiple processing steps to mask the area outside the bounding boxes, to subtract the local background and to reduce the noise in the imaging data before finding the flux peaks in either the \textit{u}-band or the \textit{g}-band image for the CLAUDS/HSC-SSP or HSC-SSP galaxies, respectively.

We started by estimating the local background flux from the galaxy image using $3\sigma$-clipping and subtracted the background fluctuation from the original galaxy image. Next, we smoothed the background subtracted galaxy image by convolving the image with a $4 \times 4$ pixel Box Filter kernel that applies a bilinear interpolation over the kernel size to assign a now averaged per pixel flux to each pixel. Local maxima of the per pixel fluxes were determined from the pixels within each bounding box of the convolved image but were required to have fluxes greater than the median plus the standard deviation of the per pixel fluxes of all pixels in the bounding box. Each maximum is also required to have a distance of at least the image-specific seeing FWHM in pixels to a next local maximum to ensure that a flux peak represents a point-like object that cannot be further resolved given the seeing limit.

The exact sub-pixel position of every local flux maximum was finally determined by calculating the flux-weighted centroid with coordinates $x_{\mathrm{centre}}$ and $y_{\mathrm{centre}}$:
\begin{equation}\label{eq:hsc_det_over_peaks_centroid}
    x_{\mathrm{centre}} = \frac{\sum_i F_{x_i} x_i}{\sum_i F_{x_i}} \qquad \mathrm{and} \qquad y_{\mathrm{centre}} = \frac{\sum_i F_{y_i} y_i}{\sum_i F_{y_i}},
\end{equation}
where $x_i$ and $y_i$ are the coordinates of the pixels that lie within a circle with radius of the image-specific seeing FWHM in pixels centred at the local flux maximum, $F_{x_i}$ and $F_{y_i}$ are the corresponding per pixel flux values.

The local flux maxima or flux peaks mark the positions of the final sample of clump candidates. If no flux peak was found for a bounding box the centroid of the bounding box was used instead. In Figure \ref{fig:hsc_det_over_peaks_example}, we show the bounding boxes of the model detections together with the extracted flux peaks for three galaxy examples to illustrate the peak finding process. Table \ref{tab:hsc_det_over_peaks_sample} lists the number of potential clump candidates after the clump positions were extracted from the model detections. The number of detected clump candidates varies with redshift and physical size of the host galaxies. Examples of galaxies with fewer clump detections are also shown in Figure \ref{fig:hsc_det_over_model_dev_eval_GRIZY_UGRIZY} and in the Appendix \ref{sec:failures} (Fig. \ref{fig:failures}).

\begin{table*}
	\centering
	\caption[Number of detected potential clumps.]{Number of potential clumpy galaxies, clump candidates and clump candidates per galaxy after the clump positions were extracted from the model detections. The counts are shown for the CLAUDS/HSC-SSP galaxy sample with 6-filter band (\textit{ugrizy}) photometry, the HSC-SSP galaxy sample with 5-filter band (\textit{grizy}) photometry and for the whole galaxy sample. The ratio of clump candidates per clumpy galaxy are shown in brackets.}
    \label{tab:hsc_det_over_peaks_sample}
	\footnotesize
        \begin{tabular}{lrrrrrrr}
		\hline
        Selection & \multicolumn{1}{l}{Galaxies} & \multicolumn{1}{l}{Pot. clump} & \multicolumn{5}{c}{Clump candidates (per galaxy)}\\
                  & & \multicolumn{1}{l}{complexes} & \multicolumn{1}{l}{$0.0\leq z \leq 0.1$} & \multicolumn{1}{l}{$0.1 < z \leq 0.2$} & \multicolumn{1}{l}{$0.2 < z \leq 0.3$} & \multicolumn{1}{l}{$z > 0.3$} & \multicolumn{1}{l}{Total} \\
        \hline
        \textit{ugrizy} &   7,135 &    30,148  &   6,928 (9.06) &  12,609 (4.99) &   6,461 (3.28) &   4,638 (2.48) &    30,636 (4.23) \\
        \textit{grizy}  & 339,719 & 1,382,337  & 362,675 (8.47) & 586,704 (4.57) & 289,462 (3.25) & 184,235 (2.32) & 1,423,076 (4.19) \\
        Total           & 346,854 & 1,412,485  & 369,603 (8.48) & 599,313 (4.58) & 295,923 (3.25) & 188,873 (2.33) & 1,453,712 (4.19) \\
        \hline
	\end{tabular}
\end{table*}

\section{Detection results}\label{sec:results}
A first evaluation of the detection performance was done during the model development process (Section \ref{sec:frcnn_development}), but the raw model predictions were further processed to extract the positions of flux peaks (Section \ref{sec:frcnn_model_peaks}). Furthermore, because we trained different object detection models that use different imaging data as input, it is important to assess how similar the detections from each of the models are. 

We first test whether the \textit{ugrizy} and the \textit{grizy} model produce similar detection results (Section \ref{sec:hsc_det_over_model_dev_eval_GRIZY_UGRIZY}). If so, then the 5-channel model can be confidently applied to the much larger set of galaxies that only have imaging data from the five \textit{grizy}-filter bands used by HSC-SSP. We then estimate the purity and completeness of both models on simulated data (Section \ref{sec:sims}) to evaluate the observational and physical parameters for which reliable clump detections can be made (Section \ref{sec:hsc_det_over_model_dev_eval_complete}).

\subsection{Comparison of the \textit{grizy} and \textit{ugrizy} model detections}\label{sec:hsc_det_over_model_dev_eval_GRIZY_UGRIZY}
As shown in Section \ref{sec:frcnn_model_performance} the detection results of the \textit{grizy} and the \textit{ugrizy} model are similar. However, the models were evaluated based on the test set of the training data and the overlap of the bounding boxes that are the direct output of the detection models. To assess how similar the detections from each of the models are after the flux peaks have been determined in each bounding box (Section \ref{sec:frcnn_model_peaks}), we applied both FRCNN models on the same set of 14,231 galaxies that have imaging data from the CLAUDS and HSC-SSP survey. The \textit{u}-band data was omitted for the 5-channel model but both detection runs were otherwise made in exactly the same way. The flux peaks representing the positions of the clump candidates were extracted from the \textit{u}-band image for the 6-channel model and from the \textit{g}-band image for the 5-channel model.

We then crossmatched the positions of the clump candidates from the 5-channel FRCNN model with the detected clump candidates from the 6-channel FRCNN model using a maximum separation distance of 0.75 times the image-specific \textit{u}-band seeing FWHM (see also Section \ref{sec:hsc_det_over_model_dev_eval_complete_UGRIZY} and Fig. \ref{fig:hsc_det_over_model_dev_eval_complete_dist}), keeping only the closest crossmatch and discarding others that are further away. The \textit{u}-band seeing was chosen as it is worse than the \textit{g}-band seeing for all images. Of the 30,636 clump candidates that were detected by the 6-channel model, 25,262 ($82.46\%$) were also detected by the 5-channel model, while the remaining 5,374 ($17.54\%$) were only detected by the 6-channel model. The 5-channel model detected in total 28,641 clump candidates of which $88.20\%$ are clump candidates found by both models. The 5-channel model detected only 3.379 ($11.80\%$) clump candidates that are not detected by the 6-channel model.

\begin{figure}
    \centering
    \subfloat[\centering 6-channel (\textit{ugrizy}) model. \label{fig:detections_5_6_channel_UGRIZY}]{{\includegraphics[width=0.45\columnwidth]{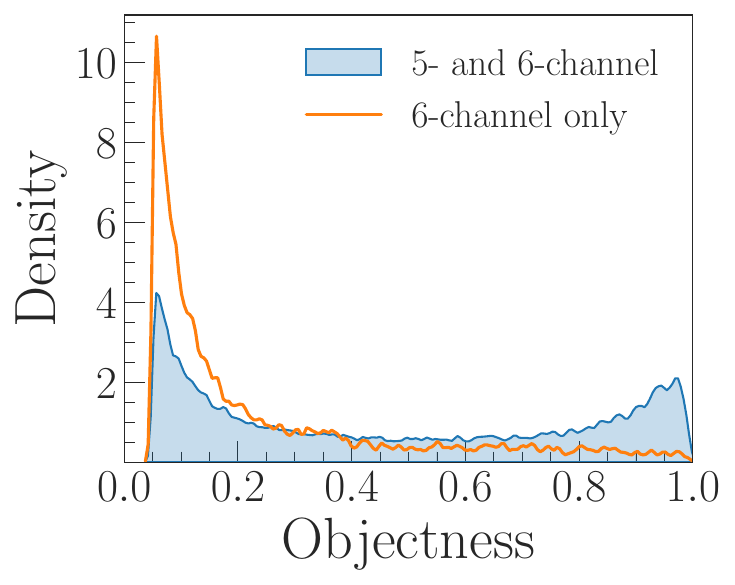} }}
    \subfloat[\centering 5-channel (\textit{grizy}) model. \label{fig:detections_5_6_channel_GRIZY}]{{\includegraphics[width=0.45\columnwidth]{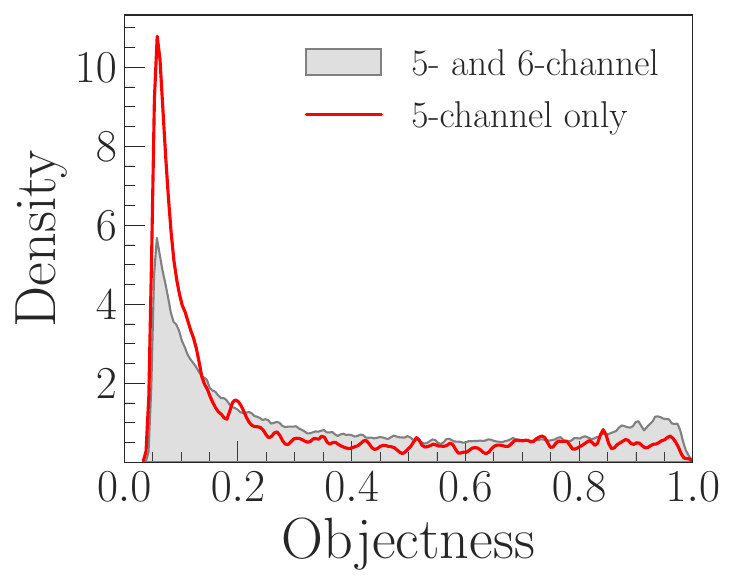} }}
    \caption[Comparison of the clump detections from both models and detections unique to either the \textit{grizy} or \textit{ugrizy} model.]{Comparison of the clump detections from both models and detections unique to either the \textit{ugrizy} or \textit{grizy} model. The kernel density estimates of the clump detections are shown separately as a function of the objectness score that is output by the 6-channel (a) and 5-channel model (b). Clump candidates that were detected by the 6-channel model and the 5-channel model are shown in blue (a) and in grey (b). Clump detections that are unique to the 6-channel model are shown in orange and unique to the 5-channel model in red.}
    \label{fig:detections_5_6_channel_UGRIZY_GRIZY}
\end{figure}

Both models show similar clump detections with an overlap of $\gtrsim 80\%$. The 6-channel model appears to detect more clump candidates that are bright in the NUV wavelength ranges of the \textit{u}-band but less pronounced in the \textit{g}- and redder bands. In addition, some of the differences are due to varying signal to noise ratios (SNRs) of the \textit{u}- and \textit{g}-filter band images that were used to extract the flux peak positions. This resulted in clump positions extracted from the \textit{g}-band images that were not matched with clump positions in the \textit{u}-band images given our threshold of 0.75 times the image-specific \textit{u}-band seeing FWHM. However, the majority of clump candidates that were detected either by only the 6-channel or only the 5-channel model have low objectness scores $\lesssim 0.2$, suggesting that those predictions unique to each model contain spurious detections (Fig. \ref{fig:detections_5_6_channel_UGRIZY_GRIZY}).

\begin{figure*}
    \centering
    \includegraphics[width=1.0\textwidth]{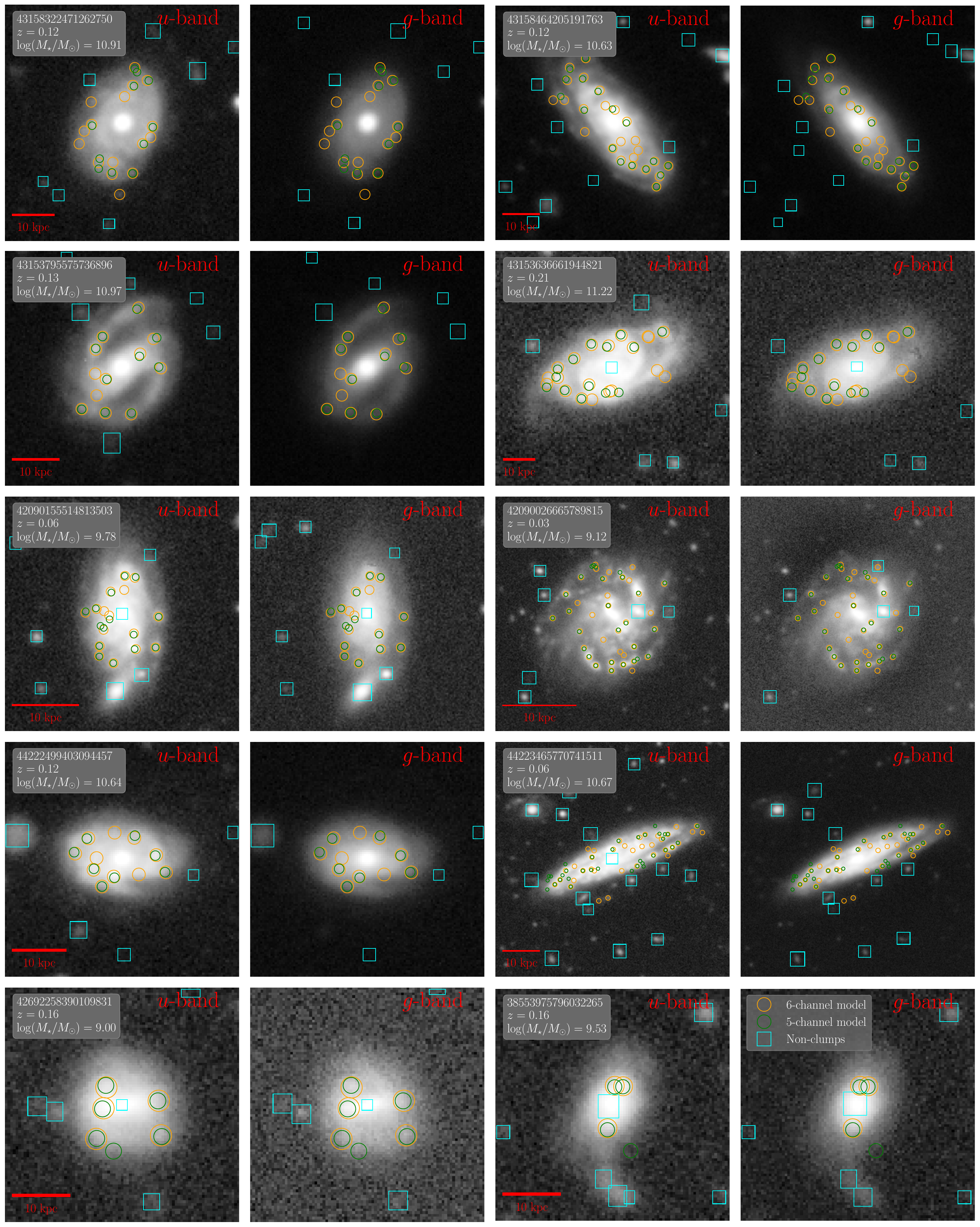}
    \caption[Detections from the 5- and 6-channel FRCNN models shown for an example of ten galaxies.]{Detections from the 5- and 6-channel FRCNN models shown for an example of ten galaxies. The galaxies are shown in pairs with the \textit{u}-band image first and followed by the \textit{g}-band image of the same galaxy. Detections from the 6-channel FRCNN model are marked with orange circles and the detections from the 5-channel model with green circles. Non-clump detections (e.g. fore-/background galaxies, stars) are marked by cyan boxes. Each galaxy is labelled with their HSC-SSP object-ID and the galaxy's redshift and stellar mass estimate from the HSC-SSP catalogue.}
    \label{fig:hsc_det_over_model_dev_eval_GRIZY_UGRIZY}
\end{figure*}

For a visual comparison, we show ten galaxy examples with the detections from both models indicated in Figure \ref{fig:hsc_det_over_model_dev_eval_GRIZY_UGRIZY}. The figure shows the \textit{u}- and \textit{g}-band images for galaxies that have noticeably more detections from the 6-channel model than the 5-channel model.

\subsection{Simulating star-forming clumps}\label{sec:sims}
\begin{figure}
    \centering
    \includegraphics[width=0.8\columnwidth]{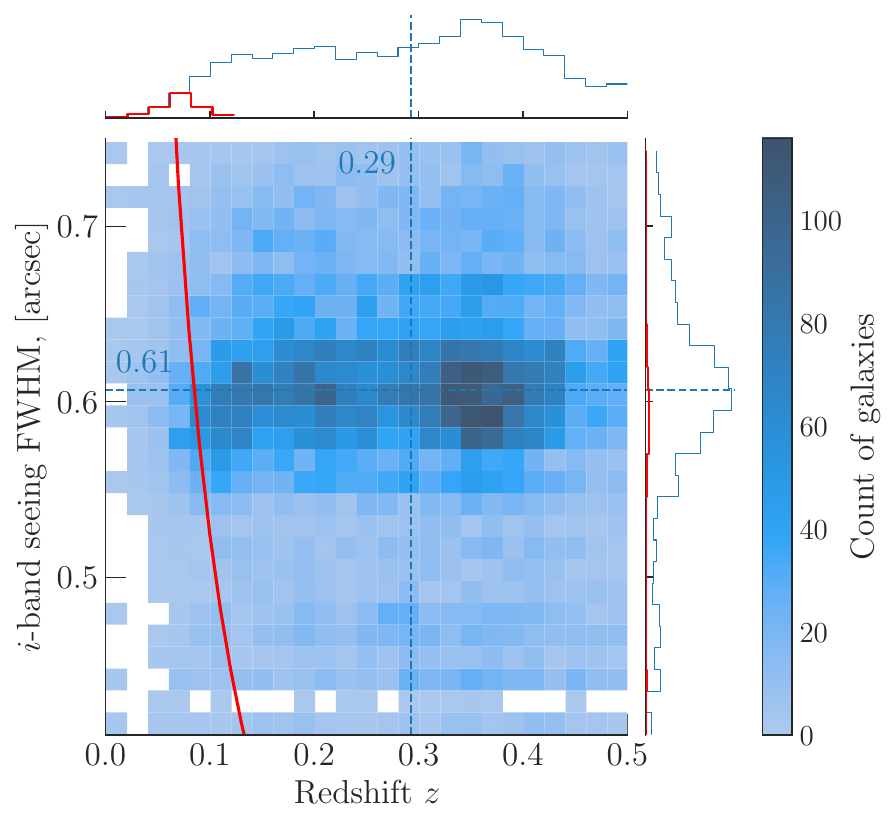}
    \caption[Redshift vs. \textit{i}-band seeing FWHM for HSC-SSP galaxies.]{Redshift vs. \textit{i}-band seeing FWHM for HSC-SSP galaxies and their image cutouts. The red line plots the theoretical seeing FWHM that is required to resolve objects with 1.0 kpc in physical size. The marginal histograms on the top and right side of the two-dimensional histogram show the univariate distributions for all galaxies in blue and galaxies for which the \textit{i}-band seeing FWHM translates to a physical size of $\leq 1$ kpc in red.}
    \label{fig:hsc_discussion2_seeing_redshift}
\end{figure}

Simulating objects that are as similar as possible to the real objects provides another way of testing whether the model is producing genuine scientific results. With the given seeing, features with physical sizes $<1$ kpc can be theoretically resolved in only a small fraction of galaxies at $z\lesssim 0.1$. This is shown in Figure \ref{fig:hsc_discussion2_seeing_redshift} where we plot the \textit{i}-band seeing FWHM as a function of redshift for the sample of HSC-SSP galaxies. As clumps have reported physical sizes of $\lesssim$1 kpc \citep[e.g.][]{Elmegreen2007a,FoersterSchreiber2011a,Guo2018,Zanella2019} and as the \textit{i}-band seeing is significantly better than the seeing of the other filter bands, the vast majority of the clumps are unresolved in our observations. Artificially generated overdensities of star-forming regions or clumps can be generated to appear as point-like objects with integrated fluxes as seen in different broadband filters by forward-modelling a predefined set of physical properties, like stellar mass, age, dust attenuation and metallicity. As their positions, fluxes and underlying physical properties are known, these simulated objects are used to evaluate the model completeness and purity.

The Faster R-CNN model (Section \ref{sec:frcnn_development}) outputs bounding boxes around potential overdensities of star-forming regions or clump complexes, where regions of higher flux likely indicate the positions of one or more regions of intense star-formation. Using the Python package \textsc{Pointpats} \citep{Rey2024}, which provides functions for randomly placing points within a polygon, we placed up to 30 clump positions randomly within the host galaxy segmentation mask (Appendix \ref{sec:hsc_data_gal_extent}) while also masking the detection bounding boxes. We then rejected any location closer than three times the largest FWHM of the seeing, over all six \textit{ugrizy}-filter bands, to any detection bounding box of those initial number of 30 positions for the simulated clumps. In doing so, we ensured that none of the simulated clumps overlap with any of the real clumps, foreground stars or fore-/background galaxies that our model detected and that the simulated clumps are not blended with preexisting structures of the host galaxy.

\begin{figure*}
    \centering
    \includegraphics[width=0.8\textwidth]{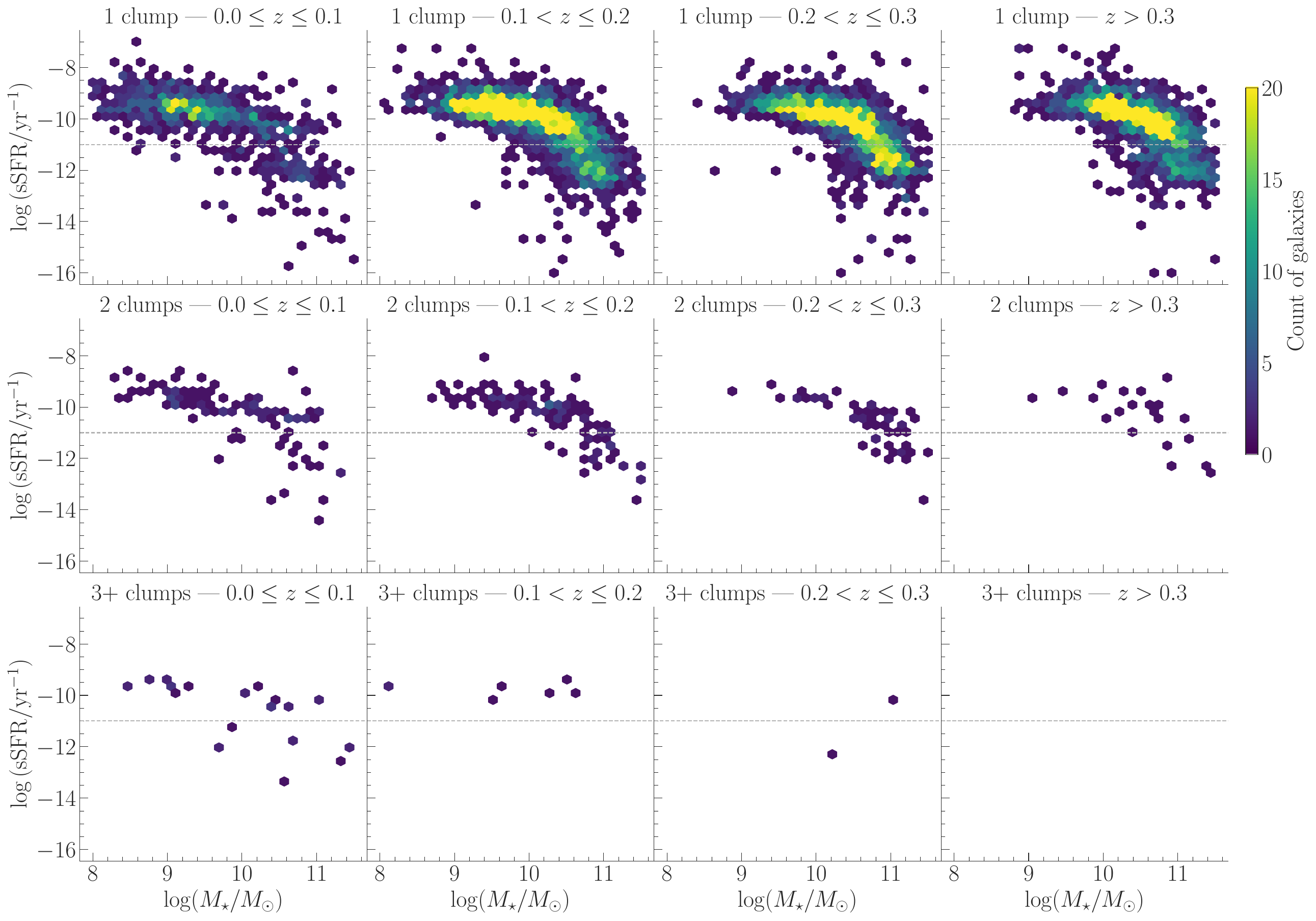}
    \caption[Stellar mass vs. sSFR of the host galaxies of detected clump candidates.]{Stellar mass vs. specific star-formation rate (sSFR) of the host galaxies with bounding box detections that contain different numbers of identified clumps (rows) and per redshift bin (columns). The distribution of the host galaxies are shown in coloured hex-bins. The horizontal dashed line marks the separation between star-forming and quiescent galaxies at $\log(\mathrm{sSFR}/\mathrm{yr}^{-1}) = -11.0$ \citep[e.g.][]{McGee2011,Wetzel2013}.}
    \label{fig:hsc_det_over_model_dev_eval_sims_peaks_per_clump}
\end{figure*}

The number of sub-clumps that can be resolved per bounding box varies with the stellar mass and star formation rate of the host galaxy but also with redshift due to limited spatial resolution in higher redshift bins (Figure \ref{fig:hsc_det_over_model_dev_eval_sims_peaks_per_clump}). We created a look-up table where the number of \textit{u}-band clump complexes per real galaxy having $1, 2, 3$ or more than $4$ clumps is calculated per redshift bin ($z \in [0.0, 0.5]$, with binwidth $0.05$), per stellar mass bin ($\log\,(M_\star/M_\odot) \in [8.0, 12.0]$, with binwidth $0.5$) and specific star-formation rate (sSFR) bin ($\log \,(\mathrm{sSFR}/\mathrm{yr}^{-1}) \in [-16.0, -6.0]$, with binwidth $0.5$) of the host galaxy. For a specific galaxy with given redshift, stellar mass and sSFR, the number of clumps per simulated clump complex was sampled from this distribution. Each random trial could result in up to three additional sub-clump positions around the initial clump complex position. These additional sub-clump positions were then randomly placed within a radius of $0.5 \leq \mathrm{r} \leq 1.0\,\mathrm{kpc}$ around the initial position. If no clump-per-clump-complex distribution in the look-up table is found for a specific galaxy, i.e. the galaxy properties are outside the parameter space of the look-up table, or the look-up table distribution shows only one clump per clump complex, only the initial position is used. 

The simulated sample of clump complexes with one or more clumps consists of 32,241 clumps in 31,777 clump complexes, hosted by 13,789 galaxies, which is almost the galaxy count of the full set of the \textit{ugrizy}-galaxies. The missing galaxies from the full set of \textit{ugrizy}-galaxies are mainly galaxies that were either too small or for which the algorithm could not find at least one suitable position for a simulated clump.

To determine the luminosity and colour of each clump within a clump complex, we simulated spectra from composite stellar populations (CSPs) using the Fortran package \textsc{FSPS}: Flexible Stellar Population Synthesis \citep{Conroy2009, Conroy2010} in conjunction with the Python wrapper PythonFSPS \citep{Johnson2024}. Here, the default stellar isochrone library \textsc{MIST} \citep{Paxton2010,Choi2016,Dotter2016} and stellar spectral library \textsc{MILES} \citep{SanchezBlazquez2006, Cenarro2007} were used. Infrared-emission from the dust heated by starlight was modelled using the \citet{Draine2007} model, nebular continuum and line emission are accounted for by using the FSPS-implementation of the \textsc{Cloudy} radiative transfer code \citep{Ferland1998,Ferland2013} and ionisation sources as described in \citet{Byler2017}.

Each clump in a clump complex was treated as a CSP with a delayed exponentially declining SFH (delayed $\tau$ model) and a \citet{Chabrier2003} IMF. We adopted a \citet{Calzetti2000} dust attenuation curve with an effective dust attenuation $A_V$ uniformly sampled from $[0.0, 4.0]\,m_{\mathrm{AB}}$. The stellar metallicity for each clump was also chosen from a uniform distribution ranging from $-2.0\,Z_\odot$ to $0.19\,Z_\odot$ and the gas phase metallicity was set to the same value. The e-folding time $\tau$ for the SFH was chosen from a log-uniform distribution with $\tau \sim \mathcal{U}(\ln{0.1}, \ln{30.0})\,\mathrm{Gyr}$. 

\begin{figure}
    \centering
    \subfloat[\centering Ratio of clump stellar mass as a function of host galaxy stellar mass. \label{fig:hsc_det_over_model_dev_eval_sims_massDistribution}]{{\includegraphics[width=0.45\columnwidth]{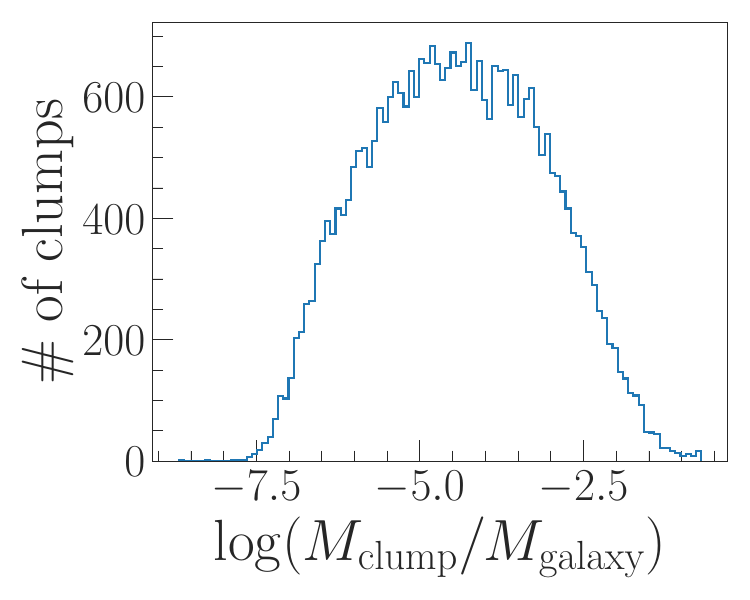} }}
    \subfloat[\centering Distribution of the apparent magnitudes. \label{fig:hsc_det_over_model_dev_eval_sims_magDistribution}]{{\includegraphics[width=0.45\columnwidth]{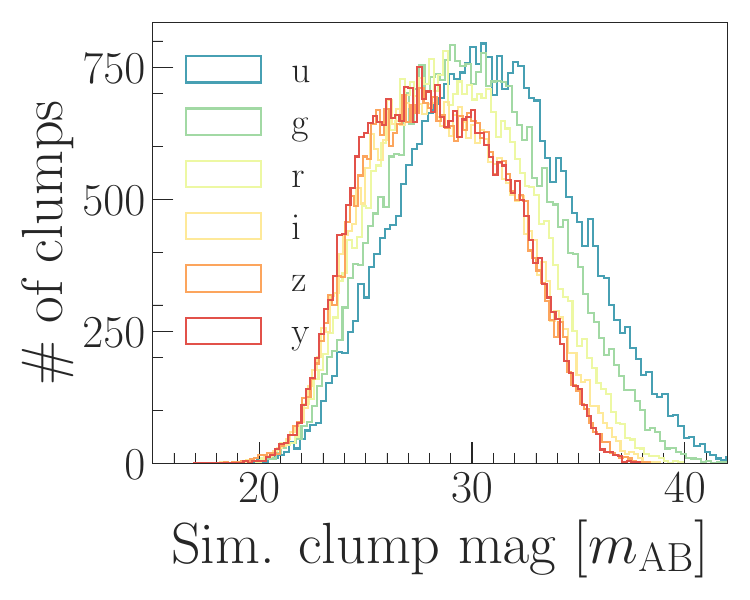} }}
    \caption[Histograms of the parameters for the final set of simulated clumps.]{Histograms showing the distribution of the sampled parameters for the final set of simulated clumps. The clump stellar masses (a) were sampled from a log-uniform distribution with a lower limit of $10^4\,M_\odot$ and upper limit of $0.1\, M_{\mathrm{galaxy}}$ (see also Table \ref{tab:hsc_det_over_model_dev_eval_sims_sps_params}). The plot (b) shows the distributions of the resulting apparent magnitudes in each filter band.}
\end{figure}

We sampled the stellar mass of each individual clump from a log-uniform distribution with a lower limit of $10^4\,M_\odot$ and the upper limit depending on the redshift of the host galaxy. We set the upper limit to $10^7\,M_\odot$ for host galaxies at redshift $z\leq0.1$ and extended the upper limit to $5\times 10^7\,M_\odot$, $10^8\,M_\odot$ and $5\times 10^8\,M_\odot$ for the redshift ranges $0.1<z\leq 0.2$, $0.2<z\leq 0.3$ and $z>0.3$, respectively. This allows for more massive clump complexes at higher redshifts for which the individual clumps cannot be resolved and are blurred into a single clump. We limited the simulated clump's stellar mass to $0.1\, M_{\mathrm{galaxy}}$ to ensure that the simulated clumps do not dominate their hosts with stellar masses of more than $10\%$ of the total galaxy. Figure \ref{fig:hsc_det_over_model_dev_eval_sims_massDistribution} shows the ratio of the stellar masses of the clumps and the host galaxy $M_{\mathrm{clump}}/M_{\mathrm{galaxy}}$ for the final set of simulated clumps. The majority of the simulated clumps are far less massive than $1\%$ of the host galaxy's mass. The details of all parameters used are listed in Table \ref{tab:hsc_det_over_model_dev_eval_sims_sps_params}.

The choice of model parameters used with \textsc{FSPS} to simulate the spectra was motivated by observed physical parameters of star-forming clumps \citep[e.g.][]{Guo2018,HuertasCompany2020,Mehta2021}, parametric SFH models used by other studies \citep[e.g.][]{Guo2018,HuertasCompany2020} and by extensive tests on how different parameter ranges affect the resulting luminosity of the stellar populations. The simulated clumps were required to match an observed magnitude distribution but also needed to extend into fainter luminosity ranges so that reliable completeness limits could be determined (Section \ref{sec:hsc_det_over_model_dev_eval_complete}).

\begin{table}
	\centering
	\caption[Parameters and sampling distribution for each simulated peak spectrum.]{Parameters and sampling distribution for each simulated peak spectrum.}
    \label{tab:hsc_det_over_model_dev_eval_sims_sps_params} 
	\footnotesize
        \begin{tabular}{lll}
		\hline
		Parameter/Unit & Range & Sampling \\ 
		\hline
		$\tau$/Gyr                       & $[0.1, 0.30]$ & Log-uniform \\ 
        $\log(Z_\star/Z_\odot)$          & $[-2.0, 0.19]$ & Uniform \\
        $\log(Z_{\mathrm{gas}}/Z_\odot)$ & same as $Z_\star$ & Uniform \\
        $\mathrm{A}_\mathrm{V}\,/\,m_{\mathrm{AB}}$ & $[0.0, 4.0]$ & Uniform \\
        $M_\star/M_\odot$                & $[10^4, 10^7]$ for $z\leq0.1$ & Log-uniform \\
                                         & $[10^4, 5\times 10^7]$ for $0.1<z\leq0.1$ & \\
                                         & $[10^4, 10^8]$ for $0.2<z\leq0.3$ & \\
                                         & $[10^4, 5\times 10^8]$ for $z>0.3$ & \\
                                         & max. $0.1\, M_{*,\mathrm{galaxy}}$ & \\
        $t_{\mathrm{age}}/\mathrm{Gyr}$  & $[0.005, 1.0]$ & Uniform \\
        $z$                              & Host galaxy redshift & fixed \\
		\hline
	\end{tabular}
\end{table}

\begin{figure*}
    \centering
    \includegraphics[width=1\textwidth]{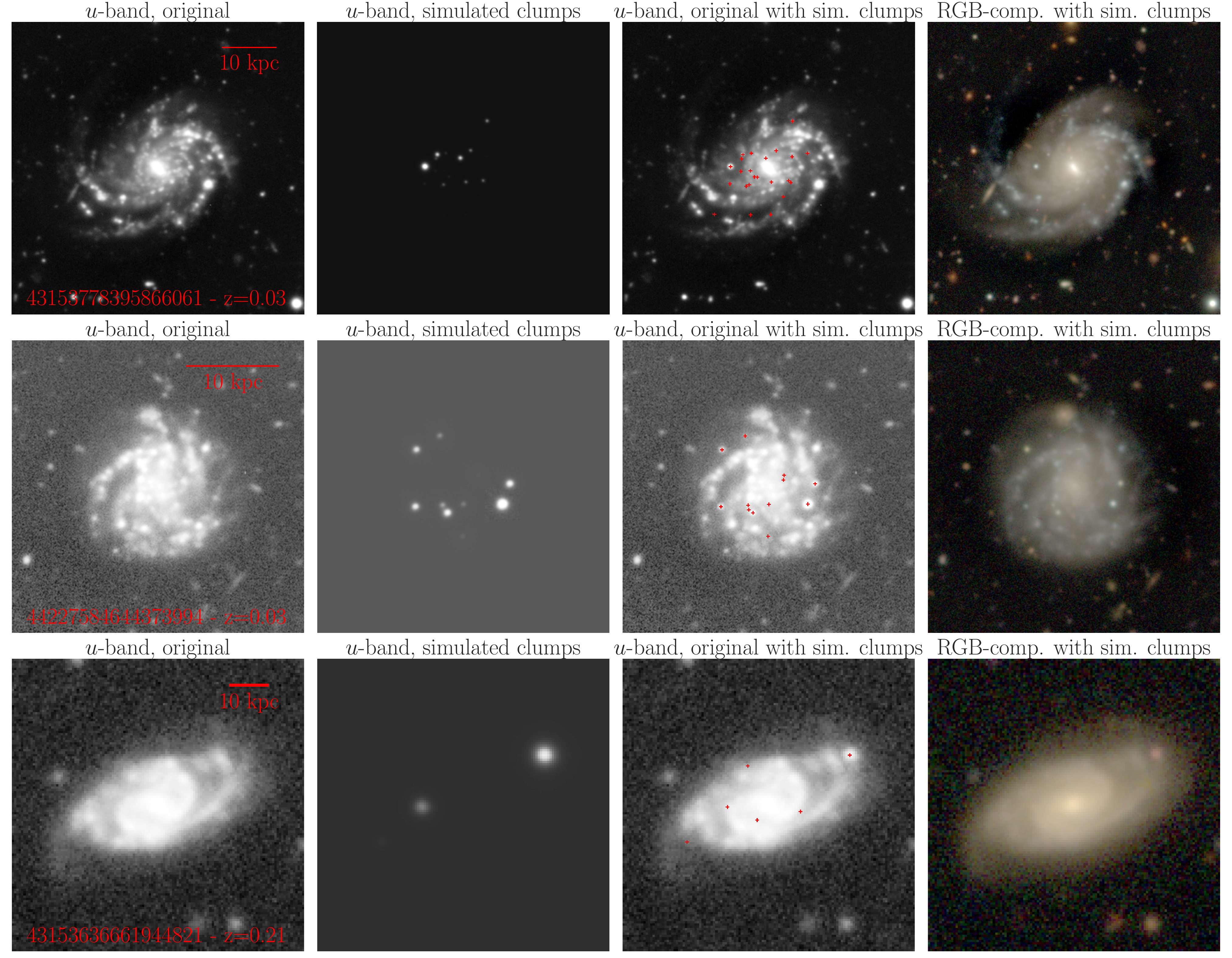}
    \caption[Galaxy examples with simulated clumps.]{Galaxy examples with simulated clumps, shown as the original \textit{u}-band science image (first column), the simulated clumps alone (second column), \textit{u}-band science image with simulated clumps injected (third column) and the RGB-composite image (generated using the GRI-bands) with injected simulated clumps (last column). The positions of the simulated clumps are indicated by small red markers in the images of the third column. Please note that some clumps are too faint to be visible in the images of the second column with the simulated clumps alone.}
    \label{fig:hsc_det_over_model_dev_eval_sims_example}
\end{figure*}

The resulting spectra were redshifted to match the redshift of the host galaxy and integrated over the wavelength range of each individual CLAUDS and HSC filter band. Once the positions of the simulated clump complexes with their individual clump(s) and per-pixel fluxes for each filter band have been determined, a separate image for each filter band with the same extent as the galaxy image cutout was created. The image was convolved with the effective PSF (ePSF) specific to each host galaxy image and filter band, which we estimated from bright stars in the vicinity of our target galaxies following the algorithm described by \citet{Anderson2000}. The process is illustrated for three galaxy examples in Figure \ref{fig:hsc_det_over_model_dev_eval_sims_example}.

\subsection{Completeness and purity of the model detections}\label{sec:hsc_det_over_model_dev_eval_complete}
Purity and completeness of the FRCNN models were first assessed as part of the model development process (Section \ref{sec:frcnn_model_performance}). During the model development process, we evaluated the model performance based on only the validation set of the labelled training data. The simulated clumps allow the same evaluation to be performed but on a completely independent data set. The advantage of using simulated data for the model evaluation is that the physical properties of the objects that the model is intended to detect are known. With that knowledge, the limits of the object detection model can be better explored and its performance measured in relation to the simulated physical properties.

Each of the galaxy images in Figure \ref{fig:hsc_det_over_model_dev_eval_complete_detect} illustrates the three possible cases that we considered when measuring the completeness and purity for the simulated clumps: (1) a true positive or successful detection, (2) a false positive detection and (3) a false negative detection. If a simulated clump (red marker) is found within one of the dotted bounding boxes, a successful detection of a simulated clump is counted. An `empty' dotted bounding box adds a spurious or false positive detection that lowers the overall purity, whereas a simulated clump (red marker) without a dotted bounding box around its position is a false negative detection that lowers the overall completeness.

\subsubsection{Completeness of the 6-channel (\textit{ugrizy}) model detections}\label{sec:hsc_det_over_model_dev_eval_complete_UGRIZY}
\begin{figure*}
    \centering
    \includegraphics[width=1\textwidth]{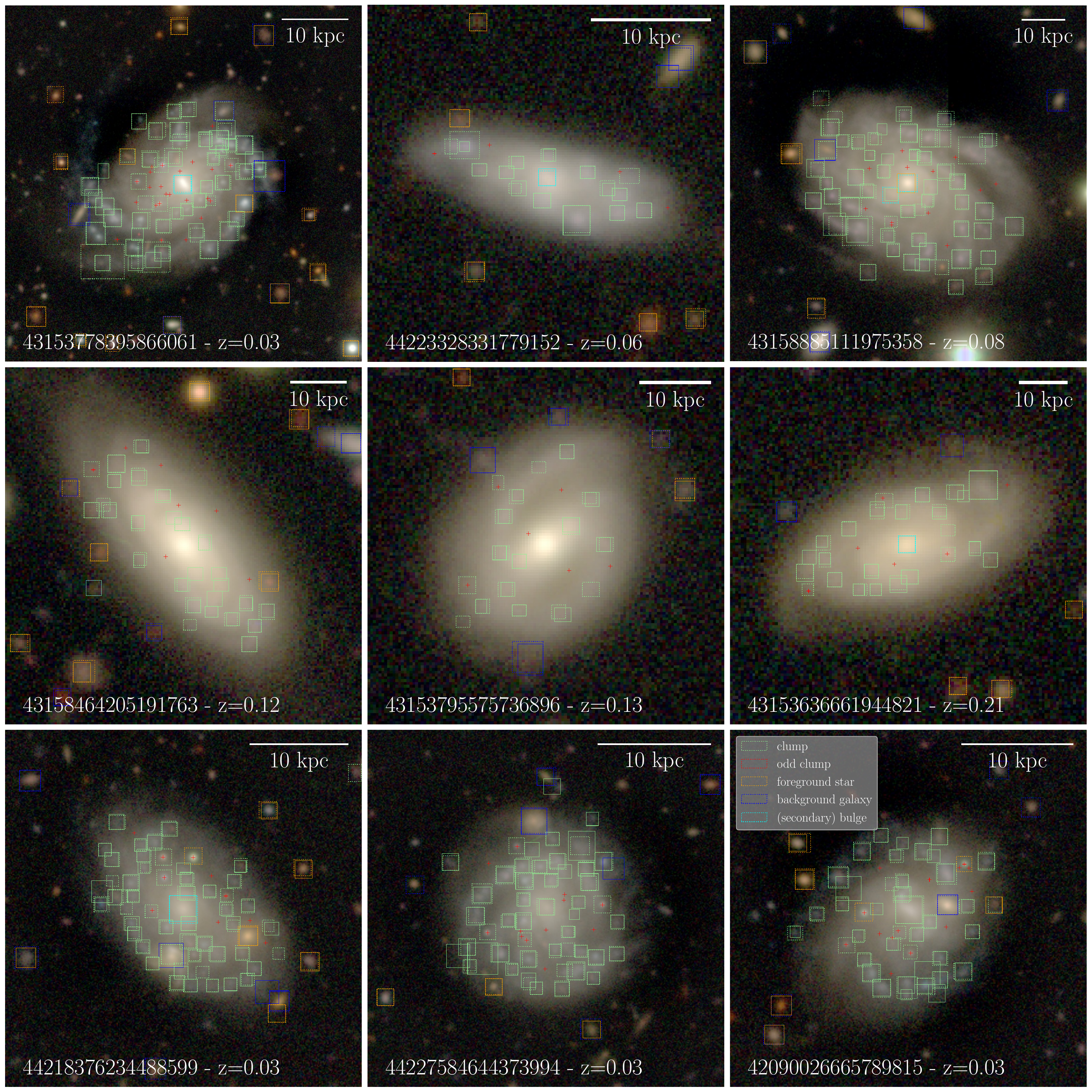}
    \caption[RGB-composite images showing the model detections with simulated clumps injected.]{RGB-composite images (generated using the GRI-bands) showing the model detections with simulated clumps injected. The positions of the simulated clumps are indicated by small red markers in the images and the model detections are shown as bounding boxes around clump complexes (solid boxes: detections on the real images, dotted boxes: additional detections from the images with injected clumps). Simulated clumps that are not detected by the model do not show a bounding box around their position. Some detections are wrongly classified as `foreground stars' by the model (e.g. objects 44218376234488599 and 42090026665789815 in the bottom row).}
    \label{fig:hsc_det_over_model_dev_eval_complete_detect}
\end{figure*}

Using the six single \textit{ugrizy}-filter band images, we applied the Faster R-CNN models on the 13,789 galaxies with injected simulated clumps. We also applied the postprocessing steps (Section \ref{sec:frcnn_model_postprocess}) and extracted the flux peaks from the model detections as described in Section \ref{sec:frcnn_model_peaks}. Figure \ref{fig:hsc_det_over_model_dev_eval_complete_detect} shows RGB-composite images with detections and the injected simulated clumps for nine examples. The detections originating from the real, unaltered image are shown by solid line bounding boxes and dotted line bounding boxes show detections for the run with the simulated clumps injected. Simulated clumps are marked in the images from Figure \ref{fig:hsc_det_over_model_dev_eval_complete_detect} with small red crosses and have dotted-line bounding boxes around them if the model detected them. 

\begin{figure}
    \centering
    \includegraphics[width=0.45\columnwidth]{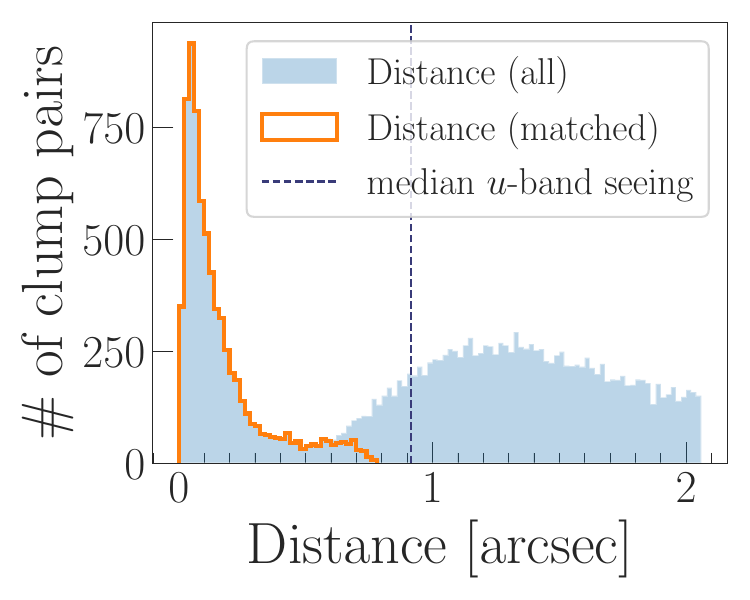}
    \caption[Distribution of the distance in arcsec between the simulated clumps and the closest detections.]{Distribution of the distance in arcsec between the simulated clumps and the closest detections. The distribution of those clumps that are considered a matching or successful detection are shown in orange.}
    \label{fig:hsc_det_over_model_dev_eval_complete_dist}
\end{figure}

The extracted clumps from the model detections were then crossmatched with the `ground truth' set of simulated clumps, for which the coordinates are known (Section \ref{sec:sims}). We counted a successful detection if the distance between the extracted flux peak of a predicted clump candidate and simulated clump is less than 0.75 of the image-specific \textit{u}-band seeing FWHM. We chose this threshold based on the distribution of the distance in arcsec between the simulated clumps and the closest detections while also allowing for some margin (Fig. \ref{fig:hsc_det_over_model_dev_eval_complete_dist}). Of the 32,241 simulated clumps, 7,161 were detected and correctly classified as a clump by the FRCNN model, which results in an overall detection ratio, or completeness, of 22.21\%. From Figure \ref{fig:hsc_det_over_model_dev_eval_complete_detect} it appears that it is mainly the fainter clumps in the simulated sample that were not detected. Also, some detections were misclassified as `foreground stars' by the model (see, for example, objects 44218376234488599 and 42090026665789815 in the bottom row of Figure \ref{fig:hsc_det_over_model_dev_eval_complete_detect}). In total, there were 1,420 clumps from the simulated sample that were detected but not classified as a `clump' by the model. These misclassified objects are predominantly the brightest simulated clump complexes with only one simulated clump, which made them indistinguishable from stars in the foreground of the host galaxy to the model.

\begin{figure}
    \centering
    \subfloat[\centering Model completeness per $m_{\mathrm{AB}}$-bin of the 6-channel (\textit{ugrizy}) FRCNN model. \label{fig:hsc_det_over_model_dev_eval_complete_mags}]{{\includegraphics[width=1\columnwidth]{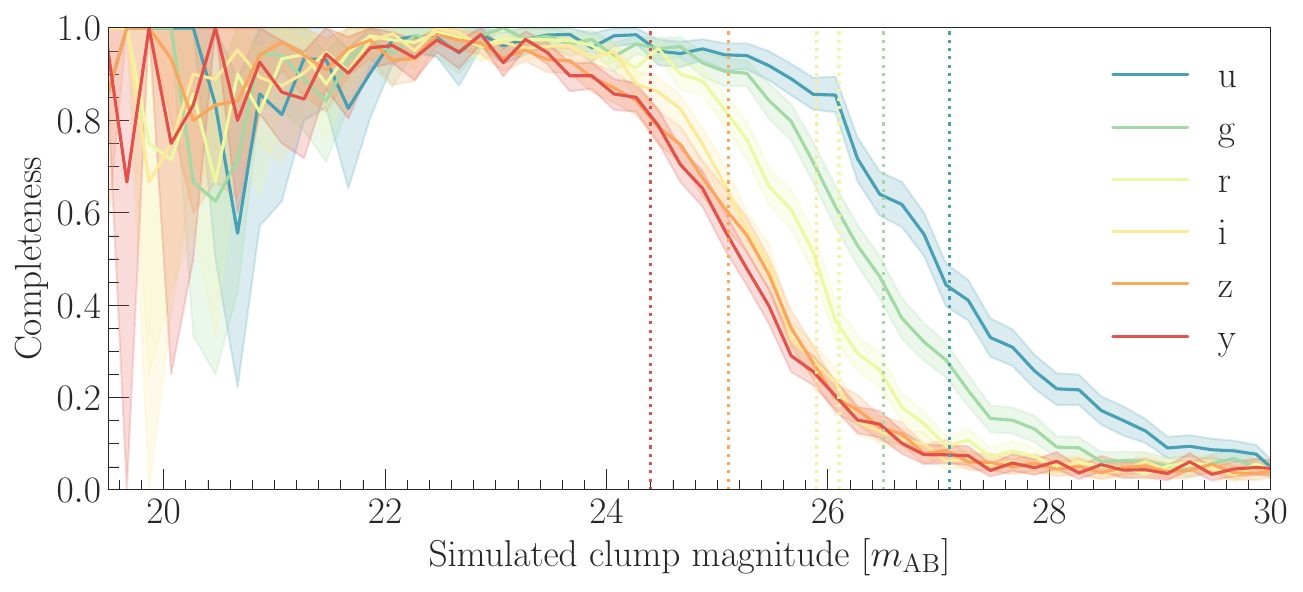} }}
    \\
    \subfloat[\centering Model completeness per $m_{\mathrm{AB}}$-bin of the 5-channel (\textit{grizy}) FRCNN model. \label{fig:hsc_det_over_model_dev_eval_complete_mags_GRIZY}]{{\includegraphics[width=1\columnwidth]{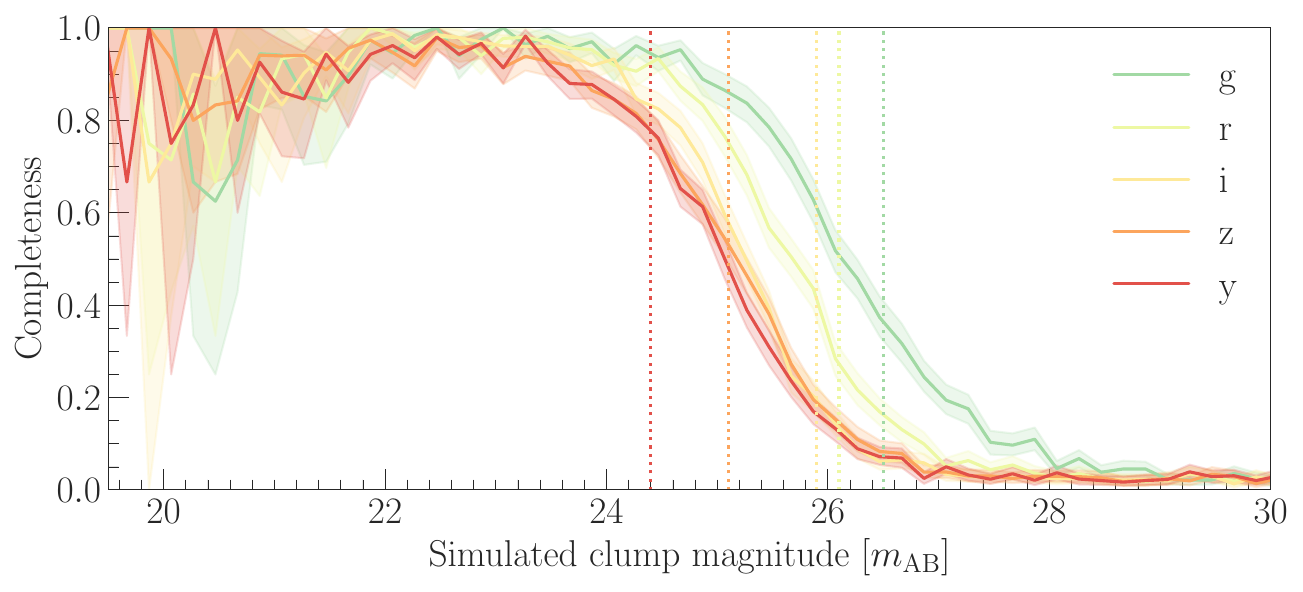} }}
    \caption[Completeness of the 5- and 6-channel FRCNN model as a function of apparent magnitude of the simulated clumps.]{Detection completeness of the 6-channel (a) and 5-channel (b) FRCNN model with respect to the simulated clumps. The plots show the fraction of detected simulated clumps in magnitude $m_{\mathrm{AB}}$-bins with bin widths of $0.2\,m_{\mathrm{AB}}$. Shaded areas show the 95\% confidence interval and the dotted vertical lines mark the detection limits for each filter band.}
    \label{fig:hsc_det_over_model_dev_eval_complete_mags_UGRIZY_GRIZY}
\end{figure}

We also calculated the completeness for apparent magnitude bins with bin widths of $0.2\,m_{\mathrm{AB}}$ in each filter band (Figure \ref{fig:hsc_det_over_model_dev_eval_complete_mags}). For objects brighter than detection limit of the \textit{y}-band, the completeness is well above 80\% and mostly close to 100\%. At the brighter end ($m_{\mathrm{AB}} \lesssim 21$), the estimated completeness varies more due to the low number of bright simulated clumps (see also Figure \ref{fig:hsc_det_over_model_dev_eval_sims_magDistribution}) and also due to the objects wrongly classified as `foreground stars'.

The sample of simulated clumps contains many low-mass clumps to specifically probe the detection rate for faint objects. However, the real observations are limited by the $5\sigma$ point-source depth limits in each of the used filter bands (for CLAUDS: $u\leq 27.1\,m_{\mathrm{AB}}$ and HSC-SSP: $g\leq 26.5\,m_{\mathrm{AB}}$, $r\leq 26.1\,m_{\mathrm{AB}}$, $i\leq 25.9\,m_{\mathrm{AB}}$, $z\leq 25.1\,m_{\mathrm{AB}}$, $y\leq 24.4\,m_{\mathrm{AB}}$). For clumps that are fainter than the detection limit, the SNR can be too low to measure the flux of the clumps accurately.

\begin{figure}
    \centering
    \subfloat[\centering \textit{u}-band apparent magnitude. \label{fig:sim_clumps_det_completeness_clumps_a}]{{\includegraphics[width=0.5\columnwidth]{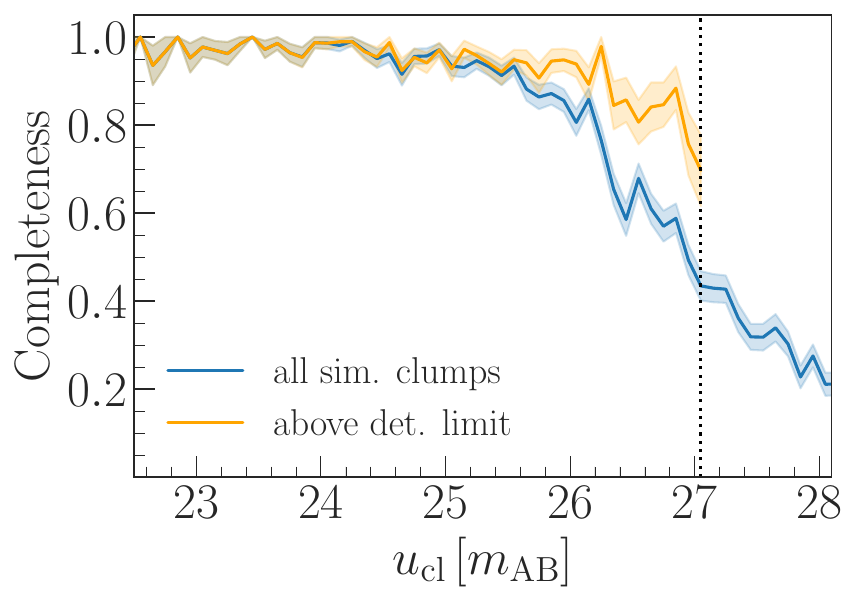} }}
    \subfloat[\centering Colour (\textit{u}-\textit{r}). \label{fig:sim_clumps_det_completeness_clumps_b}]{{\includegraphics[width=0.5\columnwidth]{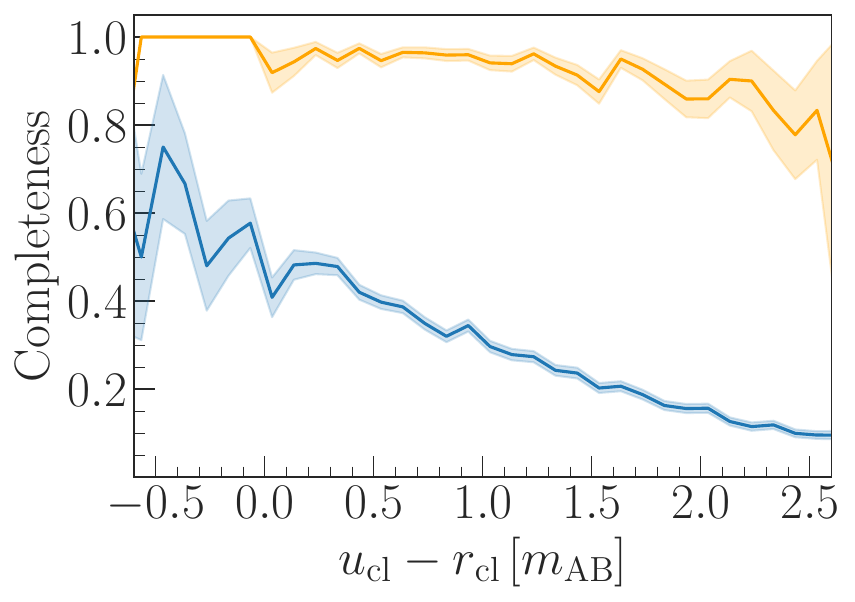} }}
    \\
    \subfloat[\centering Colour contrast \textit{u}-band. \label{fig:sim_clumps_det_completeness_clumps_c}]{{\includegraphics[width=0.5\columnwidth]{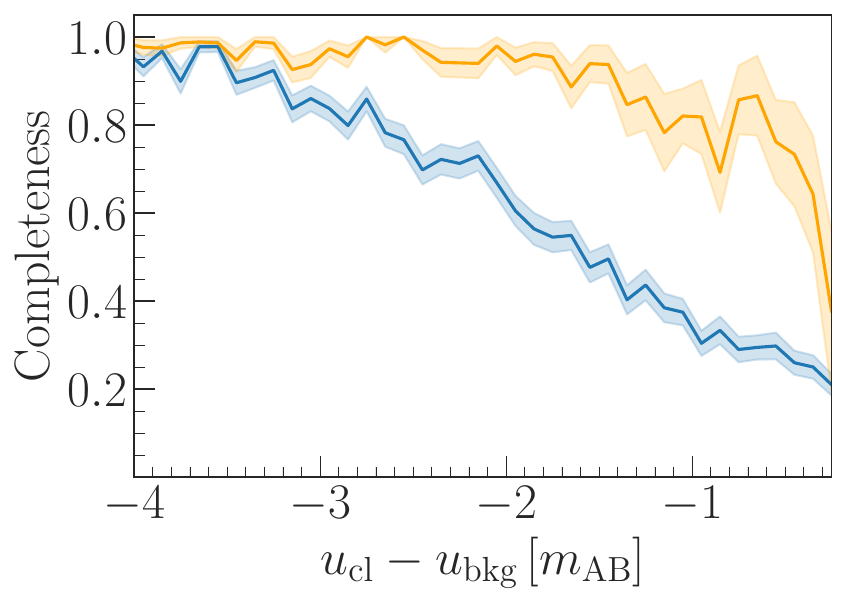} }}
    \subfloat[\centering Radial distance. \label{fig:sim_clumps_det_completeness_clumps_d}]
    {{\includegraphics[width=0.5\columnwidth]{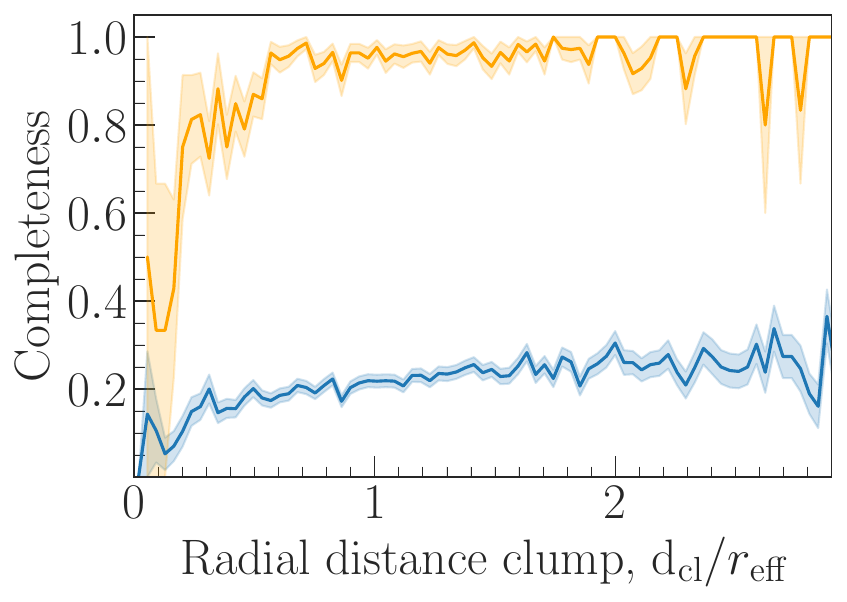} }}
    \\
    \subfloat[\centering Simulated clump stellar mass. \label{fig:sim_clumps_det_completeness_clumps_e}]{{\includegraphics[width=0.5\columnwidth]{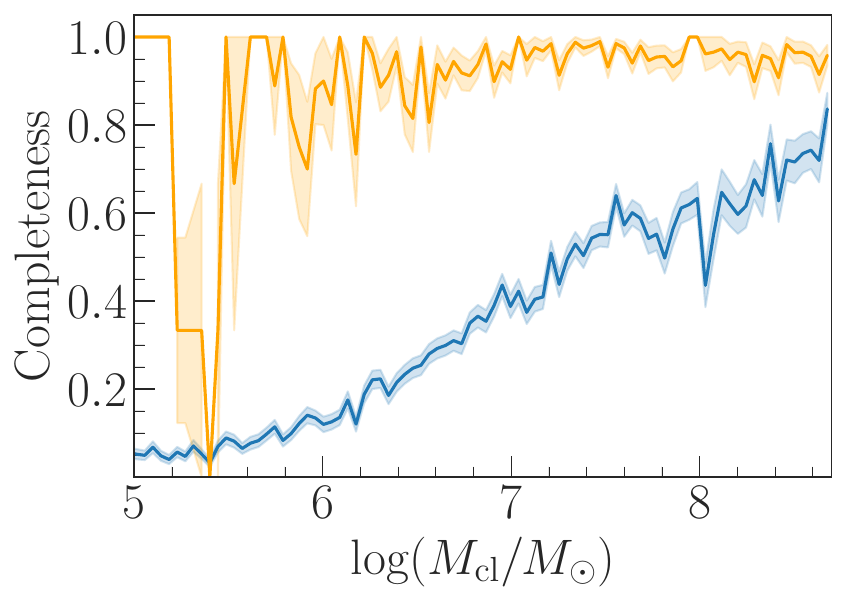} }}
    \subfloat[\centering Mass ratio clump/galaxy. \label{fig:sim_clumps_det_completeness_clumps_f}]{{\includegraphics[width=0.5\columnwidth]{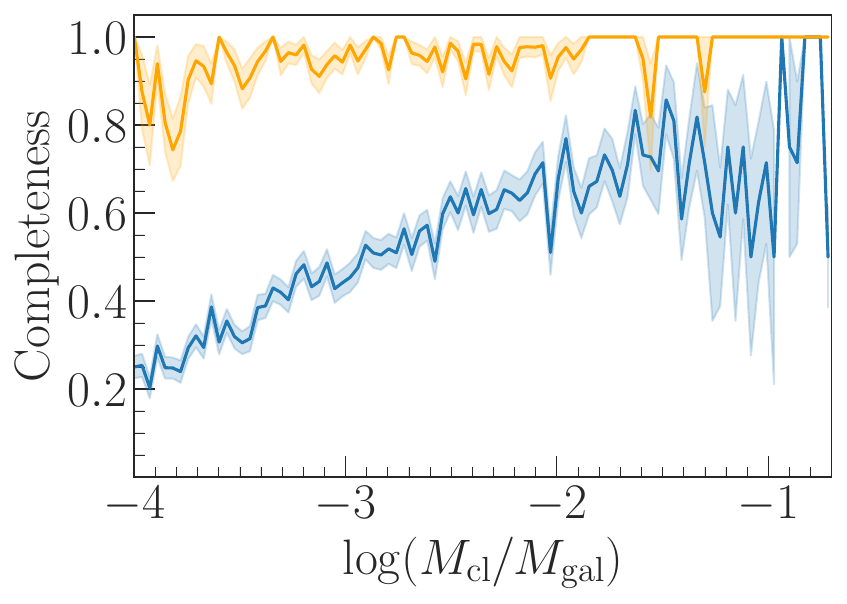} }}
    \caption[Detection completeness of the simulated clumps as a function of different physical clump properties for the 6-channel FRCNN model.]{Detection completeness of the simulated clumps as a function of different physical clump properties for the 6-channel FRCNN model. The completeness is plotted in blue for the full sample of simulated clumps and in orange for clumps that are brighter than the detection limit of the two surveys ($u\leq 27.1\,m_{\mathrm{AB}}$, $g\leq 26.5\,m_{\mathrm{AB}}$, $r\leq 26.1\,m_{\mathrm{AB}}$, $i\leq 25.9\,m_{\mathrm{AB}}$, $z\leq 25.1\,m_{\mathrm{AB}}$ and $y\leq 24.4\,m_{\mathrm{AB}}$). The shaded areas show the $1\sigma$ errors and the dotted vertical line in plot (a) marks the limiting magnitude in the \textit{u}-band for the CLAUDS survey.}
    \label{fig:sim_clumps_det_completeness_clumps}
\end{figure}

\begin{figure}
    \centering
    \subfloat[\centering Host galaxy redshift. \label{fig:sim_clumps_det_completeness_gal_a}]{{\includegraphics[width=0.5\columnwidth]{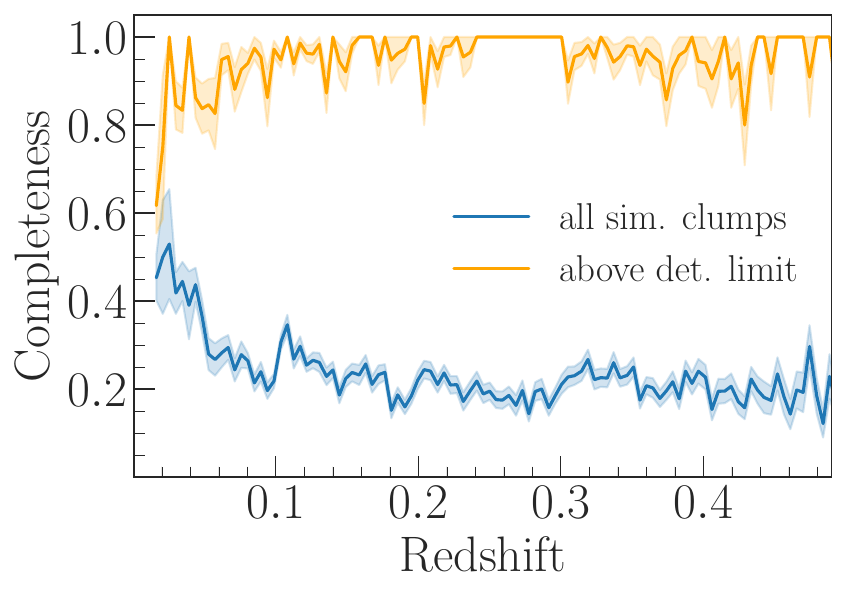} }}
    \subfloat[\centering Host galaxy stellar mass. \label{fig:sim_clumps_det_completeness_gal_b}]{{\includegraphics[width=0.5\columnwidth]{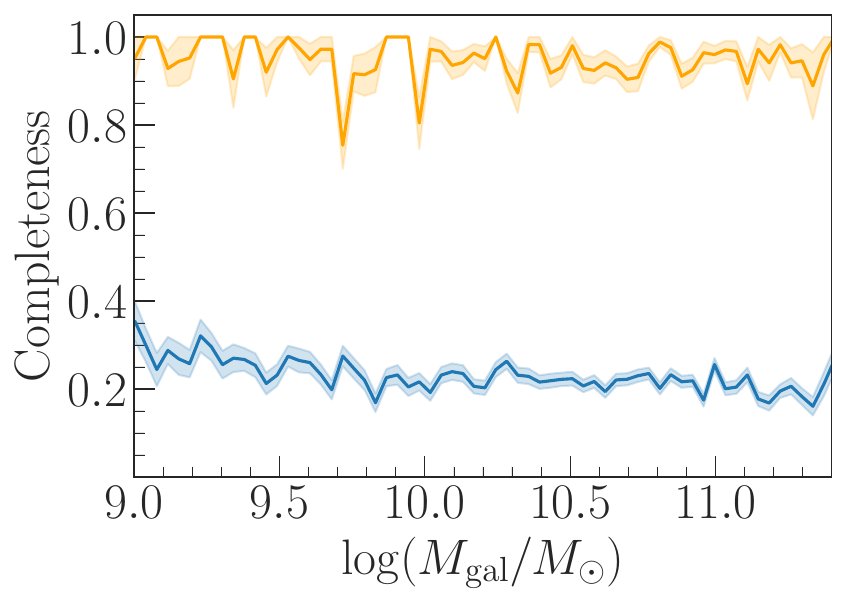} }}
    \\
    \subfloat[\centering Host galaxy sSFR. \label{fig:sim_clumps_det_completeness_gal_c}]{{\includegraphics[width=0.5\columnwidth]{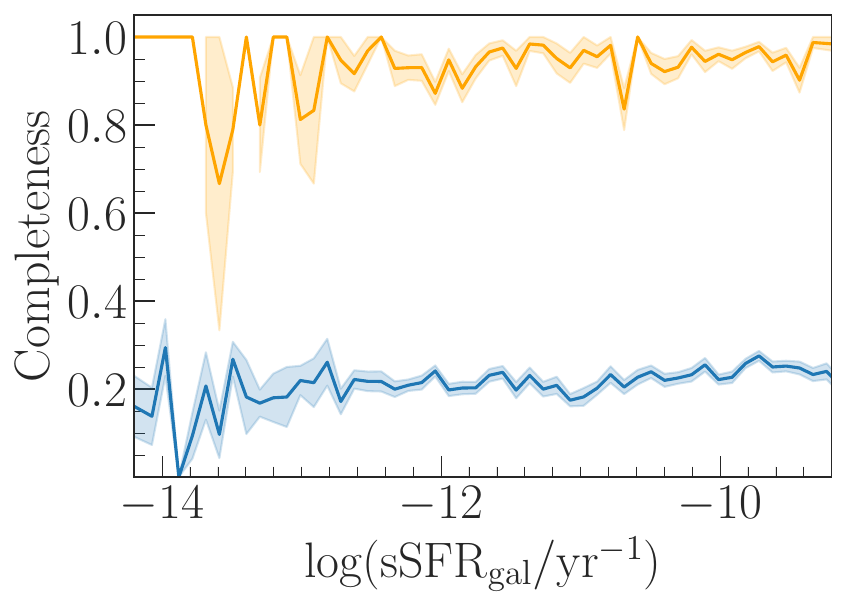} }}
    \caption[Detection completeness of the simulated clumps as a function of different host galaxy parameters.]{Similar to Figure \ref{fig:sim_clumps_det_completeness_clumps} but showing the detection completeness of the simulated clumps as a function of different host galaxy properties.}
    \label{fig:sim_clumps_det_completeness_gal}
\end{figure}

Therefore, we determined the model completeness for only those simulated clumps that are brighter than the detection limit, i.e. the apparent magnitude values in all of the six filter bands are required to be brighter than or equal to the instrument-specific magnitude limits. Figures \ref{fig:sim_clumps_det_completeness_clumps} and \ref{fig:sim_clumps_det_completeness_gal} compare the model completeness as a function of different clump and host galaxy properties of the sample that is brighter than the detection limit to the sample of all simulated clumps.

As expected, the detection completeness is increased for the sample of brighter clumps compared to the full sample containing many fainter clumps. The detection completeness is well above 0.8 and close to 1.0 over most of the ranges of the physical clump and host galaxy properties. Only if the contrast between the simulated clump and the diffuse galaxy background (measured from an annulus spanning radii of 1.5 to 2.0 PSF-FWHM around the clump) becomes weaker, i.e. $u_{\mathrm{cl}} - u_{\mathrm{bkg}} \gtrsim -1$ (Figure \ref{fig:sim_clumps_det_completeness_clumps_c}) or the clumps are close to the centre of the galaxy (Figure \ref{fig:sim_clumps_det_completeness_clumps_d}), does the detection completeness drop noticeably. The distance of the clumps from the centroid of the host galaxy (as reported from the HSC-SSP PDR3 catalogue) $\mathrm{d}_{\mathrm{cl}}$ is measured in units of the effective or half-light radius $r_{\mathrm{eff}}$ of the galaxy. Simulated clumps that are less than $\mathrm{d}_{\mathrm{cl}} \lesssim 0.5\,r_{\mathrm{eff}}$ from the galaxy centre are more likely to be outshone by the stars of the central bulge where the contrast $u_{\mathrm{cl}} - u_{\mathrm{bkg}}$ becomes too low for a successful clump detection.

Other fluctuations of the completeness are mainly due to the low number of clumps that are brighter than the detection limit (e.g. the drop in model completeness seen for clump masses below $\log(M_{\mathrm{cl}}/M_\odot) \lesssim 6.3$ in Figure \ref{fig:sim_clumps_det_completeness_clumps_e}). The model also reliably detects the brighter clumps over the full redshift range of the sample of galaxies analysed in this study and the detection completeness appears to be independent of the redshift, stellar mass or sSFR of the host galaxy (Figure \ref{fig:sim_clumps_det_completeness_gal}).

Table \ref{tab:sim_clumps_det_completeness} lists the overall completeness calculated for different selections of the simulated clump sample. Besides the completeness for only those simulated clumps that are brighter than the detection limits, we defined additional selections based on clump-galaxy \textit{u}-band flux ratio cuts \citep[similar to][for example]{Guo2015,Guo2018,Adams2022,Popp2024} and a clump stellar mass cut at $M_{\mathrm{cl}} \geq 10^7 M_\odot$ \citep[similar to][for example]{HuertasCompany2020}.

\begin{table}
	\centering
	\caption[Completeness of the 6-channel FRCNN model measured on simulated clumps.]{Completeness of the 6-channel FRCNN model measured on simulated clumps. The completeness is calculated for different selections of simulated clumps: (1) all simulated clumps, (2) only simulated clumps that have \textit{ugrizy} fluxes greater then the detection limit in all filter bands, (3) simulated clumps with \textit{u}-band fluxes $F_{u,\mathrm{cl}}$ of at least 3\% of the host galaxy \textit{u}-band flux $F_{u,\mathrm{gal}}$ ($F_{u,\mathrm{cl}}/F_{u,\mathrm{gal}} \geq 0.03$) or (4) 8\% ($F_{u,\mathrm{cl}}/F_{u,\mathrm{gal}} \geq 0.08$) and (5) simulated clumps with a stellar mass of $M_{\mathrm{cl}} \geq 10^7 M_\odot$.}
    \label{tab:sim_clumps_det_completeness}
	\footnotesize
        \begin{tabular}{clrrr}
		\hline
         & Selection & \multicolumn{1}{c}{Total} & \multicolumn{1}{c}{Detected} & \multicolumn{1}{c}{Completeness} \\
		\hline
        1 & All sim. clumps & 32,241 & 7,161 & 22.21\% \\
        2 & Above det. limit & 3,378 & 3,176 & 94.02\% \\
        3 & $F_{u,\mathrm{cl}}/F_{u,\mathrm{gal}} \geq 0.03$ & 2,534 & 2,271 & 89.62\% \\
          & \textit{thereof:} above det. limit & 1,894 & 1,833 & 96.78\% \\
        4 & $F_{u,\mathrm{cl}}/F_{u,\mathrm{gal}} \geq 0.08$ & 1,423 & 1,329 & 93.39\% \\
          & \textit{thereof:} above det. limit & 1,224 & 1,183 & 96.65\% \\
        5 & $M_{\mathrm{cl}} \geq 10^7 M_\odot$ & 7,680 & 4,143 & 53.95\% \\
          & \textit{thereof:} above det. limit & 2,468 & 2,363 & 95.75\% \\
        \hline
	\end{tabular}
\end{table}

\subsubsection{Completeness of the 5-channel (\textit{grizy}) model detections}\label{sec:hsc_det_over_model_dev_eval_complete_GRIZY}
Using the same methods as we did for the \textit{ugrizy} FRCNN model, we tested how complete the model detections are using the FRCNN model that has been trained on only the five \textit{grizy}-filter band images. The model detections were matched with the sample of simulated clumps in the same way as for the 6-channel model, such that a successful detection is counted if the detected flux peak of the clump candidate is less than 0.75 of the image-specific \textit{u}-band seeing FWHM away from the flux peak of the simulated clump. Here, we kept the maximum distance tied to the \textit{u}-band seeing for consistency.

Similarly to Figure \ref{fig:hsc_det_over_model_dev_eval_complete_mags} and Table \ref{tab:sim_clumps_det_completeness}, Figure \ref{fig:hsc_det_over_model_dev_eval_complete_mags_GRIZY} shows the completeness as a function of apparent magnitude of the simulated clump in each filter band and Table \ref{tab:sim_clumps_det_completeness_GRIZY} lists the completeness for the 5-channel model detections on the simulated clumps.

\begin{table}
	\centering
	\caption[Completeness of the 5-channel FRCNN model measured on simulated clumps.]{Completeness of the 5-channel FRCNN model measured on simulated clumps. Similar to Table \ref{tab:sim_clumps_det_completeness}, the completeness is calculated for different selections of simulated clumps: (1) all simulated clumps, (2) only simulated clumps that have \textit{grizy} fluxes greater then the detection limit in all filter bands, (3) simulated clumps with \textit{g}-band fluxes $F_{g,\mathrm{cl}}$ of at least 3\% of the host galaxy \textit{g}-band flux $F_{g,\mathrm{gal}}$ ($F_{g,\mathrm{cl}}/F_{g,\mathrm{gal}} \geq 0.03$) or (4) 8\% ($F_{g,\mathrm{cl}}/F_{g,\mathrm{gal}} \geq 0.08$) and (5) simulated clumps with a stellar mass of $M_{\mathrm{cl}} \geq 10^7 M_\odot$.}
    \label{tab:sim_clumps_det_completeness_GRIZY}
	\footnotesize
        \begin{tabular}{clrrr}
		\hline
         & Selection & \multicolumn{1}{c}{Total} & \multicolumn{1}{c}{Detected} & \multicolumn{1}{c}{Completeness} \\
		\hline
        1 & All sim. clumps & 32,241 & 6,151 & 19.08\% \\
        2 & Above det. limit & 3,580 & 3,275 & 91.48\% \\
        3 & $F_{g,\mathrm{cl}}/F_{g,\mathrm{gal}} \geq 0.03$ & 1,467 & 1,366 & 93.12\% \\
          & \textit{thereof:} above det. limit & 1,337 & 1,264 & 94.54\% \\
        4 & $F_{g,\mathrm{cl}}/F_{g,\mathrm{gal}} \geq 0.08$ & 661 & 623 & 94.25\% \\
          & \textit{thereof:} above det. limit & 650 & 612 & 94.15\% \\
        5 & $M_{\mathrm{cl}} \geq 10^7 M_\odot$ & 7,680 & 3,816 & 49.69\% \\
          & \textit{thereof:} above det. limit & 2,618 & 2,442 & 93.28\% \\
        \hline
	\end{tabular}
\end{table}

The detection completeness of the 5-channel model differs by 2-3\% relative to the completeness of the 6-channel object detection model (Section \ref{sec:hsc_det_over_model_dev_eval_complete_UGRIZY}). The omitted \textit{u}-band data cannot be fully compensated for by the five \textit{grizy}-filter bands from HSC-SSP. Of the 7,161 simulated clumps detected by the 6-channel model (Table \ref{tab:sim_clumps_det_completeness}), 5,958 were also detected by the 5-channel model and 1,203 were only detected by the 6-channel model (193 simulated clumps were only detected by the 5-channel model). The median stellar mass of the 1,203 simulated clumps that were only detected by the 6-channel model is $\bar{M}_{\star,\mathrm{sim}} = 3.09 \times 10^6 M_\odot$, which is almost by a factor of 10 lower than the median stellar mass of the remaining 5,958 detected clumps ($\bar{M}_{\star,\mathrm{sim}} = 2.01 \times 10^7 M_\odot$). The increased depth of the \textit{u}-band filter from CLAUDS, i.e. the better detection limit of $u\leq 27.1\,m_{\mathrm{AB}}$, is primarily responsible for the additional detections and the difference almost vanishes if the detections are required to be above the detection limit.

\begin{figure}
    \centering
    \subfloat[\centering \textit{g}-band apparent magnitude. \label{fig:sim_clumps_det_completeness_clumps_GRIZY_a}]{{\includegraphics[width=0.5\columnwidth]{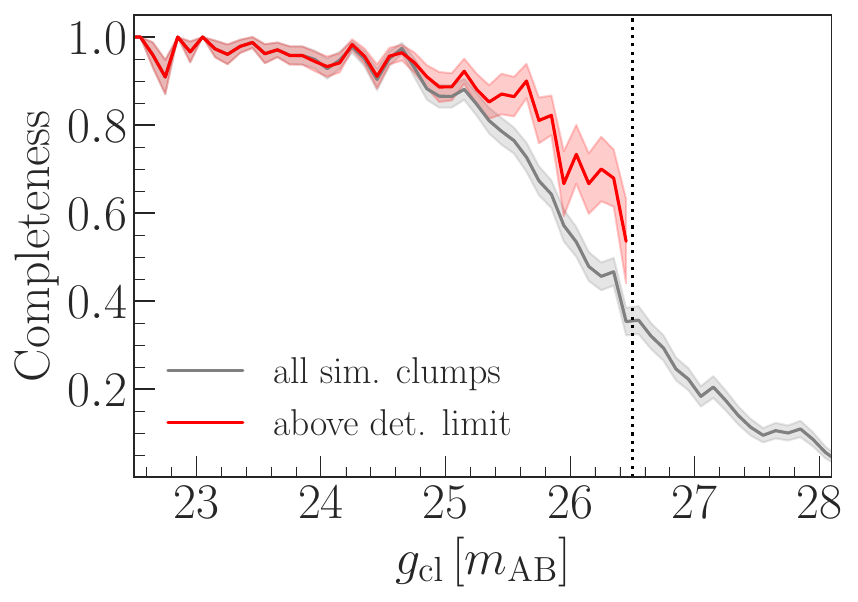} }}
    \subfloat[\centering Colour (\textit{g}-\textit{r}). \label{fig:sim_clumps_det_completeness_clumps_GRIZY_b}]{{\includegraphics[width=0.5\columnwidth]{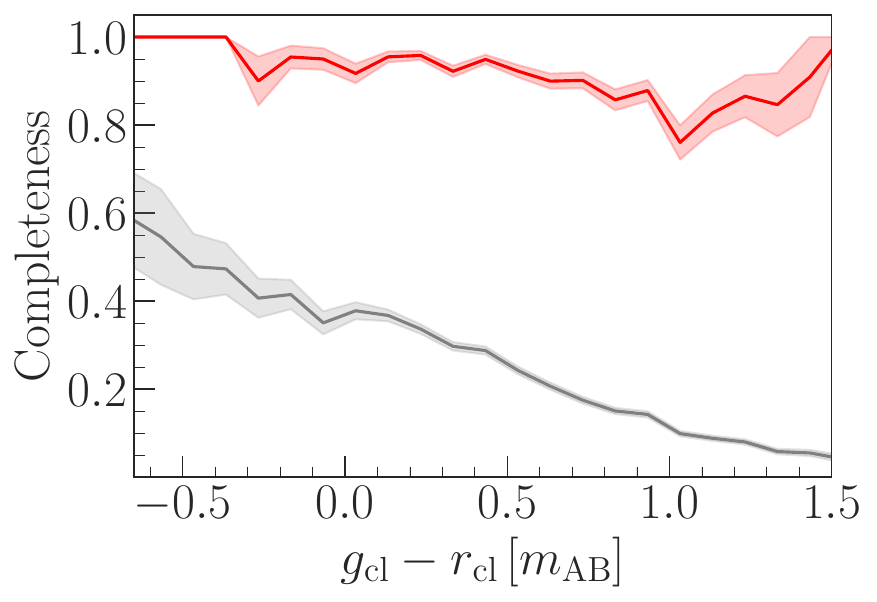} }}
    \\
    \subfloat[\centering Colour contrast \textit{g}-band. \label{fig:sim_clumps_det_completeness_clumps_GRIZY_c}]{{\includegraphics[width=0.5\columnwidth]{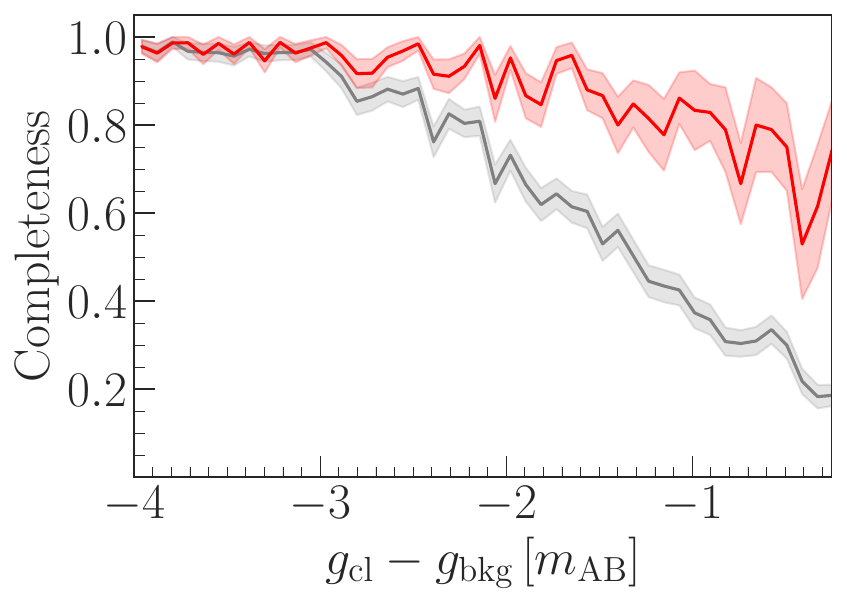} }}
    \subfloat[\centering Radial distance. \label{fig:sim_clumps_det_completeness_clumps_GRIZY_d}]{{\includegraphics[width=0.5\columnwidth]{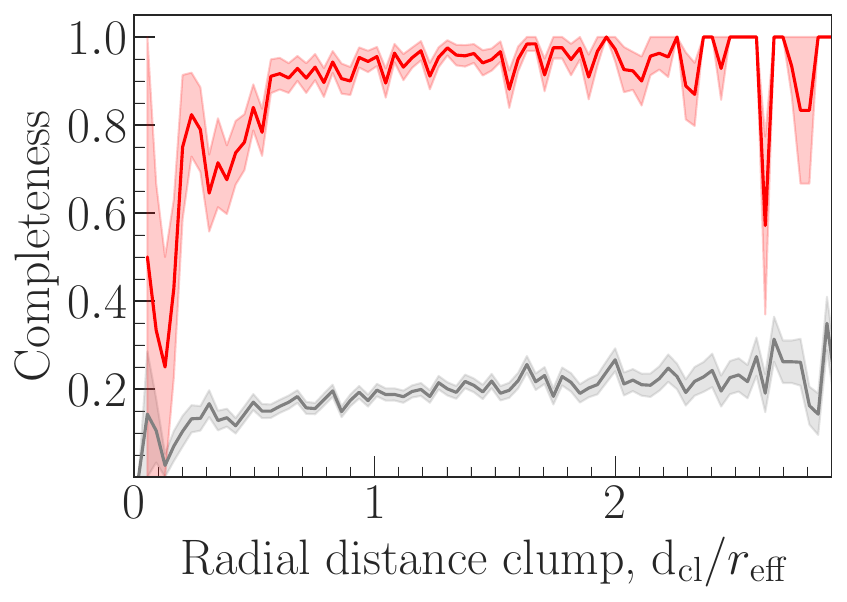} }}
    \\
    \subfloat[\centering Simulated clump stellar mass. \label{fig:sim_clumps_det_completeness_clumps_GRIZY_e}]{{\includegraphics[width=0.5\columnwidth]{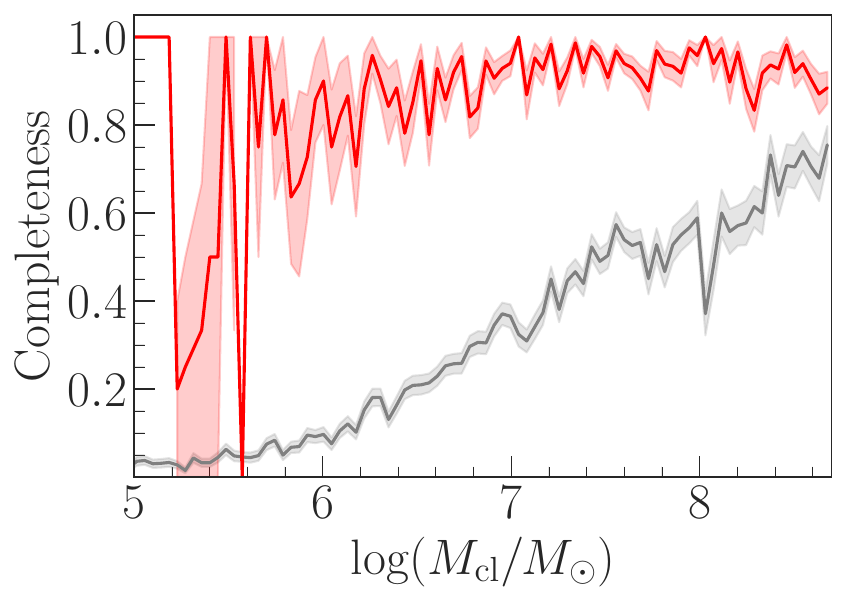} }}
    \subfloat[\centering Mass ratio clump/galaxy. \label{fig:sim_clumps_det_completeness_clumps_GRIZY_f}]{{\includegraphics[width=0.5\columnwidth]{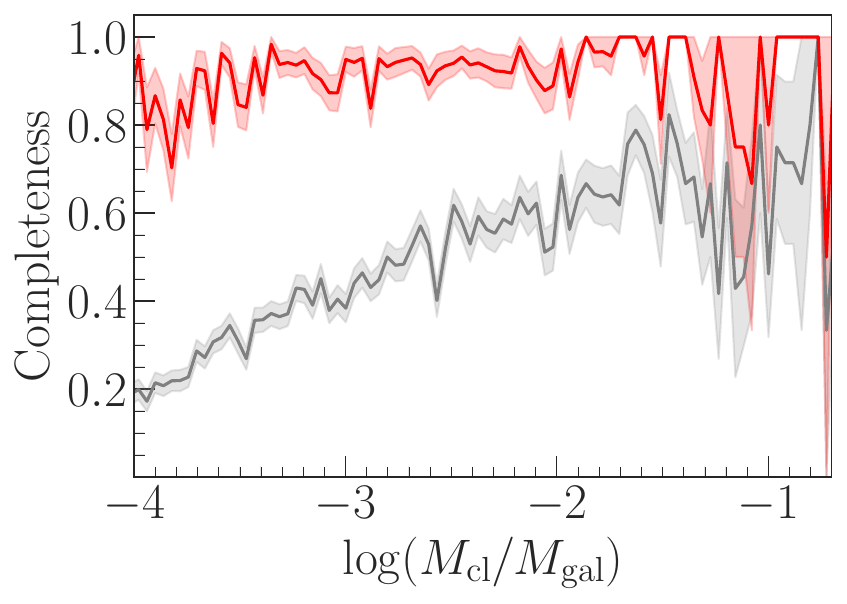} }}
    \caption[Detection completeness of the simulated clumps as a function of different physical clump properties for the 5-channel FRCNN model.]{Detection completeness of the simulated clumps as a function of different physical clump properties for the 5-channel FRCNN model. The completeness is plotted in grey for the full sample of simulated clumps and in red for clumps that are brighter than the detection limit for the HSC-SSP filter bands ($g\leq 26.5\,m_{\mathrm{AB}}$, $r\leq 26.1\,m_{\mathrm{AB}}$, $i\leq 25.9\,m_{\mathrm{AB}}$, $z\leq 25.1\,m_{\mathrm{AB}}$ and $y\leq 24.4\,m_{\mathrm{AB}}$). The shaded areas show the $1\sigma$ errors and the dotted vertical line in plot (a) marks the limiting magnitude in the \textit{g}-band for the HSC-SSP survey.}
    \label{fig:sim_clumps_det_completeness_clumps_GRIZY}
\end{figure}

\begin{figure}
    \centering
    \subfloat[\centering Host galaxy redshift. \label{fig:sim_clumps_det_completeness_gal_GRIZY_a}]{{\includegraphics[width=0.5\columnwidth]{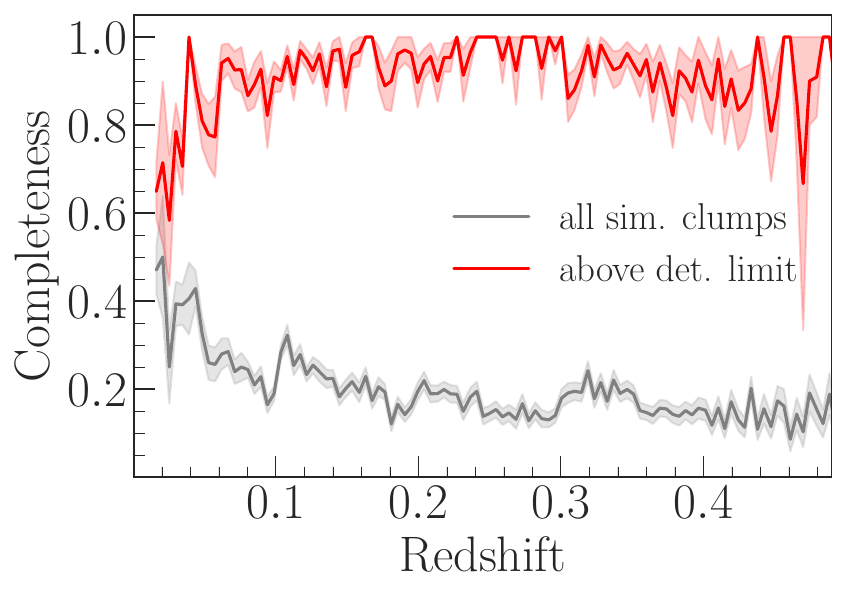} }}
    \subfloat[\centering Host galaxy stellar mass. \label{fig:sim_clumps_det_completeness_gal_GRIZY_b}]{{\includegraphics[width=0.5\columnwidth]{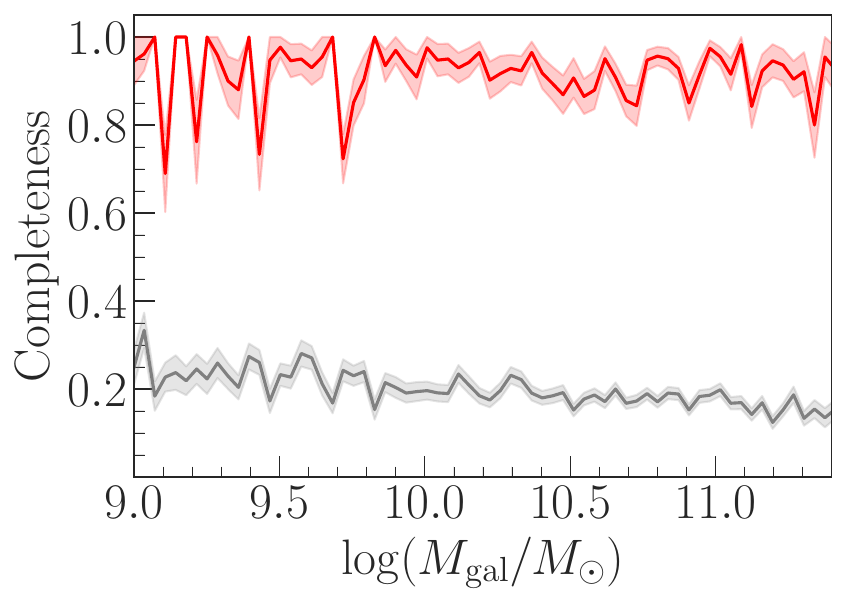} }}
    \\
    \subfloat[\centering Host galaxy sSFR. \label{fig:sim_clumps_det_completeness_gal_GRIZY_c}]{{\includegraphics[width=0.5\columnwidth]{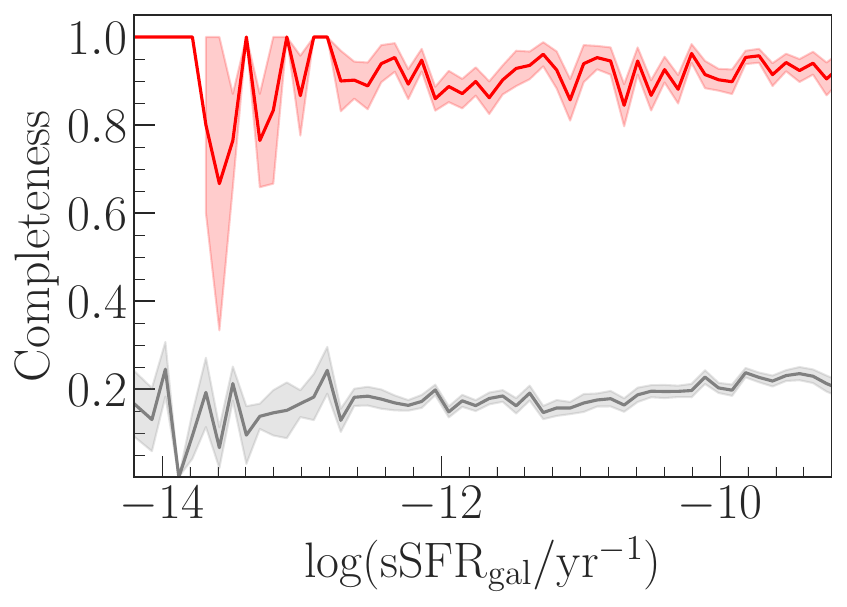} }}
    \caption[Detection completeness of the simulated clumps as a function of different host galaxy properties for the 5-channel FRCNN model.]{Similar to Figure \ref{fig:sim_clumps_det_completeness_clumps_GRIZY} but showing the detection completeness of the simulated clumps as a function of different host galaxy properties.}
    \label{fig:sim_clumps_det_completeness_gal_GRIZY}
\end{figure}

The completeness for only those simulated clumps that are brighter than the detection limit of the five \textit{grizy}-filter bands from the HSC is generally $\gtrsim 0.8$ over most of the clump property ranges (Figure \ref{fig:sim_clumps_det_completeness_clumps_GRIZY}) and host galaxy property ranges (Figure \ref{fig:sim_clumps_det_completeness_gal_GRIZY}). Also, the detection completeness appears to be unrelated to the redshift, stellar mass or sSFR of the host galaxy (Figure \ref{fig:sim_clumps_det_completeness_gal_GRIZY}). As for the 6-channel model, the completeness starts to decrease if the clump flux approaches the detection limit (Figure \ref{fig:sim_clumps_det_completeness_clumps_GRIZY_a}), the contrast between the \textit{g}-band flux of the clump and the galaxy background becomes less than $\sim 1\,m_{\mathrm{AB}}$ (Figure \ref{fig:sim_clumps_det_completeness_clumps_GRIZY_c}), or the clump is located close to the centre of the galaxy (Figure \ref{fig:sim_clumps_det_completeness_clumps_GRIZY_d}). 

\subsubsection{Purity of the 6-channel (\textit{ugrizy}) model detections}\label{sec:hsc_det_over_model_dev_eval_complete_purity_UGRIZY}
The 7,161 detected simulated clumps are only a subset of the 37,966 clump candidates that were detected by the FRCNN model. This would imply a purity of $\sim 19\%$. However, the simulated clumps were injected into real galaxy images that also contain real clumps. The remaining 24,687 detections cannot be treated as false positives as they might be genuine clumps accurately detected by a model that has been trained on real data. This means that calculating the purity of the model detections based on the simulated clumps would be not representative of the true detection performance of the FRCNN model.

Instead, we masked the 37,966 clump detections that include the detections of the simulated clumps with the predictions of the FRCNN model that were made on images without the injected simulated clumps. As we outlined in Section \ref{sec:sims}, the bounding boxes of these detections were used to mask areas in a galaxy where no simulated clump is placed. After removing all detections that lie within any of these bounding boxes, the remaining 9,420 detections are assumed to be due to the addition of the simulated clumps to the images or false positive detections. Figure \ref{fig:hsc_det_over_model_dev_eval_complete_detect} illustrates this process for nine galaxy examples. The detections marked by solid and dotted bounding boxes were removed as they are detections common to images with and without injected simulated clumps. The remaining dotted bounding boxes represent the additional detections from images with the injected clumps only.

\begin{figure}
    \centering
    \subfloat[\centering Model purity per $m_{\mathrm{AB}}$-bin of the 6-channel (\textit{ugrizy}) FRCNN model. \label{fig:hsc_det_over_model_dev_eval_complete_purity_UGRIZY}]{{\includegraphics[width=1\columnwidth]{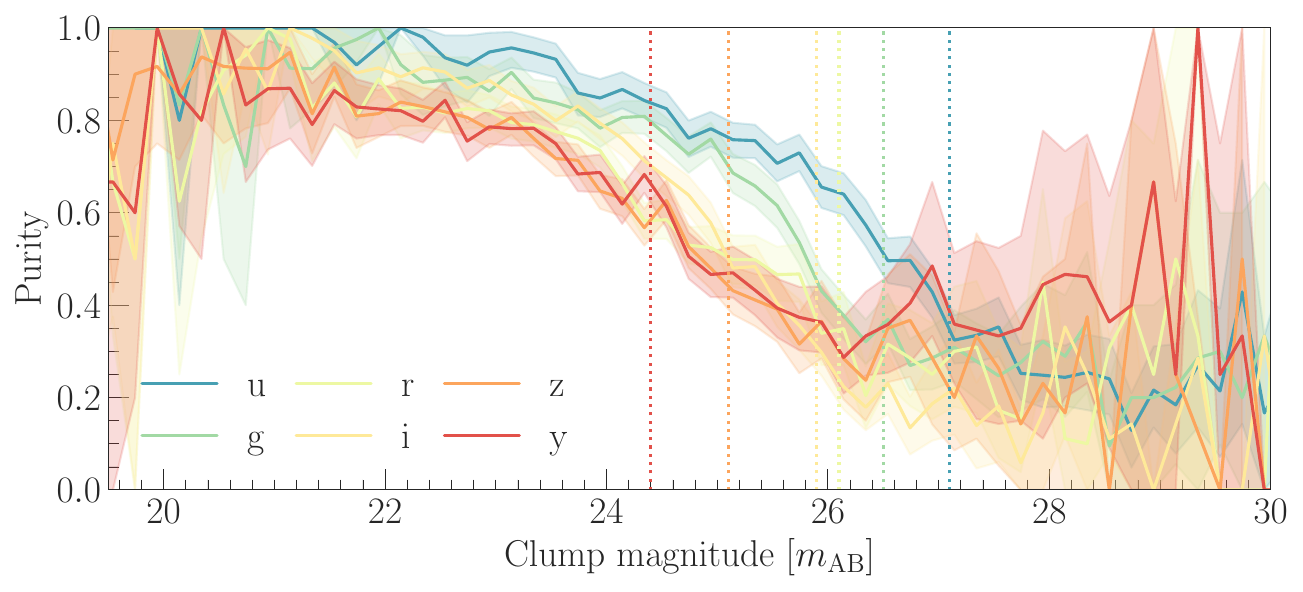} }}
    \\
    \subfloat[\centering Model purity per $m_{\mathrm{AB}}$-bin of the 5-channel (\textit{grizy}) FRCNN model. \label{fig:hsc_det_over_model_dev_eval_complete_purity_GRIZY}]{{\includegraphics[width=1\columnwidth]{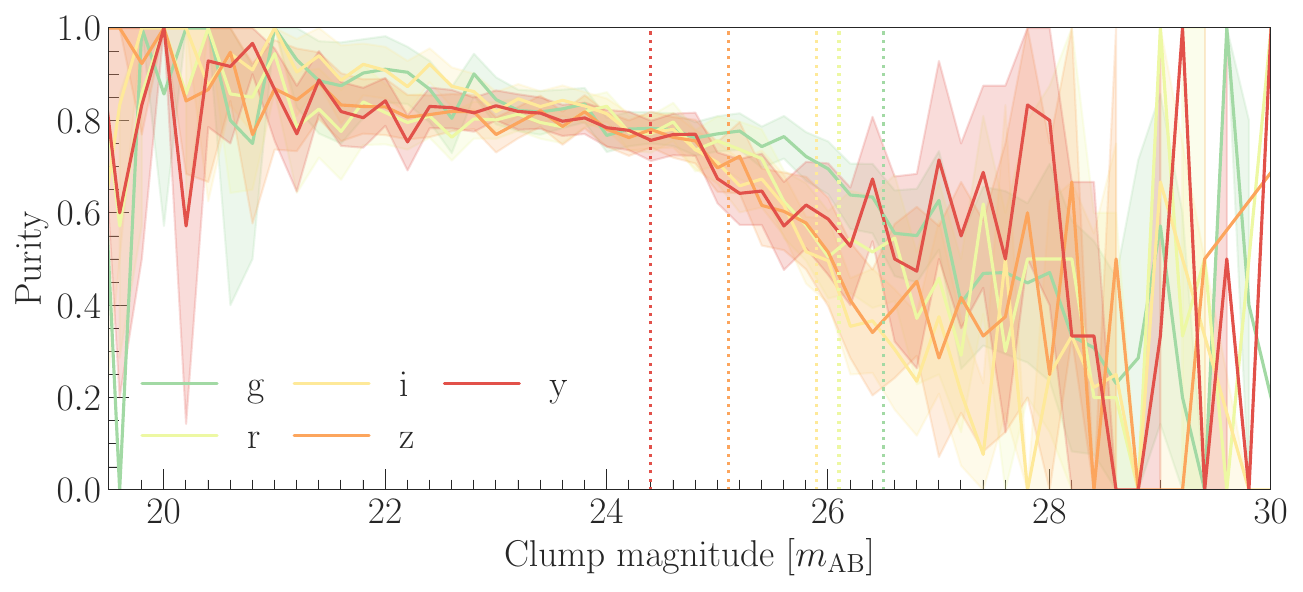} }}
    \caption[Purity of the 5- and 6-channel FRCNN model as a function of apparent magnitude of the simulated clumps.]{Purity of the 6-channel (a) and 5-channel (b) FRCNN model measured on simulated clumps as a function of apparent magnitude. The plots show the fraction of detected simulated clumps to all detections per magnitude $m_{\mathrm{AB}}$-bins with bin widths of $0.2\,m_{\mathrm{AB}}$. Shaded areas show the 95\% confidence interval and the dotted vertical lines mark the detection limits for each filter band.}
    \label{fig:hsc_det_over_model_dev_eval_complete_purity_UGRIZY_GRIZY}
\end{figure}

Based on the remaining 9,420 detections we determined the purity of the FRCNN model detections for all detected clumps and those with measured flux above the detection limits in all filter bands and above the two clump-galaxy flux ratios of $F_{u,\mathrm{cl}}/F_{u,\mathrm{gal}} \geq 0.03$ and $F_{u,\mathrm{cl}}/F_{u,\mathrm{gal}} \geq 0.08$ (Table \ref{tab:hsc_det_over_model_dev_eval_complete_purity_UGRIZY}). We measured the background-subtracted fluxes in all six filter bands for every detection using aperture photometry (see Appendix \ref{sec:photometry}) and plotted the purity of the model detections as a function of measured apparent magnitude for each \textit{ugrizy}-filter band in Figure \ref{fig:hsc_det_over_model_dev_eval_complete_purity_UGRIZY}. Compared to the completeness in Figure \ref{fig:hsc_det_over_model_dev_eval_complete_mags}, the purity is noticeably lower over most of the magnitude range. The purity remains lower than the completeness if different cuts are made on the detected sample based on either the flux directly or the clump-galaxy flux ratio in the \textit{u}-band, even though the overall purity is $\sim 76\%$, which is higher than the overall completeness (Table \ref{tab:hsc_det_over_model_dev_eval_complete_purity_UGRIZY}). 

\begin{table}
	\centering
	\caption[Purity of the 6-channel FRCNN model measured on simulated clumps.]{Purity of the 6-channel FRCNN model measured on simulated clumps. The purity is calculated for different selections: (1) all detected clumps, (2) only detected clumps that have \textit{ugrizy} fluxes greater then the detection limit in all filter bands, (3) detected clumps with \textit{u}-band fluxes $F_{u,\mathrm{cl}}$ of at least 3\% of the host galaxy \textit{u}-band flux $F_{u,\mathrm{gal}}$ ($F_{u,\mathrm{cl}}/F_{u,\mathrm{gal}} \geq 0.03$) and (4) 8\% ($F_{u,\mathrm{cl}}/F_{u,\mathrm{gal}} \geq 0.08$).}
    \label{tab:hsc_det_over_model_dev_eval_complete_purity_UGRIZY}
	\footnotesize
        \begin{tabular}{clrrr}
		\hline
         & Selection & \multicolumn{1}{c}{Total} & \multicolumn{1}{c}{Detected} & \multicolumn{1}{c}{Purity} \\
         &           &                           & \multicolumn{1}{c}{sim. clumps} & \\
		\hline
        1 & All detected clumps & 9,420 & 7,161 & 76.02\% \\
        2 & Above det. limit & 5,157 & 4,182 & 81.09\% \\
        3 & $F_{u,\mathrm{cl}}/F_{u,\mathrm{gal}} \geq 0.03$ & 3,630 & 2,843 & 78.32\% \\
          & \textit{thereof:} above det. limit & 3,045 & 2,516 & 82.63\% \\
        4 & $F_{u,\mathrm{cl}}/F_{u,\mathrm{gal}} \geq 0.08$ & 1,717 & 1,534 & 89.34\% \\
          & \textit{thereof:} above det. limit & 1,591 & 1,457 & 91.58\% \\
        \hline
	\end{tabular}
\end{table}

\begin{figure}
    \centering
    \subfloat[\centering \textit{u}-band apparent magnitude. \label{fig:hsc_det_over_model_dev_eval_complete_purity_detail_UGRIZY_a}]{{\includegraphics[width=0.5\columnwidth]{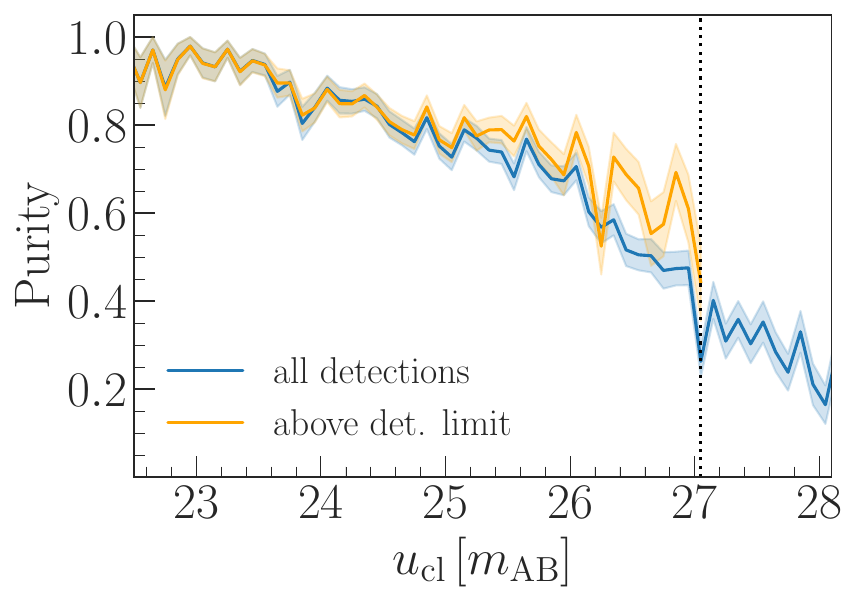} }}
    \subfloat[\centering Colour (\textit{u}-\textit{r}). \label{fig:hsc_det_over_model_dev_eval_complete_purity_detail_UGRIZY_b}]{{\includegraphics[width=0.5\columnwidth]{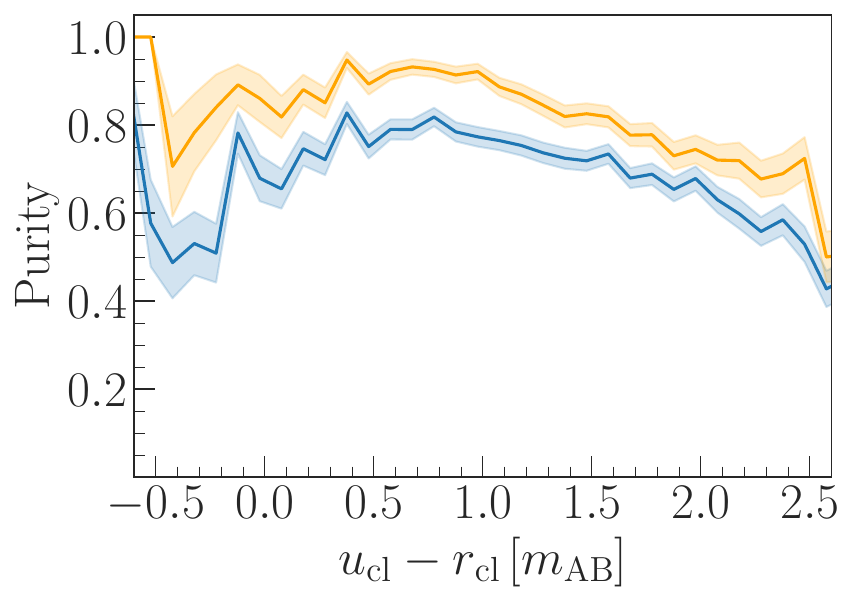} }}
    \\
    \subfloat[\centering Radial distance. \label{fig:hsc_det_over_model_dev_eval_complete_purity_detail_UGRIZY_c}]{{\includegraphics[width=0.5\columnwidth]{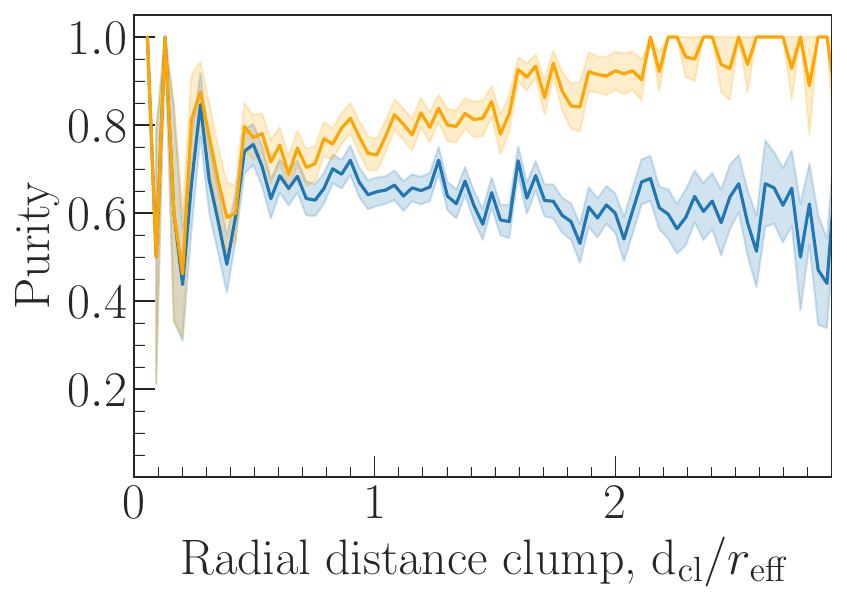} }}
    \subfloat[\centering Host galaxy redshift. \label{fig:hsc_det_over_model_dev_eval_complete_purity_detail_UGRIZY_d}]{{\includegraphics[width=0.5\columnwidth]{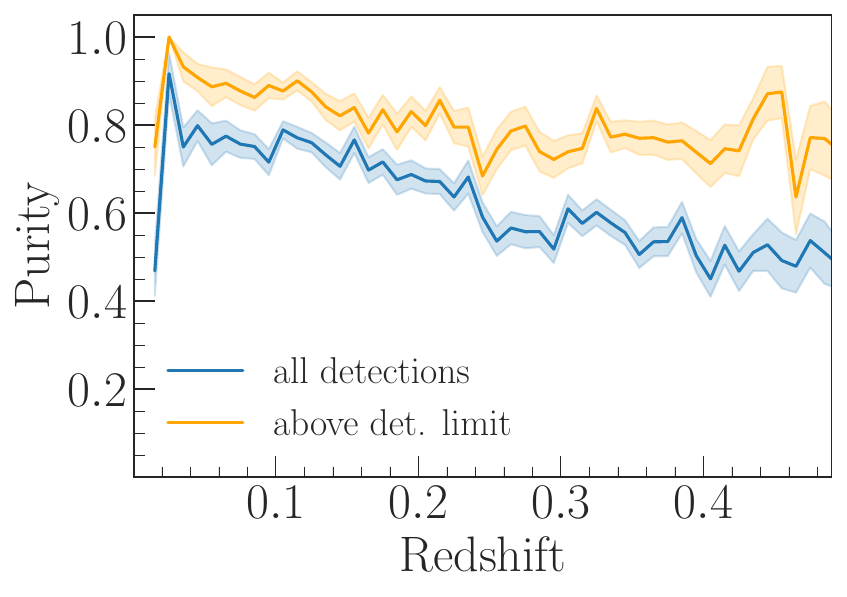} }}
    \\
    \subfloat[\centering Host galaxy stellar mass. \label{fig:hsc_det_over_model_dev_eval_complete_purity_detail_UGRIZY_e}]{{\includegraphics[width=0.5\columnwidth]{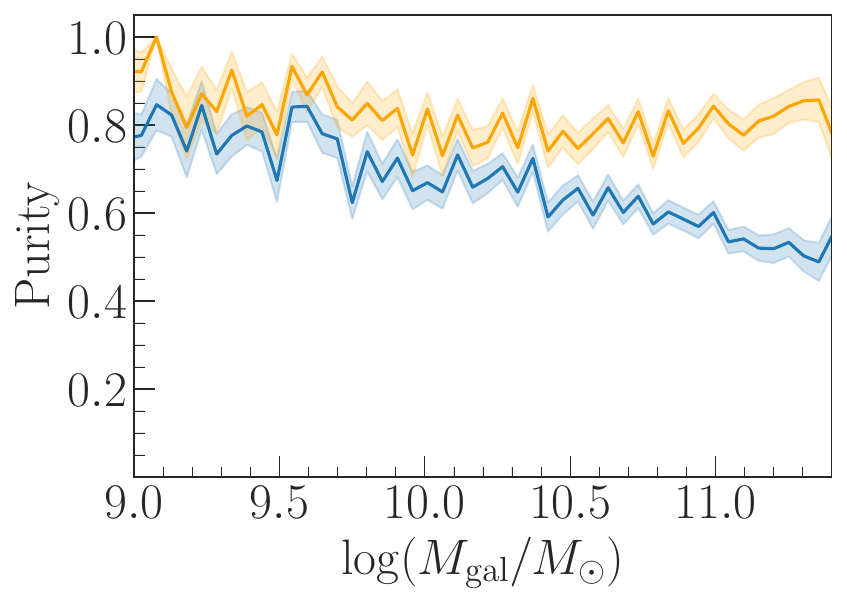} }}
    \subfloat[\centering Host galaxy sSFR. \label{fig:hsc_det_over_model_dev_eval_complete_purity_detail_UGRIZY_f}]{{\includegraphics[width=0.5\columnwidth]{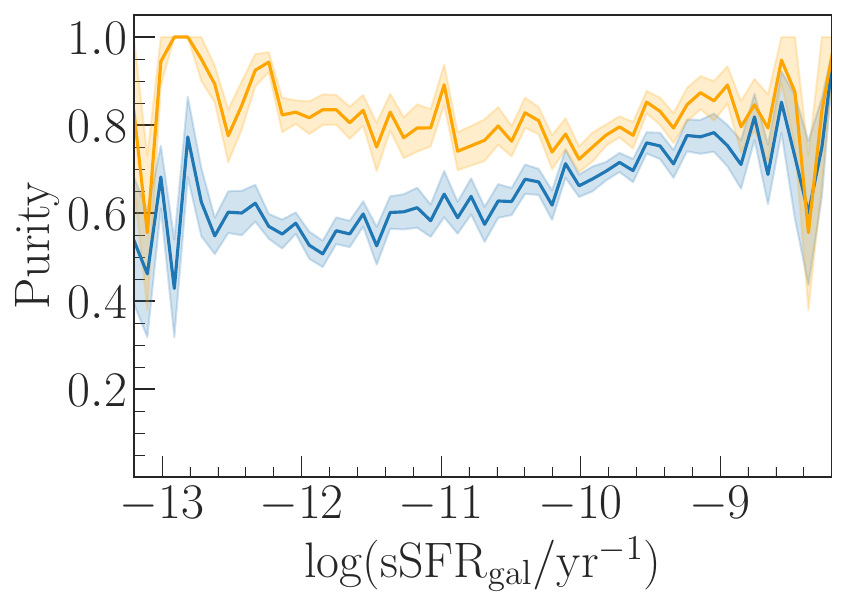} }}
    \caption[Purity as a function of different physical clump and host galaxy properties measured on simulated clumps for the 6-channel FRCNN model.]{Purity as a function of different physical clump and host galaxy properties measured on simulated clumps for the 6-channel FRCNN model. The purity is plotted in blue for the full sample of detected clumps and in orange for clumps that are brighter than the detection limit of the two surveys ($u\leq 27.1\,m_{\mathrm{AB}}$, $g\leq 26.5\,m_{\mathrm{AB}}$, $r\leq 26.1\,m_{\mathrm{AB}}$, $i\leq 25.9\,m_{\mathrm{AB}}$, $z\leq 25.1\,m_{\mathrm{AB}}$ and $y\leq 24.4\,m_{\mathrm{AB}}$). The shaded areas show the $1\sigma$ errors and the dotted vertical line in plot (a) marks the limiting magnitude in the \textit{u}-band for the CLAUDS survey.}
    \label{fig:hsc_det_over_model_dev_eval_complete_purity_detail_UGRIZY}
\end{figure}

Even with no objectness threshold applied (Section \ref{sec:frcnn_model_performance}), the purity is well above $80\%$ for those detected clumps that have flux measurements above the detection limit. A purity of $\gtrsim 80\%$ is also observed over different ranges of the physical clump and host galaxy properties (Figure \ref{fig:hsc_det_over_model_dev_eval_complete_purity_detail_UGRIZY}). Purity decreases with decreasing brightness of the clumps (Figure \ref{fig:hsc_det_over_model_dev_eval_complete_purity_detail_UGRIZY_a}) and is lower for clumps that have either very blue or red colours (Figure \ref{fig:hsc_det_over_model_dev_eval_complete_purity_detail_UGRIZY_b}). The detections become more pure if the detected clumps have fluxes above the detection limits and are further from the galaxy centre (Figure \ref{fig:hsc_det_over_model_dev_eval_complete_purity_detail_UGRIZY_c}), likely due to a more prominent contrast in the more diffuse outskirts of most galaxies.

Moreover, the purity of the detections appears to be more varied as a function of redshift (Figure \ref{fig:hsc_det_over_model_dev_eval_complete_purity_detail_UGRIZY_d}), stellar mass (Figure \ref{fig:hsc_det_over_model_dev_eval_complete_purity_detail_UGRIZY_e}) and sSFR of the host galaxy (Figure \ref{fig:hsc_det_over_model_dev_eval_complete_purity_detail_UGRIZY_f}) compared to what is observed for the completeness. Generally, the detections become less pure with increasing redshift, increasing stellar mass or decreasing sSFR of the host galaxy. This is partly compensated for if only detected clumps above the detection limit are considered and the observed purity for those clumps remains above $80\%$.

\subsubsection{Purity of the 5-channel (\textit{grizy}) model detections}\label{sec:hsc_det_over_model_dev_eval_complete_purity_GRIZY}
Using a similar method to that used when measuring the completeness for the 5-channel \textit{grizy} detection model (Section \ref{sec:hsc_det_over_model_dev_eval_complete_GRIZY}), we also measured the purity for the 5-channel FRCNN model. The procedure is identical to that applied for the 6-channel FRCNN model in Section \ref{sec:hsc_det_over_model_dev_eval_complete_purity_UGRIZY}, so that the observed purity for the \textit{grizy}-only data in Figure \ref{fig:hsc_det_over_model_dev_eval_complete_purity_GRIZY} and Table \ref{tab:hsc_det_over_model_dev_eval_complete_purity_GRIZY} can be directly compared to the purity measured for the \textit{ugrizy} data.

Overall purity for the detections from the 5-channel detection model is higher by $\sim 4.5\%$ compared to the detections from the 6-channel model. This changes if only detections with measured fluxes above the detection limits are taken into account. For those detections, the purity measurements from both models are almost identical i.e. $\gtrsim80\%$

\begin{table}
	\centering
	\caption[Purity of the 5-channel FRCNN model measured on simulated clumps.]{Purity of the 5-channel FRCNN model measured on simulated clumps. The purity is calculated for different selections: (1) all detected clumps, (2) only detected clumps that have \textit{grizy} fluxes greater then the detection limit in all filter bands, (3) detected clumps with \textit{g}-band fluxes $F_{g,\mathrm{cl}}$ of at least 3\% of the host galaxy \textit{g}-band flux $F_{g,\mathrm{gal}}$ ($F_{g,\mathrm{cl}}/F_{g,\mathrm{gal}} \geq 0.03$) and (4) 8\% ($F_{g,\mathrm{cl}}/F_{g,\mathrm{gal}} \geq 0.08$).}
    \label{tab:hsc_det_over_model_dev_eval_complete_purity_GRIZY}
	\footnotesize
        \begin{tabular}{clrrr}
		\hline
         & Selection & \multicolumn{1}{c}{Total} & \multicolumn{1}{c}{Detected} & \multicolumn{1}{c}{Purity} \\
         &           &                           & \multicolumn{1}{c}{sim. clumps} & \\
		\hline
        1 & All detected clumps & 7,619 & 6,151 & 80.73\% \\
        2 & Above det. limit & 5,115 & 4,168 & 81.49\% \\
        3 & $F_{g,\mathrm{cl}}/F_{g,\mathrm{gal}} \geq 0.03$ & 2,108 & 1,809 & 85.82\% \\
          & \textit{thereof:} above det. limit & 2,010 & 1,745 & 86.82\% \\
        4 & $F_{g,\mathrm{cl}}/F_{g,\mathrm{gal}} \geq 0.08$ & 849 & 774 & 91.17\% \\
          & \textit{thereof:} above det. limit & 835 & 767 & 91.86\% \\
        \hline
	\end{tabular}
\end{table}

\begin{figure}
    \centering
    \subfloat[\centering \textit{g}-band apparent magnitude. \label{fig:hsc_det_over_model_dev_eval_complete_purity_detail_GRIZY_a}]{{\includegraphics[width=0.5\columnwidth]{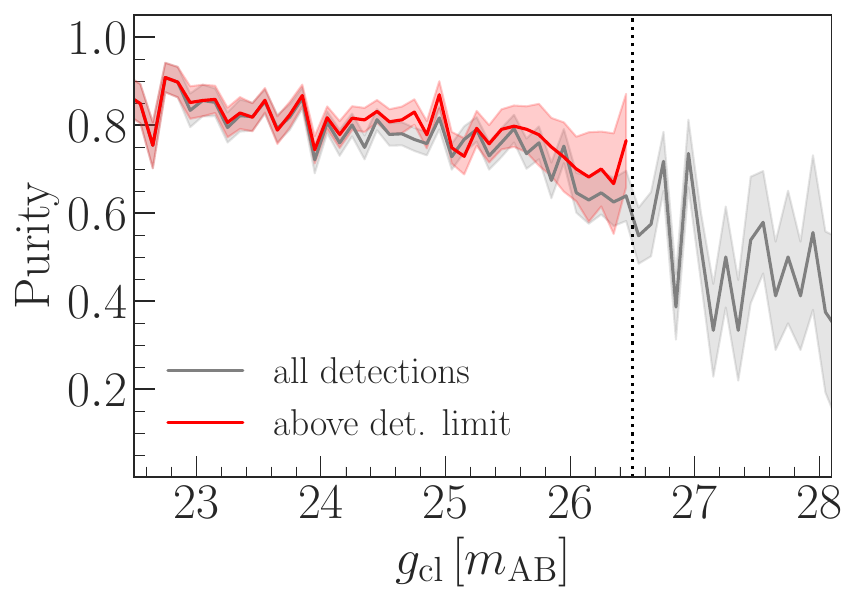} }}
    \subfloat[\centering Colour (\textit{g}-\textit{r}). \label{fig:hsc_det_over_model_dev_eval_complete_purity_detail_GRIZY_b}]{{\includegraphics[width=0.5\columnwidth]{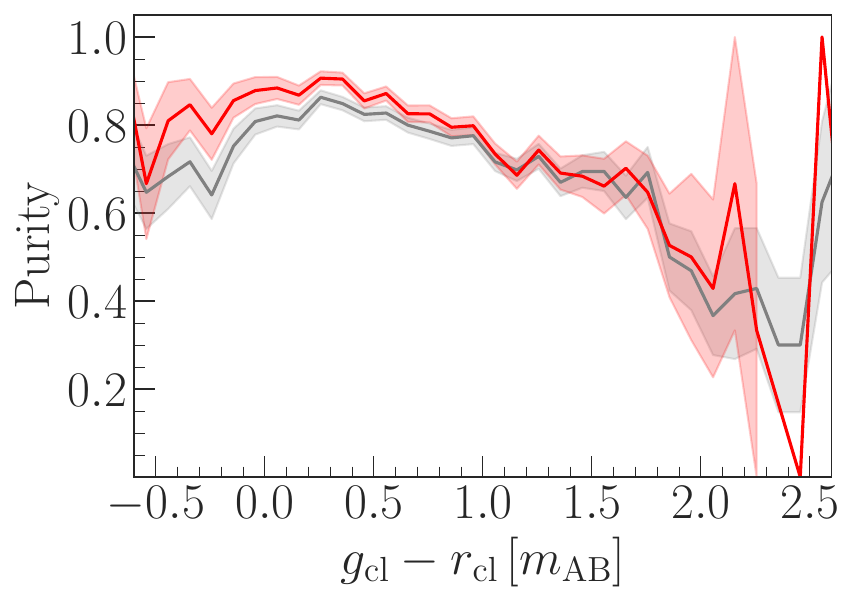} }}
    \\
    \subfloat[\centering Radial distance. \label{fig:hsc_det_over_model_dev_eval_complete_purity_detail_GRIZY_c}]{{\includegraphics[width=0.5\columnwidth]{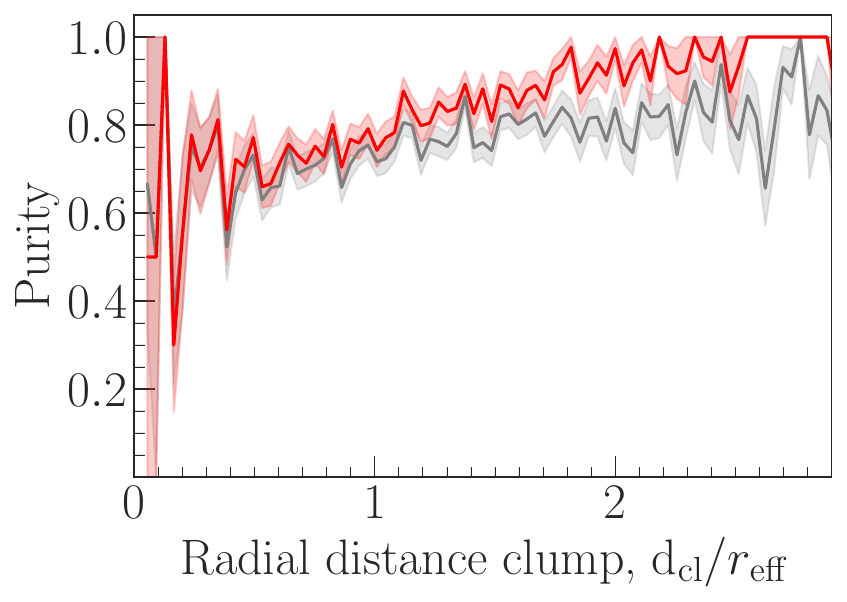} }}
    \subfloat[\centering Host galaxy redshift. \label{fig:hsc_det_over_model_dev_eval_complete_purity_detail_GRIZY_d}]{{\includegraphics[width=0.5\columnwidth]{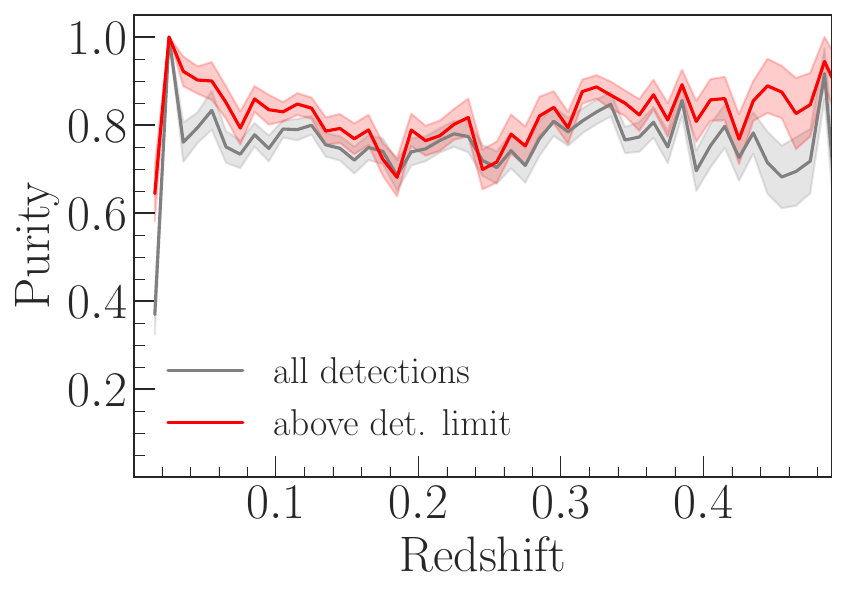} }}
    \\
    \subfloat[\centering Host galaxy stellar mass. \label{fig:hsc_det_over_model_dev_eval_complete_purity_detail_GRIZY_e}]{{\includegraphics[width=0.5\columnwidth]{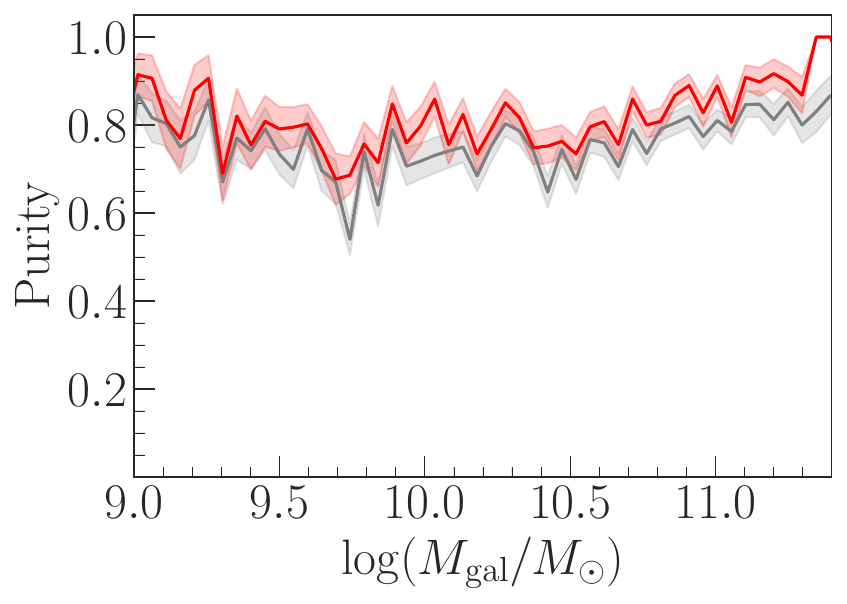} }}
    \subfloat[\centering Host galaxy sSFR. \label{fig:hsc_det_over_model_dev_eval_complete_purity_detail_GRIZY_f}]{{\includegraphics[width=0.5\columnwidth]{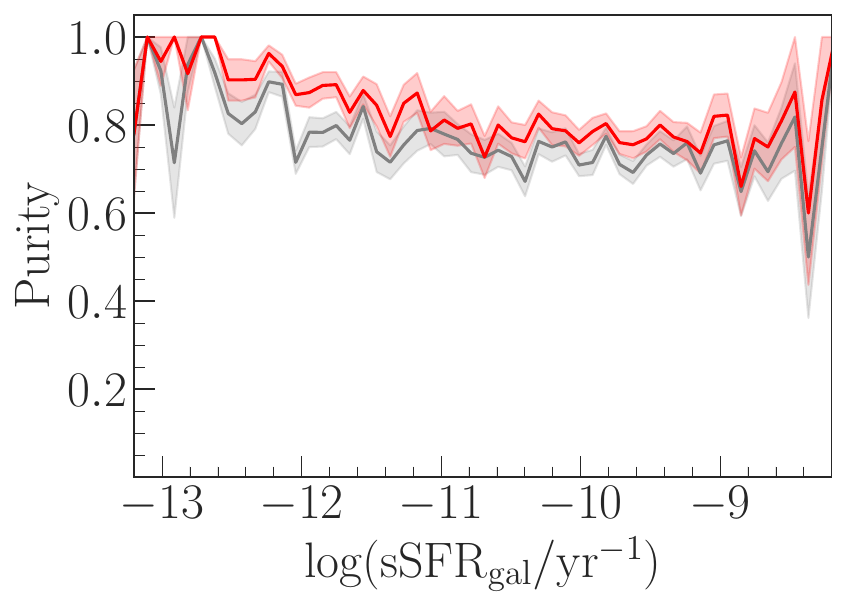} }}
    \caption[Purity as a function of different physical clump and host galaxy properties measured on simulated clumps for the 5-channel FRCNN model.]{Purity as a function of different physical clump and host galaxy properties measured on simulated clumps for the 5-channel FRCNN model. The purity is plotted in blue for the full sample of detected clumps and in orange for clumps that are brighter than the detection limit for the HSC-SSP filter bands ($g\leq 26.5\,m_{\mathrm{AB}}$, $r\leq 26.1\,m_{\mathrm{AB}}$, $i\leq 25.9\,m_{\mathrm{AB}}$, $z\leq 25.1\,m_{\mathrm{AB}}$ and $y\leq 24.4\,m_{\mathrm{AB}}$). The shaded areas show the $1\sigma$ errors and the dotted vertical line in plot (a) marks the limiting magnitude in the \textit{g}-band for the HSC-SSP survey.}
    \label{fig:hsc_det_over_model_dev_eval_complete_purity_detail_GRIZY}
\end{figure}

When measured as a function of the different clump and galaxy properties, the purity of the 5-channel model detections follows similar trends to the 6-channel detections (Figure \ref{fig:hsc_det_over_model_dev_eval_complete_purity_detail_GRIZY}). However, the decline in purity with increasing redshift, increasing stellar mass and decreasing sSFR seen in the \textit{ugrizy} results (Figs. \ref{fig:hsc_det_over_model_dev_eval_complete_purity_detail_GRIZY_d}, \ref{fig:hsc_det_over_model_dev_eval_complete_purity_detail_GRIZY_e} and \ref{fig:hsc_det_over_model_dev_eval_complete_purity_detail_GRIZY_f}) is not replicated for the 5-channel detections for which the purity measured for all detections closely follows the trend of the purity that is measured for detections with fluxes above the instrument-specific detection limits.

\section{Discussion}\label{sec:discussion}
Using galaxy images that were annotated with manually corrected clump markings from a small prototype clump detection model, we have developed a novel architecture for a DL-based object detection model that builds on the FRCNN architecture and uses six imaging channels as input instead of the usual three RGB-channels. 

While developing the final models, we also tested unmodified model architectures that use different versions of the RGB-composite images (Section \ref{sec:data_preprocess}) as input. Subsequent tests comparing models that use only three imaging channels as input with models that use five or six channels showed that precision and recall are increased by $\sim15$ to 25\% if additional information is made available to the detection model through the two or three additional filter bands. This is especially true for the added \textit{u}-band data from the CLAUDS survey, which is more sensitive to recent star-formation and increases the number of detections compared to a model without \textit{u}-band data.


The detection performance of the models was further increased by allowing the model to identify objects from more classes than the two that we used in previous works \citep[i.e. normal clump and `odd' clump, see][]{Popp2024}. If the training labels are changed to contain only those two classes and the FRCNN model from Section \ref{sec:frcnn_model_arch} is retrained to predict only those classes, precision is lower by $\sim 10$ to 12\% compared to the values in Table \ref{tab:hsc_det_over_model_dev_eval}. The recall is also significantly reduced by $\sim 2$ to 4\% but the main effect of adding additional classes is on the precision of the detections.

Another way to improve the model performance might be by using different feature extraction backbones. However, \citet{Popp2024} have shown that feature extractors that are not pretrained on astrophysical data would either require larger training samples or might not reach the same detection performance that is achieved by a domain-specific feature extractor like \textsc{Zoobot} \citep{Walmsley2023}. Currently, no alternative to \textsc{Zoobot} exists that can be used in astronomy as a foundation deep learning model for fine-tuning or downstream tasks related to galaxy morphologies. 

Furthermore, a principal challenge to supervised DL approaches like the one presented here, is the training data. The FRCNN model is trained to detect objects that are similar to those identified visually by human beings. This bears the risk that the training data might not be complete or contains wrongly labelled objects and any biases would be propagated into the fully trained model. For example, as only a few nearby and better resolved galaxies ($z\lesssim0.05$) were available for training our models, some fainter and less extended clump candidates might have been missed (see galaxy example with object-ID 42090026665789815 in Fig. \ref{fig:hsc_det_over_model_dev_eval_GRIZY_UGRIZY} and Fig. \ref{fig:failures} in the Appendix \ref{sec:failures}). Training data that will be based on galaxy images with higher spatial resolution (e.g. \textit{Euclid}) will help to improve the detections performance of a future version of our model. Alternatively, the training data for the models can be generated using simulated galaxies and/or simulated clumps that are injected into the galaxy images \citep[e.g.][]{Burke2019,HuertasCompany2020,Ginzburg2021}. However, some recent studies point out that training models on simulated data can lead to a worse model performance in comparison with models trained on real data \citep[e.g.][]{PearceCasey2025}. This is not necessarily true for ML-based approaches in general but models that use CNNs might be sensitive to the subtle differences between simulated data and real observational data. Currently, there is neither a complete picture nor consensus regarding the effectiveness of simulated or real training data and different systematic effects need to be considered, i.e. training on simulated (real) data and model application on real (simulated) data. With the availability of a foundation model like \textsc{Zoobot}, which is trained exclusively on real data, downstream tasks that include fine-tuning on simulated data might also be worth exploring.

In comparison with more traditional clump detection methods that are based on contrast-enhanced images in combination with peak finding algorithms and that are not yet suited to large scale and automated detections \citep{Haigh2021}, a DL-based object detection model is significantly faster and can be applied to large numbers of galaxy images while taking imaging data from different wavelength ranges into account simultaneously. The detection completeness of our model is high but because every clump study uses a different approach to test the completeness of their detection method, a direct comparison with previous work is difficult. However, the large sample of simulated clumps that we have used to measure the completeness of our models suggests that the main limiting factor is not the detection method but rather the sensitivity of the observing instrument. 

Given that purity and completeness for the 5-channel FRCNN model remain well above 80 to 90\% and that both metrics are also very similar to the purity and completeness measured on the simulated clumps for the 6-channel FRCNN model, we argue that clump detection with the FRCNN models can be extended onto the sample with \textit{grizy}-only imaging data without significant changes to the detection performance down to a limiting magnitude defined by the average $5\sigma$ depth limits of the CLAUDS and HSC-SSP surveys. However, we note that due to the missing \textit{u}-band some clumps, specifically those with very recent star-formation, can be missed by the 5-channel FRCNN model.

A limiting factor for our clump detection approach is the spatial resolution of the observations and the use of broadband imaging data. Our clump candidates are selected as compact systems that can potentially consist of multiple blended objects and that are specifically bright in the \textit{u}- and \textit{g}-filter bands compared to the diffuse background light of their host galaxies. As pointed out by \citet{Fisher2017,Tamburello2017,DessaugesZavadsky2017,DessaugesZavadsky2018,Meng2020} and \citet{HuertasCompany2020}, the sensitivity and spatial resolution of the imaging data affects the detection of clumps and the measurements of their physical properties and sizes. Furthermore, the star-forming nature of our detections needs to be confirmed by spectroscopic star-formation indicators, e.g. by H$\alpha$-emission lines. However, the development of our clump detection model has been motivated by the large amount of photometric data that is available from current and future wide surveys (e.g. \textit{Euclid}) while spectroscopic data with a similar spatial resolution is still limited in comparison to our sample of 710,271 HSC-SSP galaxies at $z<0.5$. A further analysis of our clump candidates based on their inferred physical properties through SED fitting (Popp et al. 2026c, submitted to MNRAS) indicate that most of our clump detections are indeed star-forming regions with elevated sSFRs. 

\section{Conclusions}\label{sec:conclusion}
In this paper, we have presented the development of an advanced version of the FRCNN object detection model for detecting possible star-forming clumps in low-redshift galaxies observed by the CLAUDS and HSC-SSP Wide surveys. The key improvements of the model were: 
\begin{enumerate}
\item extending the input to accept imaging data with more than the three channels so that up to six images from different photometric filter bands can be processed simultaneously, 
\item increasing the number of object classes and thereby upgrading from a binary classification scheme to a classification scheme that also includes multiple contaminating objects and
\item integrating the latest pretrained \textsc{Zoobot} CNN as a feature extraction backbone. 
\end{enumerate}
By using the astrophysical foundation model \textsc{Zoobot} in a downstream task for object detection, the FRCNN model only needed to be trained on a relatively small set of $\sim$3,200 labelled galaxy images. We validated the detection performance on a set of $\sim$13,800 CLAUDS/HSC-SSP galaxies into which $\sim$32,200 simulated clumps were injected. For simulated clumps that are brighter than the $5\sigma$ point source depths of the instruments, the model achieves a detection completeness of $\gtrsim 0.9$ and purity of $\gtrsim 0.8$.

We applied the model to our set of $\sim$14,000 CLAUDS/HSC-SSP galaxy images with six filter band photometry (\textit{ugrizy}) and to our set of $\sim$700,000 HSC-SSP galaxy images with only five filter band photometry (\textit{grizy}). In total, the model detected $\sim$1.5 million clump candidates that are made public as a catalogue \citep{Popp2026} containing the detected clump candidates, their measured photometry and estimated physical properties (Popp et al. 2026b, submitted to MNRAS and Popp et al. 2026c, submitted to MNRAS).

\section*{Acknowledgements}
We thank the anonymous referees for their useful and constructive comments that led to improvements in this manuscript.

JJP acknowledges funding from the Science and Technology Facilities Council (STFC) Grant Code ST/X508640/1. HD and SS acknowledge funding via the ELSA project. ``ELSA: Euclid Legacy Science Advanced analysis tools'' (Grant Agreement no. 101135203) is funded by the European Union. Views and opinions expressed are however those of the author(s) only and do not necessarily reflect those of the European Union or Innovate UK. Neither the European Union nor the granting authority can be held responsible for them. UK participation is funded through the UK Horizon guarantee scheme under Innovate UK grant 10093177. LFF acknowledges partial support from NASA awards 80NSSC24K1277 and 80NSSC20M0057. BDS acknowledges support through a UK Research and Innovation Future Leaders Fellowship [grant number MR/T044136/1] and its renewal [grant number MR/Z000076/1]. 

The Dunlap Institute is funded through an endowment established by the David Dunlap family and the University of Toronto.

This research made use of the open-source Python scientific computing ecosystem, including \textsc{NumPy} \citep{Harris2020}, \textsc{Matplotlib} \citep{Hunter2007}, \textsc{seaborn} \citep{Waskom2021} and \textsc{Pandas} \citep{McKinney2010}. This research made use of \textsc{Astropy}, a community-developed core Python package for Astronomy \citep{astropy2022} and the \textsc{Photutils} Python package \citep{LarryBradley2025}. The Python framework \textsc{PyTorch} \citep{Paszke2019} was used for the DL model development.

This publication uses data generated via the Zooniverse.org platform, development of which is funded by generous support, including a Global Impact Award from Google, and by a grant from the Alfred P. Sloan Foundation. We extend an enormous thank you to the volunteers who participated in the Galaxy Zoo: Clump Scout project. It is the
efforts of these volunteers that made all of this work possible.

The authors acknowledge the Minnesota Supercomputing Institute (MSI, \url{https://www.msi.umn.edu/}) at The University of Minnesota for providing high-performance computing (HPC) resources that have contributed to the research results reported within this paper.

\section*{Data Availability}
The training data and Python code examples for the FRCNN models and the adjusted feature extraction backbone are available from \citet{Popp2025} and a public Github repository\footnote{\url{https://github.com/ou-astrophysics/Zoobot-for-image-segmentation-and-object-detection}}. The catalogue of star-forming clumps, including the measured photometry and estimated physical properties, is available from \citet{Popp2026}.

We caution the reader that we did not apply any survey completeness limits to this catalogue. Such completeness limits need to be applied for further scientific analysis of our sample. For an example, where galaxy stellar mass and redshift limits were applied, see Popp et al. (2026b, submitted to MNRAS) and Popp et al. (2026c, submitted to MNRAS). We recommend applying a selection criteria for a mass-complete galaxy sample and excluding all galaxies that are close to edge-on (e.g. with an elongation $>3.0$). The catalogue also offers the possibility to limit the sample of host galaxies to only SFGs or to galaxies with spectroscopically derived redshift measurements to limit the effects of redshift uncertainties. Furthermore, all clumps that are too close to the galaxy centre ($\mathrm{d}_{\mathrm{cl}} \lesssim 0.3\,r_{\mathrm{eff}}$) should be excluded. Purity and completeness can be varied by filtering on the objectness or detection score, where a higher score threshold results in higher purity and vice versa.

\section*{Conflicts of Interest}
The authors declare no conflict of interest.



\bibliographystyle{mnras}
\bibliography{LIB_Clumpy_Galaxies}




\appendix
\section{Galaxy segmentation map}\label{sec:hsc_data_gal_extent}
In order to be able to analyse the star-forming regions that are located within the galaxy, we created a segmentation map of each galaxy that separates the image pixels which belong to the galaxy from image pixels that are not part of the target galaxy but can severely blend with the main object (i.e. sky background, foreground and background objects). The galaxy segmentation map was created in multiple steps on the \textit{r}-filter band image of the galaxy and following a modified version of the approach used by \citet{Galametz2013} and \citet{Sazonova2021}. In each of the following steps we used the image segmentation routines \texttt{detect\_sources} and \texttt{deblend\_sources} available from the Python library \textsc{Photutils} \citep{LarryBradley2025} with different parameters:
\begin{enumerate}
    \item `hot' mode step:
    \begin{description}
        \item \texttt{data} = science image convolved with a 2-dimensional Tophat filter kernel with a radius of 5 pixels
        \item \texttt{threshold} = 97th percentile value of all pixel values in the science image
        \item \texttt{npixels} = 1 pixel
        \item \texttt{mask} = circular mask with a radius $\mathrm{r}=R_{P,90\%}$
    \end{description}
    \item `cool' mode step:
    \begin{description}
        \item \texttt{data} = science image convolved with a 2-dimensional Tophat filter kernel with a radius of 5 pixels
        \item \texttt{threshold} = standard deviation of all pixel values in the science image
        \item \texttt{npixels} = $1\,\mathrm{arcsec}^2$
        \item \texttt{nlevels} = 32
        \item \texttt{contrast} = $10^{-6}$
    \end{description}
    \item `cold' mode step:
    \begin{description}
        \item \texttt{data} = science image convolved with a 2-dimensional Tophat filter kernel with a radius of 5 pixels
        \item \texttt{threshold} = standard deviation of all pixel values in the science image
        \item \texttt{npixels} = $0.01\times R_{P,90\%}^2$
        \item \texttt{mask} = object mask from `cool' mode
    \end{description}
\end{enumerate}

\begin{figure*}
    \centering
    \subfloat[\centering Object: 42090013780899677 ($z=0.066$). \label{fig:hsc_data_gal_extent_a}]{{\includegraphics[width=0.8\textwidth]{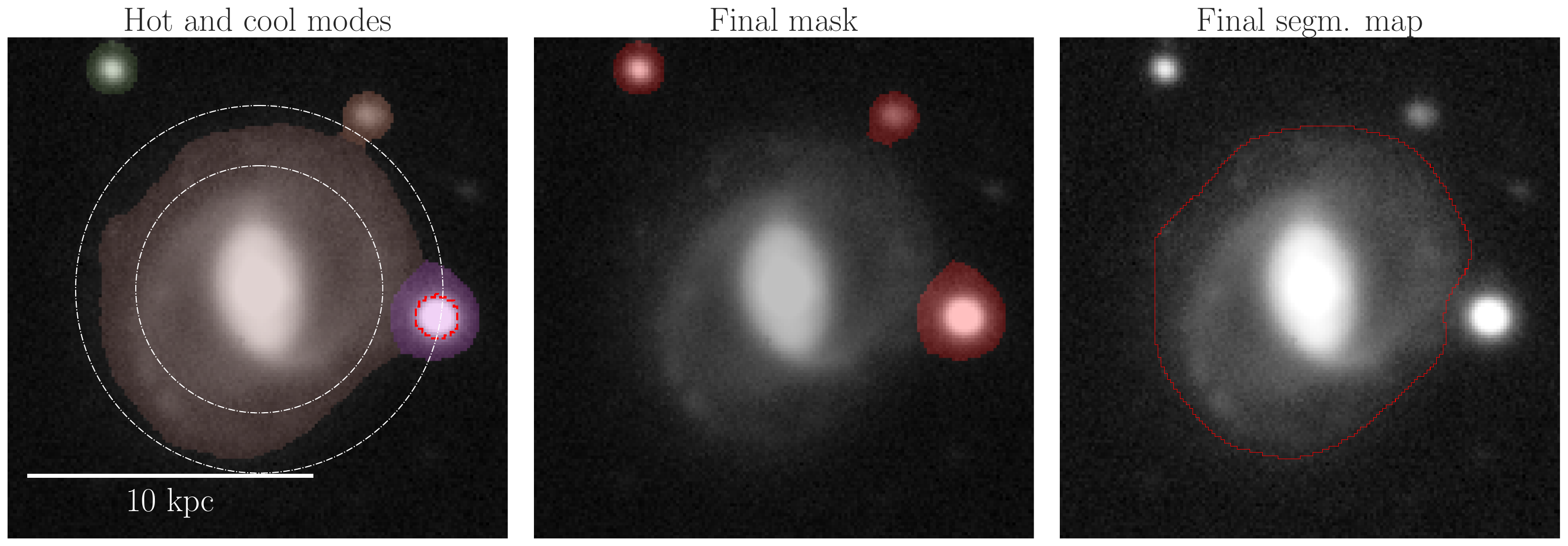} }}
    \\
    \subfloat[\centering Object: 43153778395866061 ($z=0.032$). \label{fig:hsc_data_gal_extent_b}]{{\includegraphics[width=0.8\textwidth]{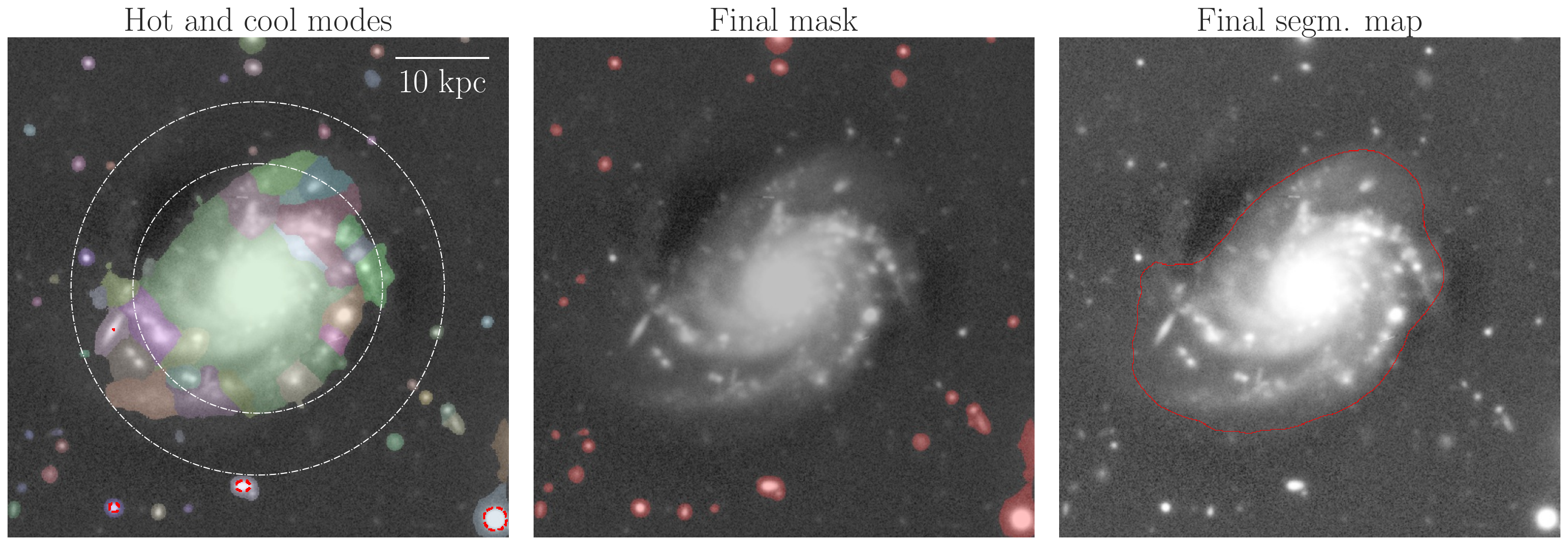} }}
    \caption[Illustration of the different steps applied to create the galaxy segmentation maps.]{Steps applied to create the galaxy segmentation maps illustrated with two galaxy examples. The first panels show the segmentation masks generated during the `hot' and `cool' mode steps. The dashed white circles are outlining the applied circular masks with radii of 1.0 and 1.5 times the SDSS \textit{r}-band 90\% Petrosian radius. Shaded areas are the detected segments in arbitrary colours and red dashed shapes outline the bright segments detected during the `hot' mode step. The final galaxy segmentation map outline is shown as a red shape in the last panels.}
    \label{fig:hsc_data_gal_extent}
\end{figure*}

In the first step (`hot' mode), bright regions in the cutout image were detected by applying a high detection threshold (\texttt{threshold}) so that only pixels with intensity values above the 97th percentile of all pixel values were selected from the input image (\texttt{data}). The input image is a modified version of the science image that is smoothed using a 2-dimensional and isotropic Tophat filter kernel with a radius of 5 pixels. The minimum area, restricted by the minimum number of connected pixels (\texttt{npixels}), is allowed to be small so that only the peak intensity regions were selected in this step. The detected regions were used to mask bright contaminating objects (e.g. foreground stars). From this mask, we excluded bright regions that fall within a circular mask with a radius equal to the SDSS \textit{r}-band 90\% Petrosian radius ($\mathrm{r}=R_{P,90\%}$) around the centre of the target galaxy to avoid excluding possible clump detections (first panels in Figs. \ref{fig:hsc_data_gal_extent_a} and \ref{fig:hsc_data_gal_extent_b}).

We then applied a second segmentation and deblending iteration to the smoothed science image (`cool' mode). The detection threshold was lowered to accept pixels that have intensity values above the mean plus standard deviation of all pixel values in the science image. The detected pixels were required to form extended areas with a size of at least $1\,\mathrm{arcsec}^2$ and were deblended into separate regions using the deblending procedure from \textsc{SExtractor} \citep{Bertin1996} that is implemented by \textsc{Photutils}. The values for \texttt{nlevels} and \texttt{contrast} were chosen to separate areas from the target galaxy that are larger than bright point-like sources but are still blended with low-surface brightness features of the main object.

From the regions detected and deblended during the `cool' mode step, the central largest segment and all segments that are located outside a radius of 1.5 times the SDSS \textit{r}-band 90\% Petrosian radius ($\mathrm{r}=1.5\,R_{P,90\%}$) were discarded but the others were kept. Furthermore, we kept all regions that fully overlap with the regions found during the `hot' mode step. The resulting set of regions are likely those areas in the image cutout that are either bright contaminants or blended areas not connected to the target galaxy (second panels in Figs. \ref{fig:hsc_data_gal_extent_a} and \ref{fig:hsc_data_gal_extent_b}).

This set of regions was used to mask the smoothed science image to which a last segment detection iteration was applied (`cold' mode). The area threshold (\texttt{npixels}) was changed to a lower value than for the `cool' mode step to include also smaller sources that were missed during the previous step. An additional deblending iteration of the detected segments was not required as all contaminants were expected to be excluded by the applied mask. We kept only the large central segmentation map that corresponds to the target galaxy's extent from the detected segments. In a final step, we smoothed the outline of the galaxy segmentation map using a majority filter kernel with a footprint equal to 10\% of the image size (third panels in Figs. \ref{fig:hsc_data_gal_extent_a} and \ref{fig:hsc_data_gal_extent_b}).

\section{Clump photometry}\label{sec:photometry}
The details of our photometry measurements of the detected simulated and real clumps are outlined in Popp et al. (2026b, submitted to MNRAS). Here, we provide a brief summary for convenience.

We used aperture photometry to recover the fluxes of the detected simulated clumps at their locations in the \textit{u}-band images, while holding the positions constant for the other \textit{g}-, \textit{r}-, \textit{i}-, \textit{z}- and \textit{y}-band images (forced photometry). We estimated the diffuse galaxy background light from an annulus around the aperture and subtracted the median value per pixel from the aperture pixel values. Furthermore, we masked the area outside the host galaxy extent and adjacent clump detections as their light would otherwise contaminate the background estimate if they are (partly) located within the annulus. The adjacent clump locations were masked out with a circular mask that blocks the per-pixel flux of the adjacent clumps. 

We set the sizes for the aperture radius ($r_{\mathrm{ap}}$), annulus radii ($r_{\mathrm{an}}$) and clump mask radius ($r_{\mathrm{mask}}$) as multiples of the filter band-specific seeing full width at half maximum (FWHM) so that $r_{\mathrm{ap}}=0.25\, \mathrm{FWHM}_{\mathrm{seeing}}$, $r_{\mathrm{an}}=1.5-2.0\, \mathrm{FWHM}_{\mathrm{seeing}}$ and $r_{\mathrm{mask}}=0.5\, \mathrm{FWHM}_{\mathrm{seeing}}$. We further corrected the measured flux for the flux that is not included within the aperture because of the extended shape of the PSF and evaluated the two-dimensional PSF models for each filter band and galaxy image over the area of the aperture. The sum of the pixel-values enclosed by the aperture in relation to the sum over the total area of the PSF model is used as the aperture correction factor.

\section{Examples of failed or misclassified detections}\label{sec:failures}
\begin{figure*}
    \centering
    \includegraphics[width=0.95\textwidth]{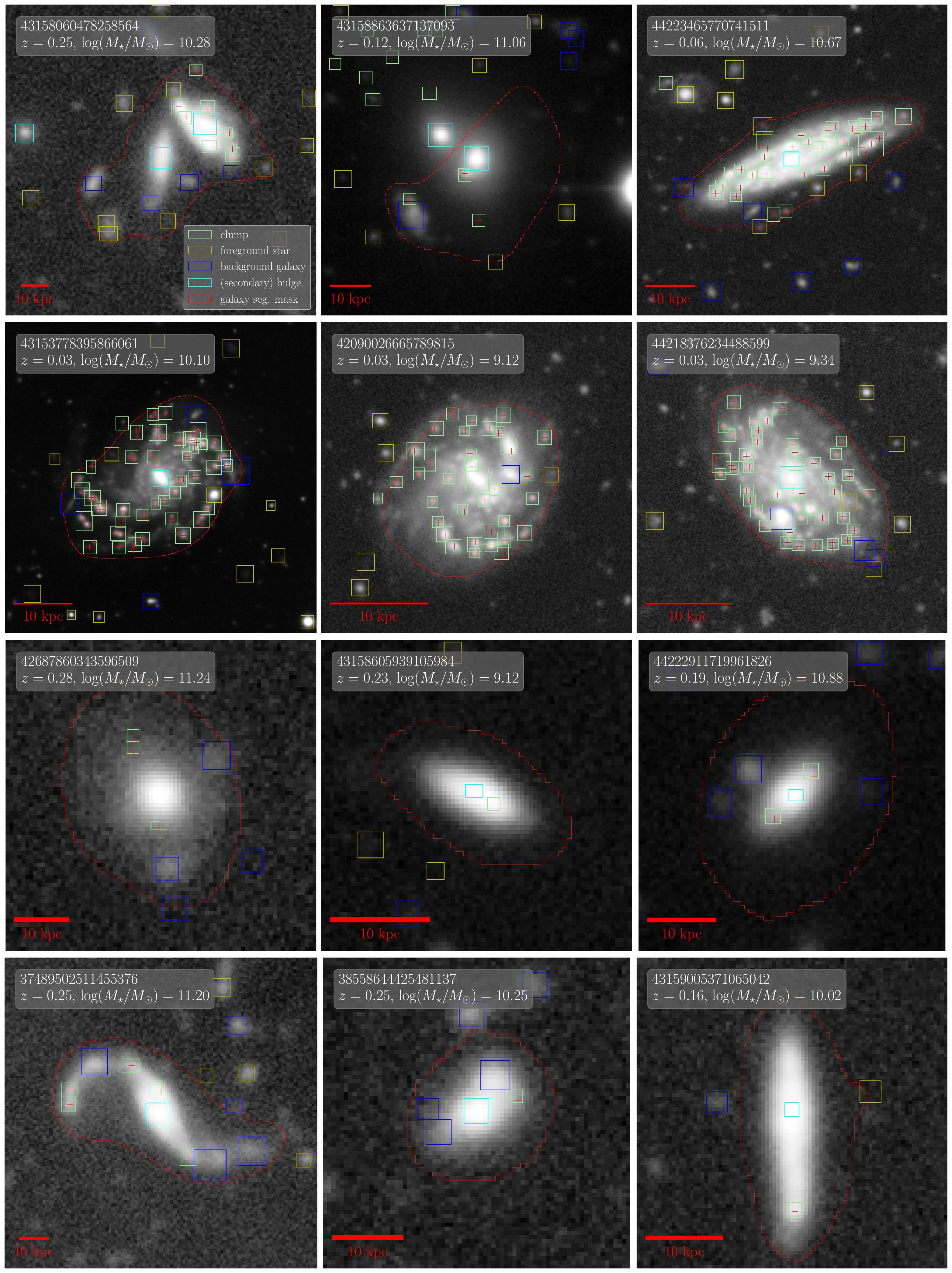}
    \caption[Examples of failed or misclassified detections.]{Examples of failed or misclassified detections: failed galaxy segmentation mask (first row), undetected potential star-forming regions that are too faint (second row) and potential misclassified and false positive detections (third and fourth row). The model detections are shown as boxes where the colour indicates the object class. Flux peaks are marked with red crosses. The galaxies are shown with their \textit{u}-band images.}
    \label{fig:failures}
\end{figure*}

We show some examples where our clump detection model and postprocessing failed to detect potential star-forming clumps or resulted in potential misclassified and false positive detections in Figure \ref{fig:failures}. To estimate the fraction of clump candidates that were either missed or included as false positives in our detection sample, we visually inspected 579 galaxies with 2,246 clump detections. We found 25 clump candidates ($1.1\%$) that are located in low-surface brightness parts of the host galaxy but were either wrongly in- or excluded by the host galaxy's segmentation map (first row of galaxy images in Fig. \ref{fig:failures}). We also found 37 clump candidates ($1.6\%$) that are either located in fore-/background galaxies which are blended with the target galaxy or in galaxies that appear to be in the process of merging with the target galaxy. In such cases, the algorithm to segment the target galaxy from surrounding objects can fail and result in clump detections that do not belong to the target galaxy. Furthermore, we found 45 clump candidates ($2.0\%$) that appear to spurious or misclassified detections (third and fourth row of galaxy images in Fig. \ref{fig:failures}). The fraction of false negative detections are difficult to determine. However, these false negative detection occur predominantly in resolved, very nearby galaxies and are limited to the 183 galaxies ($1.3\%$) at redshifts $z\lesssim0.05$ of our sample (second row of galaxy images in Fig. \ref{fig:failures}).

\bsp	
\label{lastpage}
\end{document}